\documentclass[10pt,twocolumn,showpacs,preprintnumbers,amsmath,amssymb,aps,prb,longbibliography,superscriptaddress]{revtex4-2}
\usepackage{mathrsfs}
\usepackage{graphicx}
\usepackage{dcolumn}
\usepackage{bm}
\usepackage{amsmath}
\usepackage{amsfonts}
\usepackage{color}
\usepackage{hyperref}

\usepackage[utf8]{inputenc}
\usepackage{array}
\usepackage{verbatim}
\usepackage{multirow}
\usepackage{amsmath}
\usepackage{amssymb}
\usepackage{dsfont}
\usepackage{graphicx}
\usepackage{esint}
\usepackage{wasysym}

\newcommand{\normord}[1]{:\mathrel{#1}:}

\newcommand{\PreserveBackslash}[1]{\let\temp=\\#1\let\\=\temp}
\newcolumntype{C}[1]{>{\PreserveBackslash\centering}p{#1}}
\newcolumntype{R}[1]{>{\PreserveBackslash\raggedleft}p{#1}}
\newcolumntype{L}[1]{>{\PreserveBackslash\raggedright}p{#1}}

\usepackage{cleveref}
\crefname{equation}{Eq.}{Eqs.}
\crefname{figure}{Fig.}{Figs.}
\crefname{table}{Table}{Tables}
\crefname{section}{Section}{Sections}

\renewcommand{\paragraph}[1]{\vspace{0.2cm}{\bf \textit{#1}}}
\def\ie{{\it i.e.},\ }

\usepackage{cleveref}
\crefname{equation}{Eq.}{Eqs.}
\crefname{figure}{Fig.}{Figs.}
\crefname{table}{Table}{Tables}
\crefname{section}{Section}{Sections}

\newcommand{\mbf}{\mathbf}

\newcommand{\mcl}{\mathcal}

\newcommand{\mrm}{\mathrm}

\newcommand{\td}{\widetilde}

\def\beq#1\eeq{\begin{equation}#1\end{equation}}
\def\beqs#1\eeqs{\begin{align}#1\end{align}}

\def\pare#1{\left( #1 \right)}

\def\bra#1{\langle #1 |}
\def\ket#1{| #1 \rangle}

\def\inn#1{\langle #1 \rangle}

\def\nono{\nonumber}

\def\dg{\dagger}
\def\up{\uparrow}
\def\down{\downarrow}
\def\pr{\prime}

\def\Re{\mathrm{Re}}
\def\Im{\mathrm{Im}}

\def\kk{\mathbf{k}}
\def\qq{\mathbf{q}}
\def\pp{\mathbf{p}}
\def\pp{\mathbf{p}}
\def\GG{\mathbf{G}}
\def\QQ{\mathbf{Q}}
\def\RR{\mathbf{R}}
\def\tt{\mathbf{t}}
\def\dd{\mathbf{d}}
\def\aa{\mathbf{a}}

\def\ee{\epsilon}
\def\CC{\mathcal{P}}

\makeatother

\begin{document}

\title{Twisted bilayer graphene III. Interacting Hamiltonian and exact symmetries}

\author{B. Andrei Bernevig \thanks{\href{bernevig@princeton.edu}{bernevig@princeton.edu}}}
\affiliation{Department of Physics, Princeton University, Princeton, New Jersey 08544, USA}
\author{Zhi-Da Song}
\affiliation{Department of Physics, Princeton University, Princeton, New Jersey 08544, USA}
\author{Nicolas Regnault}
\affiliation{Department of Physics, Princeton University, Princeton, New Jersey 08544, USA}
\affiliation{Laboratoire de Physique de l'Ecole normale superieure, ENS, Universit\'e PSL, CNRS,
Sorbonne Universit\'e, Universit\'e Paris-Diderot, Sorbonne Paris Cit\'e, Paris, France}
\author{Biao Lian
\thanks{\href{biao@princeton.edu}{biao@princeton.edu}}}
\affiliation{Department of Physics, Princeton University, Princeton, New Jersey 08544, USA}

\begin{abstract}
We derive the explicit Hamiltonian of twisted bilayer graphene (TBG) with Coulomb interaction projected into the flat bands, and study the symmetries of the Hamiltonian. First, we show that all projected TBG Hamiltonians can be written as Positive Semidefinite Hamiltonian, the first  example of which was  found in \cite{kang_strong_2019}. We then prove that the interacting TBG Hamiltonian exhibits an exact U(4) symmetry in the exactly flat band (nonchiral-flat) limit. 
We further define, besides a first chiral limit where the AA stacking hopping is zero, a new second chiral limit where the AB/BA stacking hopping is zero. In the first chiral-flat limit (or second chiral-flat limit) with exactly flat bands, the TBG is enhanced to have an exact U(4)$\times$U(4) symmetry, whose generators are different between the two chiral limits. While in the first chiral limit and in the non-chiral case these symmetries have been found in Ref.~\cite{bultinck_ground_2020} for the $8$ lowest bands, we here prove that they are valid for projection into \emph{any} $8 n_\text{max}$ particle-hole symmetric TBG bands, with $n_\text{max}>1$ being the practical case for small twist angles $<1^\circ$. 
Furthermore, in the first or second chiral-nonflat limit without flat bands, an exact U(4) symmetry still remains.  We also elucidate the link between the U(4) symmetry presented here and the similar but different U(4) of \cite{kang_strong_2019}. Furthermore, we show that our projected Hamiltonian can be viewed as the normal-ordered Coulomb interaction plus a Hartree-Fock term from passive bands, and exhibits a \emph{many-body} particle-hole symmetry which renders the physics symmetric around charge neutrality. We also provide an efficient parameterization of the interacting Hamiltonian. The existence of two chiral limits, with an enlarged symmetry suggests a possible duality of the model  yet undiscovered. 
\end{abstract}

\date{\today}

\maketitle

\section{Introduction}\label{sec:introduction}

Twisted bilayer graphene (TBG) near the magic angle $\theta\approx 1.1^\circ$ hosts flat electron bands, and exhibits remarkable interacting phases including correlated insulators, Chern insulators and superconductors \cite{bistritzer_moire_2011,cao_correlated_2018,cao_unconventional_2018, lu2019superconductors, yankowitz2019tuning, sharpe_emergent_2019, saito_independent_2020, stepanov_interplay_2020, liu2020tuning, arora_2020, serlin_QAH_2019, cao_strange_2020, polshyn_linear_2019,  xie2019spectroscopic, choi_imaging_2019, kerelsky_2019_stm, jiang_charge_2019,  wong_cascade_2020, zondiner_cascade_2020,  nuckolls_chern_2020, choi2020tracing, saito2020,das2020symmetry, wu_chern_2020,park2020flavour, saito2020isospin,rozen2020entropic, lu2020fingerprints, burg_correlated_2019,shen_correlated_2020, cao_tunable_2020, liu_spin-polarized_2019, chen_evidence_2019, chen_signatures_2019, chen_tunable_2020, burg2020evidence, tarnopolsky_origin_2019, zou2018, fu2018magicangle, liu2019pseudo, Efimkin2018TBG, kang_symmetry_2018, song_all_2019,po_faithful_2019,ahn_failure_2019,Slager2019WL, hejazi_multiple_2019, lian2020, hejazi_landau_2019, padhi2020transport, xu2018topological,  koshino_maximally_2018, ochi_possible_2018, xux2018, guinea2018, venderbos2018, you2019,  wu_collective_2020, Lian2019TBG,Wu2018TBG-BCS, isobe2018unconventional,liu2018chiral, bultinck2020, zhang2019nearly, liu2019quantum,  wux2018b, thomson2018triangular,  dodaro2018phases, gonzalez2019kohn, yuan2018model,kang_strong_2019,bultinck_ground_2020,seo_ferro_2019, hejazi2020hybrid, khalaf_charged_2020,po_origin_2018,xie_superfluid_2020,julku_superfluid_2020, hu2019_superfluid, kang_nonabelian_2020, soejima2020efficient, pixley2019, knig2020spin, christos2020superconductivity,lewandowski2020pairing, xie_HF_2020,liu2020theories, cea_band_2020,zhang_HF_2020,liu2020nematic, daliao_VBO_2019,daliao2020correlation, classen2019competing, kennes2018strong, eugenio2020dmrg, huang2020deconstructing, huang2019antiferromagnetically,guo2018pairing, ledwith2020, repellin_EDDMRG_2020,abouelkomsan2020,repellin_FCI_2020, vafek2020hidden, fernandes_nematic_2020,  Wilson2020TBG,wang2020chiral, ourpaper1, ourpaper2,ourpaper4,ourpaper5,ourpaper6}. Both transport \cite{cao_correlated_2018,cao_unconventional_2018, lu2019superconductors, yankowitz2019tuning, sharpe_emergent_2019, saito_independent_2020, stepanov_interplay_2020, liu2020tuning, arora_2020, serlin_QAH_2019, cao_strange_2020, polshyn_linear_2019,  saito2020,das2020symmetry, wu_chern_2020,park2020flavour} and scanning tunneling microscope \cite{xie2019spectroscopic, choi_imaging_2019, kerelsky_2019_stm, jiang_charge_2019,  wong_cascade_2020, zondiner_cascade_2020,  nuckolls_chern_2020, choi2020tracing} experiments show the correlated insulators and Chern insulators originate from strong many-body interactions. Extensive theoretical studies have been devoted to understanding the electron interactions in TBG \cite{xu2018topological,  koshino_maximally_2018, ochi_possible_2018, xux2018, guinea2018, venderbos2018, you2019,  wu_collective_2020, Lian2019TBG,Wu2018TBG-BCS, isobe2018unconventional,liu2018chiral, bultinck2020, zhang2019nearly, liu2019quantum,  wux2018b, thomson2018triangular,  dodaro2018phases, gonzalez2019kohn, yuan2018model,kang_strong_2019,bultinck_ground_2020,seo_ferro_2019, hejazi2020hybrid, khalaf_charged_2020,po_origin_2018,xie_superfluid_2020,julku_superfluid_2020, hu2019_superfluid, kang_nonabelian_2020, soejima2020efficient, pixley2019, knig2020spin, christos2020superconductivity,lewandowski2020pairing, xie_HF_2020,liu2020theories, cea_band_2020,zhang_HF_2020,liu2020nematic, daliao_VBO_2019,daliao2020correlation, classen2019competing, kennes2018strong, eugenio2020dmrg, huang2020deconstructing, huang2019antiferromagnetically,guo2018pairing, ledwith2020, repellin_EDDMRG_2020,abouelkomsan2020,repellin_FCI_2020, vafek2020hidden, fernandes_nematic_2020}. Kang and Vafek \cite{kang_strong_2019} first proposed that, by projecting in a non-maximally-symmetric Wannier basis, a non-negative interaction Hamiltonian can be obtained, whose ground state at $\nu=\pm2$ electrons per unit cell (with respect to charge neutrality) is an exactly solvable insulator with some mild approximation. A U(4) symmetry was also identified for the TBG interaction \cite{kang_strong_2019,bultinck_ground_2020,seo_ferro_2019} (both Refs.~\cite{kang_strong_2019,bultinck_ground_2020} identified a U(4), which we show here to be similar but different), which was shown to enlarge into a U(4)$\times$U(4) symmetry in the chiral limit $w_0=0$ \cite{bultinck_ground_2020}. However, these symmetries were proposed only for the $8$ lowest bands (2 bands per valley-spin) around the charge neutrality point, which applies for the first magic angle; while the TBG theoretically and experimentally exhibits, for example, $32$ low-energy "active" bands (8 bands per valley-spin) around charge neutrality at lower angles $\theta =0.45^\circ$ \cite{lu2020fingerprints}.

In this paper, we derive the explicit  TBG Hamiltonian Coulomb Hamiltonian projected within any number of $8n_\text{max}$ ($2n_\text{max}$ per spin per valley, $n_\text{max}\ge1$) particle-hole symmetric low-lying moir\'e bands. For the first magic angle, the number of bands where the projection makes sense is 8 ($2$ per spin-valley) moir\'e bands in momentum space; for smaller angles, the number increases. We show the exact projected Coulomb interaction Hamiltonian can \emph{always} be written into a Kang-Vafek type \cite{kang_strong_2019} nonnegative form, which we hereby call Positive Semidefinite Hamiltonian (PSDH). The projected Hamiltonian we derived can be understood as the normal-ordered Coulomb interaction in the active bands plus a Hartree-Fock potential from the passive bands. Furthermore, the projected Hamiltonian has a many-body particle-hole symmetry, which ensures that all the physics are particle-hole symmetric about charge neutrality, in agreement with the overall picture of the experimental observations. We then study the TBG symmetries in the flat band limit. We prove the existence of not one but two (first and second)  chiral  limits defined by zero hopping at either AA or AB/BA stackings. We prove that the projected TBG Hamiltonian in the nonchiral-flat limit has an exact U(4) symmetry, which breaks to a U(2) $\times$ U(2) when kinetic energy is added (nonchiral-nonflat case). This symmetry is enhanced into an exact U(4)$\times$U(4) symmetry in either the first chiral-flat limit or the second chiral-flat limit. The U(4)$\times$U(4) symmetry for the first chiral limit, and for projection into two low-lying active bands was obtained in Ref.~\cite{bultinck_ground_2020}, but we here extend it to any number of projected bands, as well as to a second chiral limit. In the first chiral-nonflat limit or the second chiral-nonflat limit, a kinetic term is also considered, and the bands are not flat; however, we show that an exact U(4) symmetry still remains. All these symmetries, in all limits,  are shown to be not only valid for the 8 active bands at the first magic angle \cite{bultinck_ground_2020}, but also for the projected Hamiltonian within any number of particle-hole symmetric bands. This is relevant at smaller twist angles: in Ref.~\cite{lu2020fingerprints} it was experimentally and theoretically found that $32$ bands (8 bands per valley/spin) contribute to the low energy physics. Besides, for Hamiltonian projected in the lowest 8 bands (2 bands per spin per valley), we reveal that the Hamiltonian in the first or second chiral limit can be enhanced into a stabilizer code Hamiltonian under certain assumptions. Furthermore, we elucidate the similarities and differences between the U(4) symmetry of Kang and Vafek \cite{kang_strong_2019} and the U(4) in the current paper. The explicit form and symmetries of Hamiltonian here greatly simplify the study of TBG many-body states, as we will discuss in Refs.~\cite{ourpaper4} and~\cite{ourpaper5}.

\section{Bistritzer-MacDonald Model and Coulomb Interaction}\label{sec:bmmodel}

\begin{figure}
\begin{centering}
\includegraphics[width=\linewidth]{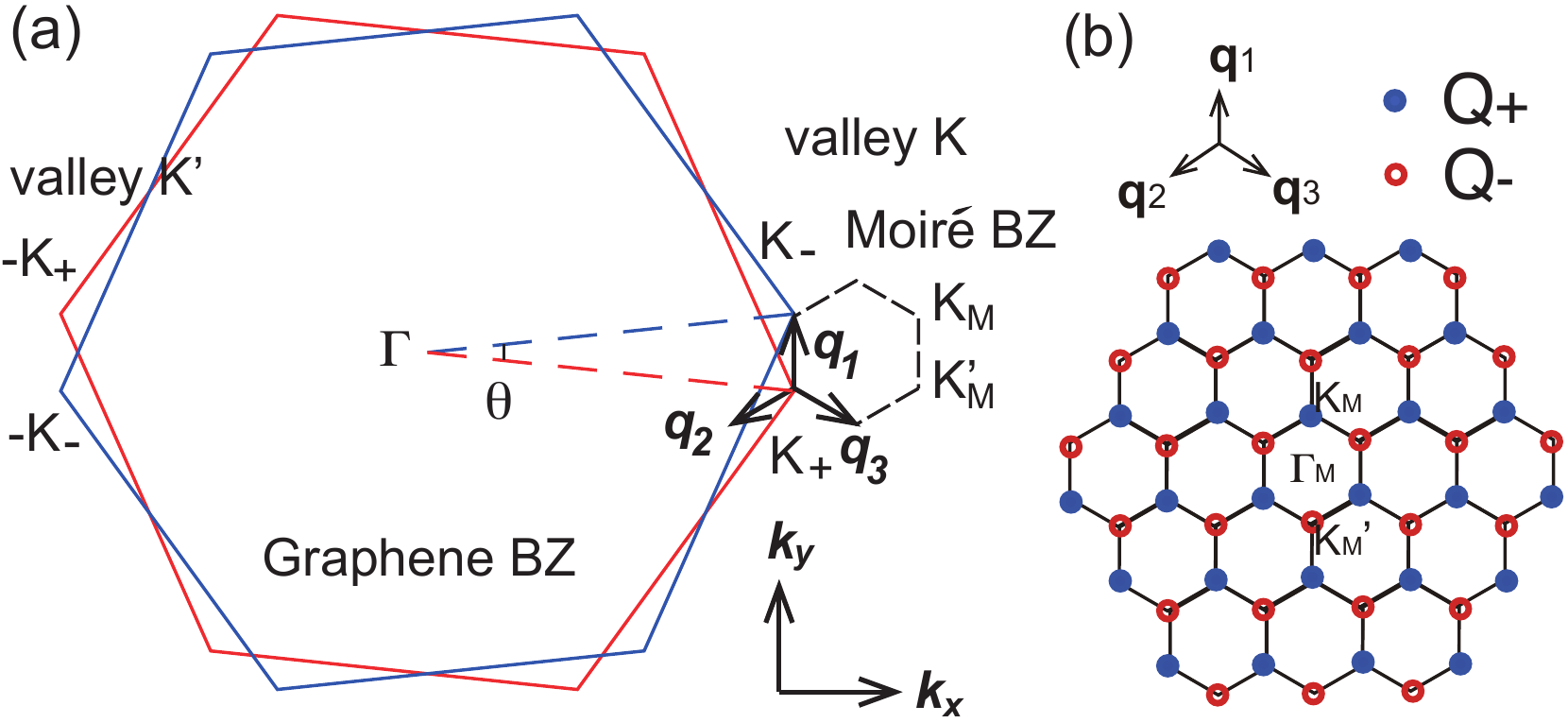}
\par\end{centering}

\protect\caption{\label{fig:MBZ} Illustration of the relation between the graphene BZs of two layers and the moir\'e BZ (MBZ). Blue solid and red empty circles represent
$\mathcal{Q}_{+}$ and $\mathcal{Q}_{-}$, respectively.}

\end{figure}

We first present a short overview of the Bistritzer-MacDonald (BM) model \cite{bistritzer_moire_2011} to define our notations. The reader might refer to Refs. \cite{ourpaper1, ourpaper2} for a in-depth discussion. For convenience, we also provide a detailed summary in App.~\ref{app:onebodyhamiltonian}. To begin, we assume $c^\dag_{\mathbf{p},\alpha,s,l}$ denotes the creation operator of the spin $s=\uparrow,\downarrow$ electron at momentum $\mathbf{p}$ in the graphene sublattice $\alpha=A,B$ and layer $l=\pm$ (denoting top and bottom) of TBG, where $\mathbf{p}$ is measured from the $\Gamma$ point of the graphene Brillouin zone (BZ) of layer $l$.
The low-energy physics of TBG is concentrated at the two graphene valleys $K,K'$ (which we denote as valleys $\eta=\pm$) at momenta $\mathbf{p}=\pm \mathbf{K}_\ell$ in layer $\ell$, respectively \cite{bistritzer_moire_2011}. We further define $\mathbf{q}_j=C_{3z}^{j-1}(\mathbf{K}_--\mathbf{K}_+)$ ($j=1,2,3$), where $C_{3z}$ is the 3-fold rotation about $z$ axis (see Fig. \ref{fig:MBZ}(a)). The kinetic Hamiltonian of TBG is then given by the continuum model \cite{bistritzer_moire_2011,ourpaper2} as
\begin{equation}\label{eq-H0}
\hat{H}_{0}=\sum_{\mathbf{k}\in\text{MBZ}}\sum_{\eta\alpha\beta s}\sum_{\mathbf{Q}\mathbf{Q}^{\prime}}\left[ h_{\mathbf{Q},\mathbf{Q}^{\prime}}^{\left(\eta\right)}\left(\mathbf{k}\right)\right]_{\alpha\beta} c_{\mathbf{k},\mathbf{Q},\eta,\alpha, s}^{\dagger}c_{\mathbf{k},\mathbf{Q}^{\prime},\eta,\beta, s},
\end{equation}
where $\eta=\pm$ and $s=\uparrow,\downarrow$ are the valley and spin indices, and the momentum $\mathbf{k}$ is measured from the center ($\Gamma_M$ point) of the moir\'e BZ (MBZ). The momenta $\mathbf{Q},\mathbf{Q'}\in\{\mathcal{Q}_+,\mathcal{Q}_-\}$ as shown in Fig. \ref{fig:MBZ}(b), where we have defined $\mathcal{Q}_\pm=\mathcal{Q}_0\pm\mathbf{q}_1$, and $\mathcal{Q}_0$ is the moir\'e reciprocal lattice generated by reciprocal vectors $\mathbf{b}_{Mj}=\mathbf{q}_3-\mathbf{q}_j$ ($j=1,2$). The electron basis $c_{\mathbf{k},\mathbf{Q},\eta,\alpha, s}^{\dagger}$ is defined as $c^\dag_{\mathbf{\eta K}_{\eta\cdot l}+\mathbf{k-Q},\alpha,s,\eta\cdot l}$ if $\mathbf{Q}\in\mathcal{Q}_l$. The detailed kinetic term $h_{\mathbf{Q},\mathbf{Q}^{\prime}}^{\left(\eta\right)}\left(\mathbf{k}\right)$ at valley $\eta=\pm$ is given in App. \ref{app:onebodyhamiltonian}. In particular, there are two parameters $w_0$ and $w_1$ in the single-particle Hamiltonian $h_{\mathbf{Q},\mathbf{Q}^{\prime}}^{\left(\eta\right)}\left(\mathbf{k}\right)$ which correspond to the interlayer hoppings at AA and AB/BA stacking centers, respectively (see Eq. (\ref{seq-Tj})):
\begin{equation}\label{eq-Tj}
\begin{split}
&w_0\ge0:\ \  \text{AA hopping},\\ &w_1\ge0:\ \ \text{AB/BA hopping}.
\end{split}
\end{equation}
Generically, $w_0<w_1$ due to the lattice relaxation and corrugation \cite{ourpaper2,Uchida_corrugation,Wijk_corrugation,dai_corrugation, jain_corrugation}.

The Coulomb interaction term in TBG takes the form (for details, see App.~\ref{app:interaction})
\begin{equation}\label{eq-HI}
\hat{H}_I=\frac{1}{2\Omega_{\text{tot}}}\sum_{\mathbf{G}\in\mathcal{Q}_0}\sum_{\mathbf{q}\in \text{MBZ}}V(\mathbf{q+G})\delta\rho_{\mathbf{-q-G}}\delta\rho_{\mathbf{q+G}}\ ,
\end{equation}
where 
\beq
\delta\rho_{\mathbf{q+G}}=\sum_{\eta,\alpha,s,\mathbf{k},\mathbf{Q}\in\mathcal{Q}_\pm} (c_{\mathbf{k+q},\mathbf{Q-G},\eta,\alpha, s}^\dag c_{\mathbf{k},\mathbf{Q},\eta,\alpha, s}-\frac{1}{2}\delta_{\mathbf{q,0}}\delta_{\mathbf{G,0}}) 
\eeq
is the total electron density at momentum $\mathbf{q+G}$ relative to the charge neutral point (CNP) of the uncoupled twisted bilayer graphene without interlayer couplings (which has a density $\langle c_{\mathbf{k+q},\mathbf{Q-G},\eta,\alpha, s}^\dag c_{\mathbf{k},\mathbf{Q},\eta,\alpha, s}\rangle = \frac{1}{2}\delta_{\mathbf{q,0}}\delta_{\mathbf{G,0}}$), and $\Omega_{\text{tot}}$ is the total area of TBG. The interaction coefficient
\begin{equation}
V(\mathbf{q})=\pi \xi^2 U_\xi \frac{\tanh (\xi |\mathbf{q}|/2)}{ \xi|\mathbf{q}|/2}
\end{equation}
is the Fourier transform of the Coulomb potential with dielectric constant $\epsilon$ screened by top and bottom gates at distance $\xi$ away, where $U_\xi=e^2/\epsilon\xi$ (see App. \ref{app:HI}). Typical TBG experiments have a screening length $\xi\approx10$nm \cite{stepanov_interplay_2020,saito_independent_2020}, and dielectric constant $\epsilon\sim 6$ as estimated from the hBN substrates. This yields a $U_\xi\approx 24$meV.

Due to the absence of spin-orbit coupling, the total Hamiltonian 
\begin{equation}
\hat{H}=\hat{H}_0+\hat{H}_I
\end{equation}
of TBG has the spinless symmetries 
\beq
[C_{3z},\hat{H}]=[C_{2z},\hat{H}]=[T,\hat{H}]=0\ ,
\eeq
where $C_{3z}$ is the 3-fold $z$-axis rotation symmetry satisfying $C_{3z}c_{\mathbf{k},\mathbf{Q},\eta,\alpha, s}^{\dagger}C_{3z}^{-1}=(e^{i\eta2\pi\sigma_z/3})_{\beta\alpha}c_{C_{3z}\mathbf{k},C_{3z}\mathbf{Q},\eta,\beta, s}^{\dagger}$, $C_{2z}$ is the 2-fold $z$-axis rotation symmetry  satisfying $C_{2z}c_{\mathbf{k},\mathbf{Q},\eta,\alpha, s}^{\dagger}C_{2z}^{-1}=(\sigma_x)_{\beta\alpha}c_{-\mathbf{k},-\mathbf{Q},-\eta,\beta, s}^{\dagger}$, and $T$ is the anti-unitary time-reversal symmetry satisfying $T iT^{-1}=-i$ and $Tc_{\mathbf{k},\mathbf{Q},\eta,\alpha, s}^{\dagger}T^{-1}=c_{-\mathbf{k},-\mathbf{Q},-\eta,\alpha, s}^{\dagger}$. Besides, each graphene valley exhibits a charge U(1) symmetry and a spin rotational SU(2) symmetry, leading to a global U(2)$\times$U(2) symmetry of two valleys (see App. \ref{app:tbg-symmetry}).

There also exists a unitary single-particle particle-hole (PH) transformation $P$ \cite{song_all_2019,ourpaper2} which anti-commutes with $\hat{H}_0$ in Eq.~(\ref{eq-H0})  (see App.~\ref{app:tbg-symmetry}) and commutes with $\hat{H}_I$ in Eq.~(\ref{eq-HI}):
\begin{equation}
\{P,\hat{H}_0\}=0\ ,\qquad [P,\hat{H}_I]=0\ ,
\end{equation}
where $P$ is defined by $Pc_{\mathbf{k},\mathbf{Q},\eta,\alpha, s}^{\dagger} P^{-1}=\zeta_\mathbf{Q}c_{-\mathbf{k},-\mathbf{Q},\eta,\alpha, s}^{\dagger}$, with $\zeta_\mathbf{Q}=\pm1$ for $\mathbf{Q}\in\mathcal{Q}_\pm$. In particular, $[P,\hat{H}_I]=0$ can be seen by noting that $\delta\rho_{\mathbf{q+G}}$ in Eq.~(\ref{eq-HI}) satisfies $P\delta\rho_{\mathbf{q+G}}P^{-1}=\delta\rho_{\mathbf{-q-G}}$. We note that an antiunitary PH transformation $\mathcal{P}=PC_{2z}T$ can also be defined, which is adopted in some literature \cite{bultinck_ground_2020,ourpaper2}.

\section{Projected Hamiltonian}\label{sec:flatbandprojham}

\begin{figure}
\begin{centering}
\includegraphics[width=\linewidth]{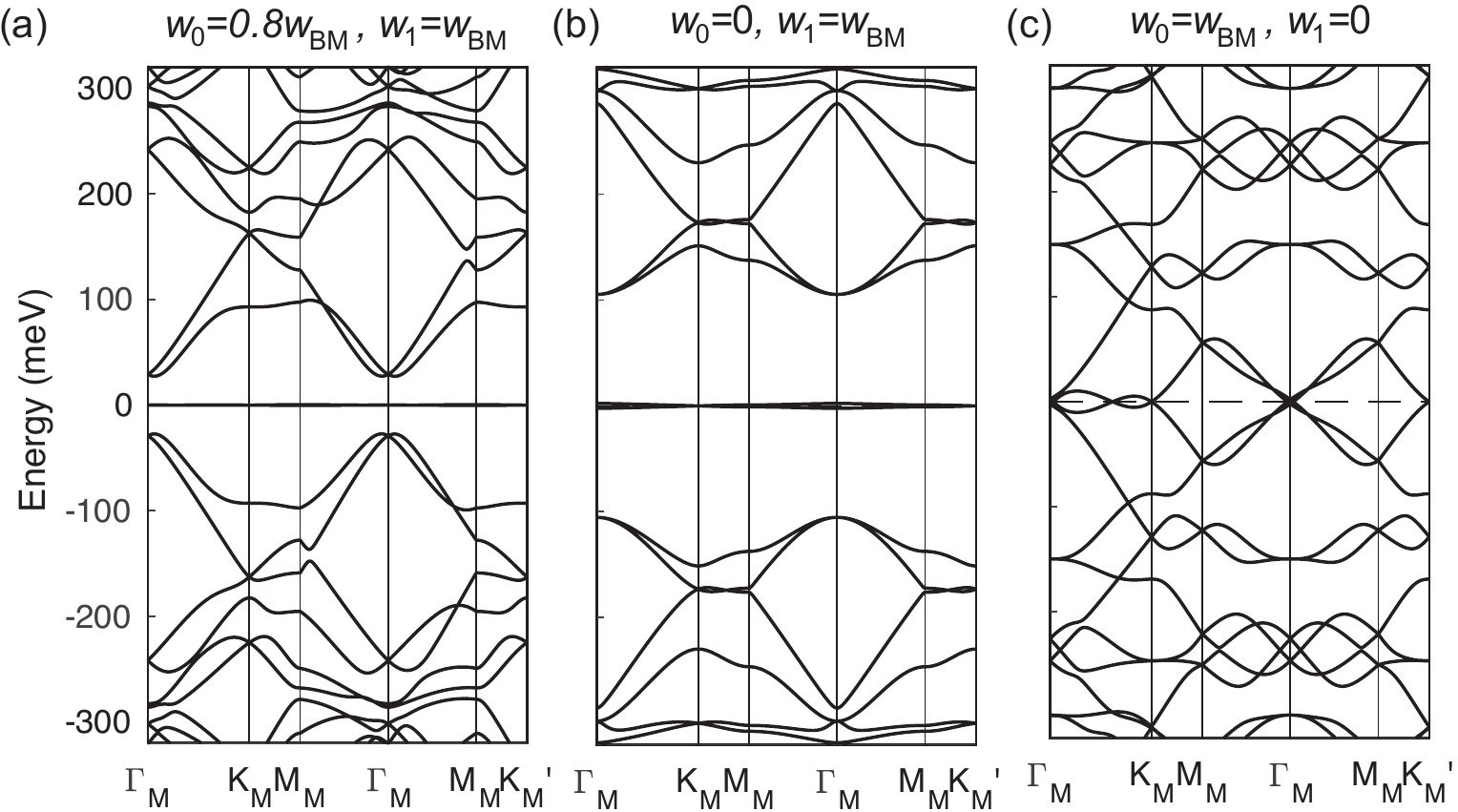}
\par\end{centering}

\protect\caption{\label{fig:band} The single-valley TBG band structure at $\theta=1.05^{\circ}$ (with exact PH symmetry $P$) for (a) the nonchiral-nonflat limit with $w_0=0.8w_{\text{BM}}$ and $w_1=w_{\text{BM}}$, (b) the first chiral limit with $w_0=0$ and $w_1=w_{\text{BM}}$, and (c) the second chiral limit $w_0=w_{\text{BM}}$ and $w_1=0$, where $w_{\text{BM}}=110$meV is the hopping in the original Bistritzer-Macdonald TBG model\cite{bistritzer_moire_2011}. In particular, in the second chiral limit, the band structure is a perfect metal where all bands are connected (proof given in Ref.~\cite{ourpaper2}).}

\end{figure}

We denote the eigenstates and eigenvectors of $h_{\mathbf{Q},\mathbf{Q}^{\prime}}^{\left(\eta\right)}\left(\mathbf{k}\right)$ in Eq.~(\ref{eq-H0}) as $\epsilon_{n,\eta}(\mathbf{k})$ and $u_{\mathbf{Q} \alpha n \eta}(\mathbf{k})$ (which are spin independent), where the integer $n\neq0$ is the band index so defined that $n>0$ ($n<0$) labels the $|n|$-th conduction (valence) band of valley $\eta$.

Near the first magic angle $\theta\approx 1.1^\circ$, the lowest conduction and valence bands ($n=\pm1$) of 2 spins and 2 valleys of TBG form 8 extremely flat bands which are energetically isolated from the higher bands (Fig. \ref{fig:band}(a)). Therefore, it is appropriate to project the Hamiltonian into the 8 flat bands for low-energy physics at the first magic angle. At higher magic angles, the number of low energy bands increase; for instance, around the second magic angle  $\theta\approx 0.5^\circ$ \cite{bistritzer_moire_2011}, the lowest conduction and valence bands form 32 (8 per spin/valley, $|n|\le4$) low energy bands \cite{lu2020fingerprints}. In this case, the projection of Hamiltonian into more PH symmetric bands is needed for studying low energy physics. Therefore, to keep our discussions generic, we consider the projection into a set of $8n_\text{max}$ number of PH symmetric low-energy bands $|n|\le n_{\text{max}}$ with any $n_{\text{max}}\ge1$. As we will show, since the symmetries $C_{2z}, T, P$ which we will study are closed within each pair of bands $\pm n$, it is sufficient to focus on the 2-dimensional band space of each pair of bands $\pm n$ when examining the symmetries of the projected Hamiltonian. 

The projection of the kinetic Hamiltonian $\hat{H}_0$ in the set of bands $|n|\le n_{\text{max}}$ bands is thus (which we denote by $H_0$ without hat)
\begin{equation}\label{eq-pH0}
H_0=\sum_{|n|\le n_{\text{max}}}\sum_{\mathbf{k} \eta s} \epsilon_{n,\eta}(\mathbf{k}) c^\dag_{\mathbf{k},n,\eta, s} c_{\mathbf{k},n,\eta, s}\ ,
\end{equation}
where $c^\dag_{\mathbf{k},n,\eta, s}=\sum_{\mathbf{Q},\alpha}u_{\mathbf{Q} \alpha n \eta}(\mathbf{k}) c_{\mathbf{k},\mathbf{Q},\eta,\alpha, s}^{\dagger}$ gives the band basis of electrons, and $\pm n_{\text{max}}$ are the highest/lowest bands we project into. 
Meanwhile, the projection of Coulomb interaction $\hat{H}_I$ in the flat bands can be written as (denoted by $H_I$ without hat, see App. \ref{app:proj-H})
\begin{equation}\label{eq-pHI}
H_I=\frac{1}{2\Omega_{\text{tot}}}\sum_{\mathbf{q}\in\text{MBZ}}\sum_{\mathbf{G}\in\mathcal{Q}_0} O_{\mathbf{-q,-G}} O_{\mathbf{q,G}}\ ,
\end{equation}
where
\begin{equation}\label{eq-OqG}
\begin{split}
O_{\mathbf{q,G}} =&\sum_{\mathbf{k}\eta s}\sum_{|m|,|n|\le n_{\text{max}}} \sqrt{V(\mathbf{q+G})} M_{m,n}^{\left(\eta\right)} \left(\mathbf{k},\mathbf{q}+\mathbf{G}\right) \\
&\quad \times \left(\rho_{\mathbf{k,q},m,n,s}^\eta-\frac{1}{2}\delta_{\mathbf{q,0}}\delta_{m,n}\right)\ .
\end{split}
\end{equation}
Here we have defined the coefficient called the \emph{form factors (overlaps)}:
\begin{equation}\label{eq:M-def0}
M_{m,n}^{\left(\eta\right)} \left(\mathbf{k},\mathbf{q}+\mathbf{G}\right)=\sum_{\alpha,\mathbf{Q}\in\mathcal{Q}_\pm}u^*_{\mathbf{Q-G},\alpha m\eta}(\mathbf{k+q})u_{\mathbf{Q},\alpha n\eta}(\mathbf{k}), 
\end{equation}
and $\rho_{\mathbf{k,q},m,n,s}^\eta=c^\dag_{\mathbf{k+q},m,\eta, s}c_{\mathbf{k},n,\eta, s}$ is the density operator. The form factors (overlaps) $M_{m,n}^{\left(\eta\right)} \left(\mathbf{k},\mathbf{q}+\mathbf{G}\right)$  was shown to exhibit properties such as exponential decay in the magnitude of $\GG$ in Ref.~\cite{ourpaper1}. As such, only $|\GG|=0$ and $|\GG| = |\mathbf{b}_{M_1}|$ momentum vectors will contribute to $O_{\mathbf{q,G}}$, all other $\GG$ leading to exponentially smaller form factors (overlaps). We now note that $O_{\mathbf{-q,-G}}=O_{\mathbf{q,G}}^\dag$, such that $ O_{\mathbf{-q,-G}} O_{\mathbf{q,G}}$ is a positive semidefinite operator for any $ \qq, \GG$.  Thus the interaction Hamiltonian $H_I$, being a sum of positive semidefinite operators, is also positive semidefinite. We call such Hamiltonians positive semidefinite Hamiltonians (PSDH). 

Below, we investigate the symmetries of the projected Hamiltonian $H=H_0+H_I$ in various different limits. Without loss of generality, we will consider the subspace of a particular pair of PH symmetric bands $n=\pm n_\text{B}$ with $1\le n_\text{B}\le n_{\text{max}}$, since all the single-particle symmetries we will be discussing are closed within the band pair $n=\pm n_\text{B}$. 

Hereafter we shall use $\zeta^a$, $\tau^a$, $s^a$ to denote the identity matrix ($a=0$) and Pauli matrices ($a=x,y,z$) in the energy band $n=\pm n_\text{B}$, valley $\eta=\pm$ and spin $s=\uparrow,\downarrow$ bases, respectively. In particular, when $n_\text{B}=1$, our discussion applies to the projected Hamiltonian in the lowest 8 flat bands near the first magic angle.

\section{Symmetries in the generic nonchiral-nonflat case}

The projected Hamiltonian $H=H_0+H_I$ preserves all the discrete TBG symmetries $C_{3z}$, $C_{2z}$, $T$ (see App.~\ref{app:HI}). Moreover, the projected Hamiltonian respects the global U(2)$\times$U(2) spin-charge rotational symmetry of two valleys, which has 8 group generators $S^{ a b}=\sum_{\mathbf{k}} (s^{ab})_{m,\eta,s;n,\eta',s'}c_{\mathbf{k},m,\eta,s}^\dag c_{\mathbf{k},n,\eta',s'}$ ($a=0,z$ and $b=0,x,y,z$, repeated indices are summed automatically hereafter). Within each band pair $n=\pm n_\text{B}$, the matrices $s^{ab}$ are given by
\begin{equation}\label{eq-s0zb}
s^{0b}=\zeta^0\tau^0 s^b,\quad s^{zb}=\zeta^0\tau^z s^b, \quad (b=0,x,y,z).
\end{equation}
We note that $s^{ab}$ has no nonzero matrix elements between different pairs of PH symmetric bands $n_\text{B}\neq n_\text{B}'$.
Also, note that the operators $S^{ab}$ preserve the electron momentum $\mathbf{k}$. 

Another $\mathbf{k}$-preserving transformation is given by the combined unitary operator $C_{2z}P$  ($P$ is the PH transformation), which acts as
\begin{equation}\label{eq-c2zPoncq}
(C_{2z}P)c_{\mathbf{k},\mathbf{Q},\eta,\alpha, s}^{\dagger}(C_{2z}P)^{-1}=\zeta_{\mathbf{Q}}(\sigma_x)_{\beta\alpha}c_{\mathbf{k},\mathbf{Q},-\eta,\beta, s}^{\dagger},
\end{equation}
and thus satisfies $(C_{2z}P)^2=1$. Since $(C_{2z}P)H_0(C_{2z}P)^{-1}=-H_0$, the single-particle band energies satisfy $\epsilon_{n,\eta}(\mathbf{k})=-\epsilon_{-n,-\eta}(\mathbf{k})$, and the eigenstate wavefunctions satisfy $\zeta_{\mathbf{Q}}(\sigma_x)_{\beta\alpha} u_{\mathbf{Q}, \alpha, n, \eta}(\mathbf{k})=[B^{C_{2z}P}(\mathbf{k})]_{-n,-\eta;n\eta}u_{\mathbf{Q}, \beta, -n, -\eta}(\mathbf{k})$, where $B^{C_{2z}P}(\mathbf{k})$ is the unitary sewing matrix of $C_{2z}P$. This implies
\begin{equation}\label{eq-c2zPoncn}
(C_{2z}P)c^\dag_{\mathbf{k},n,\eta,s}(C_{2z}P)^{-1}=[B^{C_{2z}P}(\mathbf{k})]_{-n,-\eta;n\eta}c^\dag_{\mathbf{k},-n,-\eta, s},
\end{equation}
Using the explicit form of $B^{C_{2z}P}(\mathbf{k})$, one can prove that $[C_{2z}P, O_{\mathbf{q,G}}]=0$ (see App.~\ref{app:U4-nc-f}),
and thus
\begin{equation}\label{eq-C2P}
\{C_{2z}P,H_0\}=0\ ,\quad [C_{2z}P,H_I]=0\ .
\end{equation}
Therefore, $C_{2z}P$ is a commuting symmetry of $H_I$ but not $H_0$.

Furthermore, there is a many-body charge conjugation symmetry $\mathcal{P}_c$ defined by $C_{2z}PT$ followed by the interchange of annihilation and creation operators, namely, $\mathcal{P}_cc^\dag_{\mathbf{k},n,\eta, s}\mathcal{P}_c^{-1}=(C_{2z}PT)c_{\mathbf{k},n,\eta, s}(C_{2z}PT)^{-1}$
(see App. \ref{app:manybody-cc}). 
By showing that $\mathcal{P}_cO_{\mathbf{q,G}}\mathcal{P}_c^{-1}=-O_{\mathbf{q,G}}$, one can prove that the projected Hamiltonian within bands $|n|\le n_{\text{max}}$ satisfies (see Eq. (\ref{seq-Pc-transformation}) in App. \ref{app:manybody-cc})
\begin{equation}
\mathcal{P}_c(H_0+H_I)\mathcal{P}_c^{-1}=H_0+H_I.
\end{equation}
In particular, $\mathcal{P}_c$ maps a many-body state from filling $\nu$ to $-\nu$, where $\nu$ is the number of electrons per moir\'e unit cell relative to the CNP. Therefore, $\mathcal{P}_c$ ensures that the eigenstates of the projected Hamiltonian $H=H_0+ H_I$ is PH symmetric about $\nu=0$, in agreement with the (big picture) experimental observations.

We note that $H_I$ in Eq.~(\ref{eq-pHI}) is not normal ordered. We can rewrite $H_I=\normord{H_I}+\Delta H_I+E_I$, where $\normord{H_I}$ is the normal ordered 4-fermion interaction, $\Delta H_I$ is a quadratic fermion term and $E_I$ is a constant. One can then show that $\Delta H_I=\frac{1}{2}\left(H_{HF}^{\nu=-4 n_{\text{max}}}-H_{HF}^{\nu=4n_{\text{max}}}\right)$, where $H_{HF}^{\nu}$ is the Hartree-Fock potential in the projected bands contributed by all the occupied bands below filling $\nu$ (see App.~\ref{app:passivebandcontributions}), and the factor of $4$ comes from 2 spins and 2 valleys. Note that $H_{HF}^{\nu}$ sums over all the bands below filling $\nu$, instead of only the projected active bands (see derivation in App.~\ref{app:passivebandcontributions}). Therefore, $\Delta H_I$ can be understood as the mean field Hartree-Fock potential from the remote bands projected away symmetrized about the CNP. We note that $\normord{H_I}$ alone does not have the $\mathcal{P}_c$ symmetry, and thus $\Delta H_I$ is indispensable as an effective background Hartree-Fock potential.

\section{U(4) symmetry in the nonchiral-flat limit}\label{sec:U4symnonchiralflat}

In the limit of exactly flat $|n|\le n_{\text{max}}$ bands, we have $H_0=0$, so the projected Hamiltonian is simply $H=H_I$. By Eq.~(\ref{eq-C2P}), $C_{2z}P$ becomes a symmetry of $H$. Note that $C_{2z}P$ preserves the electron momentum $\mathbf{k}$, thus is a local unitary symmetry. Accordingly, the $C_{2z}P$ symmetry and the spin-charge U(2)$\times$U(2) symmetry together generate a global U(4) symmetry of the Hamiltonian $H=H_I$. To see this, we define an operator
\begin{equation}
S^{y0}=\sum_{\kk,s} \sum_{nn^\pr \eta \eta^\pr} [B^{C_{2z}P}(\kk)]_{n \eta, n^\pr \eta^\pr} 
c^\dg_{\kk,n,\eta,s} c_{\kk,n^\pr,\eta^\pr,s}
\end{equation}
with sewing matrix $B^{C_{2z}P}(\kk)$ of $C_{2z}P$. It can be proved 
that $[S^{y0},H_I]=0$ (see App. \ref{app:U4-nc-f}). Note that $S^{y0}$ is identical to $C_{2z}P$ when acting on single-electron states. For many-body states, one can show that $C_{2z}P=e^{i\pi S^{y0}/2}$ (up to a phase factor). 
With the 8 generators $S^{0b},S^{zb}$ of U(2)$\times$U(2) ($b=0,x,y,z$), we can define another $8$ operators $S^{xb}=-\frac{i}{2}[S^{y0},S^{zb}]$ and $S^{yb}=\frac{i}{2}[S^{x0},S^{zb}]$. The $16$ operators $S^{ab}$ then satisfy the Lie algebra of U(4):
\begin{equation}\label{eq-U(4)generator}
[S^{ab},S^{cd}]=\sum_{ef}f^{ab,cd}_{ef}S^{ef}\ ,\quad (a,b=0,x,y,z)
\end{equation}
where $f^{ab,cd}_{ef}$ are U(4) group structure constants defined by $[\tau^a s^b,\tau^c s^d]=\sum_{ef}f^{ab,cd}_{ef} \tau^e s^f$. 

It is useful to fix the gauge of wavefunctions to obtain an explicit form of $S^{ab}$. We do this by requiring 
\begin{equation}\label{eq-fix1}
(C_{2z}T)c^\dag_{\mathbf{k},n,\eta, s}(C_{2z}T)^{-1}=c^\dag_{\mathbf{k},n,\eta, s}\ , 
\end{equation}
which imposes $(\sigma_{x})_{\alpha\beta}u_{\mathbf{Q}, \beta, n, \eta}(\mathbf{k})=u^*_{\mathbf{Q}, \alpha, n, \eta}(\mathbf{k})$. 
$(\sigma_{x})_{\alpha\beta}u_{\mathbf{Q}, \beta, n, \eta}(\mathbf{k})=u^*_{\mathbf{Q}, \alpha, n, \eta}(\mathbf{k})=u_{-\mathbf{Q}, \alpha, n, -\eta}(-\mathbf{k})$. 
A consistent $\mathbf{k}$-independent gauge for $C_{2z}P$ is then 
\begin{equation}\label{eq-fix2}
(C_{2z}P)c^\dag_{\mathbf{k},n,\eta, s}(C_{2z}P)^{-1}=-\text{sgn}(n)\eta c^\dag_{\mathbf{k},-n,-\eta, s}. 
\end{equation}
In addition, we require a $\kk$-space continuous gauge (which is crucial for the useful bases Eqs.~(\ref{eq-irrepbasis}) and~(\ref{eq-Chernbasis})) defined below to have well-defined Berry curvature, see Sec. \ref{app:chern}):
\begin{equation}\label{eq-fix3}
\lim_{\qq\rightarrow\mathbf{0}}|u^\dag_{n,\eta}(\kk+\qq) u_{n,\eta}(\kk)- u^\dag_{-n,\eta}(\kk+\qq) u_{-n,\eta}(\kk)|=0.
\end{equation}
Under this gauge, we can rewrite the 16 U(4) generators as $S^{ a b}=\sum_{\mathbf{k}} (s^{ab})_{m,\eta,s;n,\eta',s'}c_{\mathbf{k},m,\eta,s}^\dag c_{\mathbf{k},n,\eta',s'}$ ($a,b=0,x,y,z$), where the matrices $s^{ab}$ within each PH pair of bands $n=\pm n_\text{B}$ read
\begin{equation}\label{eq-sxyb}
s^{ab}=\{\zeta^0\tau^0 s^b,\ \zeta^y\tau^x s^b,\ \zeta^y\tau^y s^b,\ \zeta^0\tau^z s^b\}.
\end{equation}
We note that $s^{ab}$ has no nonzero matrix elements between different pairs of PH symmetric bands $n_\text{B}\neq n_\text{B}'$.
Meanwhile, the form factors (overlaps) $M_{m,n}^{\left(\eta\right)} \left(\mathbf{k},\mathbf{q}+\mathbf{G}\right)=[M\left(\mathbf{k},\mathbf{q}+\mathbf{G}\right)]_{m\eta,n\eta}$ are gauge fixed into the following matrix form in the band and valley basis (App. \ref{app:Mfixing}):
\begin{equation}\label{eq-Mmn}
M\left(\mathbf{k},\mathbf{q}+\mathbf{G}\right)=\sum_{j=0}^3 M_j \alpha_j(\mathbf{k,q+G})\ ,
\end{equation}
where $\alpha_j(\mathbf{k,q+G})$ are \emph{real} $n_\text{max}\times n_\text{max}$ matrices, and we have defined $M_0=\zeta^0\tau^0$, $M_1=\zeta^x\tau^z$, $M_2=i\zeta^y\tau^0$, and $M_3=\zeta^z\tau^z$ in the space of each pair of band basis $n=\pm n_\text{B}$ ($1\le n_\text{B}\le n_\text{max}$), all of which are real matrices. Here $M_j\alpha_j$ means the Kronecker direct product of matrices $M_j$ and $\alpha_j$.

We note that we could further fix the gauges of the $\kk$ non-preserving symmetries $C_{2z}$, $T$ and $P$ in a $\kk$-independent way in consistency with Eqs.~(\ref{eq-fix1}-\ref{eq-fix3}) (see App.~\ref{app:gauge-fixing} and Eq.~(\ref{eq:gauge-1})). Under a further gauge fixing $C_{2z}c^\dag_{\mathbf{k},n,\eta, s}C_{2z}^{-1}=c^\dag_{-\mathbf{k},n,-\eta, s}$, one can show that the functions $\alpha_j(\kk,\qq+\GG)$ ($0\le j\le 4$) satisfy the conditions in Eqs.~(\ref{eq:alpha-cond1}) and~(\ref{eq:alpha-cond2}). In particular, these conditions require
\begin{equation}
\begin{split}
&\alpha_0(\kk,\GG)=\alpha_0^T(-\kk,\GG)\ , \\
&\alpha_j(\kk,\GG)=-\alpha_j^T(-\kk,\GG)\ ,\quad (j=1,2,3)
\end{split}
\end{equation}
at $\qq=\mathbf{0}$ (see App.~\ref{app:Mfixing}). 

With the gauge fixing of Eqs.~(\ref{eq-fix1}-\ref{eq-fix3}), we can define a new basis within the pair of bands $n=\pm n_\text{B}$ as
\begin{equation}\label{eq-irrepbasis}
d^{(n_\text{B})\dag}_{\mathbf{k},e_Y,\eta, s}=\frac{c^\dag_{\mathbf{k},n_\text{B},\eta, s}+ie_Y c^\dag_{\mathbf{k},-n_\text{B},\eta, s}}{\sqrt{2}}\ , \quad (e_Y=\pm1)
\end{equation}
which we show in App. \ref{app:chern} have well-defined Berry curvatures. The reason for the notation $e_Y=\pm1$ is because this basis is the eigenbasis of the Pauli matrix $\zeta_y$ with eigenvalue $e_Y$ in the 2-dimensional energy band basis of $n=\pm n_\text{B}$. We shall call the basis (\ref{eq-irrepbasis}) the \emph{irrep basis}, for the reason below. 

At each $\mathbf{k}$ and Chern number $e_Y$, as shown in App. \ref{sec:nc-f-irrep}, the 4 irrep basis creation operators $d^{(n_\text{B})\dag}_{\mathbf{k},e_Y,\eta, s}$ of valleys $\eta=\pm$ and spins $s=\uparrow,\downarrow$ form the basis of a fundamental U(4) irreducible representation (irrep), where the generators $S^{ab}$ have $4\times4$ representation matrices
\begin{equation}\label{eq-SabeY}
s^{ab}(e_Y)=\{ \tau^0s^b,\ e_Y\tau^x s^b,\ e_Y\tau^y s^b,\ \tau^zs^b \}.
\end{equation}
This can be seen by observing that $d^{(n_\text{B})\dag}_{\mathbf{k},e_Y,\eta, s}$ diagonalizes the matrix $\zeta^y$ in Eq.~(\ref{eq-sxyb}) with the eigenvalue being $e_Y$. Note that the two irreps $s^{ab}(e_Y)$ with $e_Y=\pm1$ differ by a unitary transformation $\tau^z$, namely, a $\pi$ valley rotation about $z$ axis. Despite of this difference by a unitary transformation, the two irreps $s^{ab}(e_Y)$ are both the fundamental irrep of U(4). In Young-tableaux accepted notations, we shall denote the fundamental irrep of U(4) as $[1]_4$, and the trivial identity irrep of U(4) as $[0]_4$ (see App. \ref{app:rev-U4}, and see \cite{ourpaper4} for a detailed explanation of the Young tableaux notations).  
An electron with a fixed $e_Y=\pm1$ and $\mathbf{k}$ thus occupies a U(4) irrep $[1]_4$.

For $n_\text{B}=1$, namely, for the lowest conduction and valence bands $n=\pm1$, we denote the basis in Eq. (\ref{eq-irrepbasis}) in simplified notations without upper index as
\begin{equation}\label{eq-Chernbasis}
d^\dag_{\mathbf{k},e_Y,\eta, s}=\frac{c^\dag_{\mathbf{k},+1,\eta, s}+ie_Y c^\dag_{\mathbf{k},-1,\eta, s}}{\sqrt{2}}\ , \quad (e_Y=\pm1)
\end{equation}
which will be extensively used for solving the projected Hamiltonian within the lowest 8 flat bands in Refs.~\cite{ourpaper4,ourpaper5,ourpaper6}. As proved in \cite{ourpaper2} (see also similar discussions in \cite{bultinck_ground_2020, hejazi2020hybrid}) and briefly reviewed in App. \ref{app:chern}, if a pair of energy bands $n=\pm n_\text{B}$ are disconnected with other bands, the irrep band we defined in Eq.~(\ref{eq-irrepbasis}) will carry a Chern number $e_Ye_{2,n_\text{B}}$, where $e_{2,n_\text{B}}$ is the Wilson loop winding number of the two bands $n=\pm n_\text{B}$. 
Due to the nontrivial topological winding number $e_{2,1}=1$ in the $n=\pm1$ bands \cite{po_origin_2018,song_all_2019,po_faithful_2019,ahn_failure_2019,lian2020,po_fragile_2018,cano_fragile_2018,Slager2019WL}, the irrep basis 
$d^\dag_{\mathbf{k},e_Y,\eta, s}$ in Eq. (\ref{eq-Chernbasis}) of all $\mathbf{k}$ for each fixed $e_Y,\eta,s$ form the basis of a Chern band of Chern number $e_Y=\pm1$ (see proof in details in Ref.~\cite{ourpaper2}, see also a brief review in App. \ref{app:chern}), provided the $n=\pm1$ energy bands are gapped from the higher bands (which is true near the first magic angle). 
For this reason, we shall call $d^\dag_{\mathbf{k},e_Y,\eta, s}$ (within the $n=\pm1$ energy band space) the \emph{Chern band electron basis}, or simply the \emph{Chern basis}.  We note that our Chern basis in Eq. (\ref{eq-Chernbasis}) is (adiabatically) equivalent to the Chern bands defined in Refs. \cite{bultinck_ground_2020,hejazi2020hybrid}.

If the $|n|\le n_{\text{max}}$ bands are gapped from higher bands, but are connected among themselves, we would expect the net Chern number of the $n_\text{max}$ irrep basis $d^{(n_\text{B})\dag}_{\mathbf{k},e_Y,\eta, s}$ ($1\le n_\text{B}\le n_{\text{max}}$) to be equal to $e_Y$ (see App.~\ref{app:chern}).

\section{U(4)$\times$U(4) symmetry in the (first) chiral-flat limit}\label{sec:U4symchiralflat}

The symmetry of flat-band TBG is enhanced when $w_0=0<w_1$ in Eq.~(\ref{eq-Tj}), which is known as the \emph{chiral limit} \cite{tarnopolsky_origin_2019}. In this paper we shall also call it the \emph{first chiral limit}, to distinguish with the second chiral limit defined below in Sec. \ref{sec:2ndnonchiralflatlimit}. In this first chiral limit, there is a unitary chiral transformation $C$ acting as $Cc_{\mathbf{k},\mathbf{Q},\eta,\alpha, s}^{\dagger}C^{-1}=(\sigma_z)_{\beta\alpha}c_{\mathbf{k},\mathbf{Q},\eta,\beta, s}^{\dagger}$, which satisfies $C\hat{H}_0 C^{-1}=-\hat{H}_0$ and $C^2=1$. Therefore, the energy band eigenstates satisfy $\epsilon_{n,\eta}(\mathbf{k})=-\epsilon_{-n,\eta}(\mathbf{k})$, and $(\sigma_z)_{\beta\alpha} u_{\mathbf{Q}, \alpha, n, \eta}(\mathbf{k})=[B^{C}(\kk)]_{-n,\eta;n\eta}u_{\mathbf{Q}, \beta, -n, \eta}(\mathbf{k})$, where $B^{C}(\kk)$ is the unitary sewing matrix of $C$. This implies $Cc^\dag_{\mathbf{k},n,\eta, s}C^{-1}=[B^{C}(\kk)]_{-n,\eta;n\eta}c^\dag_{\mathbf{k},-n,\eta, s}$. 
 
When projected into the flat bands $|n|\le n_\text{max}$, by Eq.~(\ref{eq-OqG}), one can prove that $[C,O_{\mathbf{q,G}}]=0$, and thus
\begin{equation}\label{eq-Chiral}
\{C,H_0\}=0\ ,\quad [C,H_I]=0\ .
\end{equation}
Therefore, in the first chiral-flat limit where $H_0=0$ and thus $H=H_I$, the chiral transformation $C$ becomes a symmetry. Note that $C$ preserves the electron momentum $\kk$, thus is a local unitary symmetry.

We can then define a Hermitian operator
\begin{equation}
S'^{z0}=\sum_{\kk,s} \sum_{nn^\pr \eta \eta^\pr} [B^{C}(\kk)]_{n \eta, n^\pr \eta^\pr} 
c^\dg_{\kk,n,\eta,s} c_{\kk,n^\pr,\eta^\pr,s}\ ,
\end{equation}
which commutes with $H_I$. Note that $S'^{z0}$ is identical to $C$ when acting on single-electron states. For many-body states, one can verify that $C=e^{i\pi S'^{z0}/2}$ (up to a phase factor). Its commutations with the $16$ U(4) generators $S^{ab}$ in Eq.~(\ref{eq-U(4)generator}) yield another $16$ new operators $S'^{ab}$, and one can prove that $S^{ab}$ and $S'^{ab}$ form the 32 generators of a U(4)$\times$U(4) group (App. \ref{app:chiral-flat}). This can be seen explicitly under the gauge fixing of Eqs. (\ref{eq-fix1}) and (\ref{eq-fix2}), for which the only $\mathbf{k}$-independent gauge choice (up to a global sign) for $C$ is $Cc^\dag_{\mathbf{k},n,\eta, s}C^{-1}=i\text{sgn}(n)\eta c^\dag_{\mathbf{k},-n,\eta, s}$ (App. \ref{app:chiral-flat}). We note that this gauge choice is also consistent with the $\kk$-independent gauge fixings of $C_{2z}$, $T$ and $P$ in Eq. (\ref{eq:gauge-1}). The $16$ new generators can then be expressed as $S'^{ab}=\sum_{\mathbf{k}} (s'^{ab})_{m,\eta,s;n,\eta',s'}c_{\mathbf{k},m,\eta,s}^\dag c_{\mathbf{k},n,\eta',s'}$ ($a,b=0,x,y,z$), where $s'^{ab}$ within each pair of bands $n=\pm n_\text{B}$ are given by
\begin{equation}\label{eq-spxyb}
s'^{ab}=\{\zeta^y\tau^0 s^b,\ \zeta^0\tau^x s^b, \ \zeta^0\tau^y s^b,\ \zeta^y\tau^z s^b\}.
\end{equation}
We note that $s'^{ab}$ has no nonzero matrix elements between different pairs of PH symmetric bands $n_\text{B}\neq n_\text{B}'$. We can further linear combine $S^{ab}$ and $S'^{ab}$ into operators $S^{ab}_{\pm}=\sum_{\mathbf{k}} (s_\pm^{ab})_{m,\eta,s;n,\eta',s'}c_{\mathbf{k},m,\eta,s}^\dag c_{\mathbf{k},n,\eta',s'}$ ($a,b=0,x,y,z$), where
\begin{equation}\label{eq-Sabpm}
s_\pm^{ab}=\left(\zeta^0\pm\zeta^y\right)\tau^a s^b/2\ .
\end{equation}
One can then verify that
\begin{equation}
[S_{e_Y}^{ab},S_{e_Y'}^{cd}]=\delta_{e_Y,e_Y'}\sum_{ef}f^{ab,cd}_{ef}S_{e_Y}^{ef}\ , \quad (e_Y=\pm1)
\end{equation}
where $f^{ab,cd}_{ef}$ are the U(4) structure constants in Eq.~(\ref{eq-U(4)generator}). Therefore, each set of $S_{e_Y}^{ab}$ ($e_Y=\pm1$) generates a U(4) group, leading to a total U(4)$\times$U(4) symmetry. We note that the nonchiral-flat U(4) in Eq.~(\ref{eq-sxyb}) is \emph{not} one of the two U(4)s with fixed $e_Y$ here, although it is a subgroup of the first chiral-flat U(4)$\times$U(4) here.

The 4 irrep band (Chern band if $n_\text{B}=1$) basis creation operators $d^{(n_\text{B})\dag}_{\mathbf{k},e_Y,\eta, s}$ (of valley-spin flavors $\eta=\pm,s=\uparrow,\downarrow$) at a fixed $\mathbf{k}$ and $e_Y$ in Eq. (\ref{eq-irrepbasis}) occupy a fundamental irrep of the U(4) generated by $S^{ab}_{e_Y}$, and a trivial identity irrep of the U(4) generated by $S^{ab}_{-e_Y}$ ($e_Y=\pm1$). The corresponding representation matrices of $S^{ab}_{\pm}$ are
\begin{equation}
s_\pm^{ab}=\left(1\pm e_Y\right)\tau^a s^b/2\ ,
\end{equation}
which can be derived by replacing matrix $\zeta^0$ ($\zeta^y$) by its eigenvalue $1$ ($e_Y$) in the irrep band basis $d^{(n_\text{B})\dag}_{\mathbf{k},e_Y,\eta, s}$. If we use $([\lambda_1]_4,[\lambda_2]_4)$ to represent a U(4)$\times$U(4) irrep which is the tensor product of an irrep $[\lambda_1]_4$ of the first U(4) and an irrep $[\lambda_2]_4$ of the second U(4), we see that the irrep basis $d^{(n_\text{B})\dag}_{\mathbf{k},+1,\eta, s}$ at a fixed $\mathbf{k}$ occupy an irrep $([1]_4,[0]_4)$, while the irrep basis $d^{(n_\text{B})\dag}_{\mathbf{k},-1,\eta, s}$ at a fixed $\mathbf{k}$ occupy an irrep $([0]_4,[1]_4)$. 

Furthermore, in App. \ref{app:chiral-flat} we proved that (see Eq. (\ref{eq:M-para-chiral})) the $C$ symmetry restricts 
\begin{equation}\label{eq:chiral-alpha13}
\alpha_1(\mathbf{k,q+G})=\alpha_3(\mathbf{k,q+G})=0
\end{equation}
in Eq.~(\ref{eq-Mmn}). This makes $O_{\qq,\GG}$ in Eq.~(\ref{eq-OqG}) diagonal in index $e_Y$ in the basis $d^{(n_\text{B})\dag}_{\mathbf{k},e_Y,\eta, s}$ (see Eq. (\ref{eq:chiral-OqG})), thus the number of electrons in the $n_\text{max}$ irrep bands (particularly, Chern band if $n_\text{max}=1$) with a fixed $e_Y$ is conserved.

\section{U(4) symmetry in the (first) chiral-nonflat limit}\label{sec:U4symchiralnonflat}

We now turn to the first chiral-nonflat case which is in the first chiral limit $w_0=0$ (thus Eq.~(\ref{eq-Chiral}) holds), but does not have exactly flat bands ($H_0\neq0$). Since the chiral symmetry implies $\epsilon_{n,\eta}(\mathbf{k})=-\epsilon_{-n,\eta}(\mathbf{k})$, the projected kinematic term in Eq.~(\ref{eq-pH0}) within each pair of bands $n=\pm n_\text{B}$ can be rewritten as 
\begin{equation}
H_0^{(n_\text{B})}=\sum_{\mathbf{k}} \epsilon_{+n_\text{B},\eta}(\mathbf{k}) (\zeta^z\tau^0 s^0)_{m,\eta,s;n,\eta',s'}c_{\mathbf{k},m,\eta,s}^\dag c_{\mathbf{k},n,\eta',s'}.
\end{equation}
As a result, $H_0$ only commutes with 16 out of the 32 U(4)$\times$U(4) generators $S^{ab}$ and $S'^{ab}$ in Eqs. (\ref{eq-sxyb}) and (\ref{eq-spxyb}). We denote these 16 generators commuting with $H_0$ as $\widetilde{S}^{ab}=\sum_{\mathbf{k}} (\tilde{s}^{ab})_{m,\eta,s;n,\eta',s'}c_{\mathbf{k},m,\eta,s}^\dag c_{\mathbf{k},n,\eta',s'}$, where $\tilde{s}^{ab}$ within each pair of bands $n=\pm n_\text{B}$ read
\begin{equation}\label{eq-tilde-sab}
\tilde{s}^{ab}=\zeta^0\tau^a s^b\ ,\quad (a,b=0,x,y,z).
\end{equation}
They form the 16 generators of a U(4) group. In particular, the representation matrix $\tilde{s}^{x0}$ of generator $\widetilde{S}^{x0}$ at each $\kk$ is given by the sewing matrix of $iC C_{2z}P$, and thus $\widetilde{S}^{x0}$ is identical to $iC C_{2z}P$ when acting on single-electron states. For many-body states, one has $iCC_{2z}P=e^{i\pi \widetilde{S}^{x0}/2}$ (up to a phase factor). Therefore, in the first chiral-nonflat limit with $H_0\neq0$, there is a global U(4) symmetry generated by $\widetilde{S}^{ab}$, which is reduced from the U(4)$\times$U(4) symmetry of the first chiral-flat limit. We note that this first chiral-nonflat U(4) here (Eq. (\ref{eq-tilde-sab})) \emph{is different} from the nonchiral-flat U(4) (Eq.~(\ref{eq-sxyb}).

Since $\widetilde{S}^{ab}$ is proportional to $\zeta^0$ in the band basis, the energy band creation operators $c_{\mathbf{k},n,\eta,s}^\dag$ in each band $n$ at a fixed $\mathbf{k}$ occupy a fundamental irrep $[1]_4$ of the first chiral-nonflat U(4) group. Equivalently, the irrep band (Chern band if $n_\text{B}=1$) creation operators $d_{\mathbf{k},e_Y,\eta,s}^{(n_\text{B})\dag}$ for fixed $e_Y$, $n_\text{B}$ and $\mathbf{k}$ also occupy a fundamental U(4) irrep $[1]_4$. For the irrep of either $c_{\mathbf{k},n,\eta,s}^\dag$ or $d_{\mathbf{k},e_Y,\eta,s}^{(n_\text{B})\dag}$, the representation matrices of $\widetilde{S}^{ab}$ are given by 
\begin{equation}
\tilde{s}^{ab}(n)=\tilde{s}^{ab}(e_Y)=\tau^as^b\qquad (a,b=0,x,y,z). 
\end{equation}
Note that the representation matrices $\tilde{s}^{ab}(n)$ (or $\tilde{s}^{ab}(e_Y)$) are independent of $n$ (or $e_Y$). 
This is in contrast to the nonchiral-flat limit, where the representation matrices of $S^{ab}$ for $e_Y=\pm1$ differ by a unitary transformation $\tau_z$ (although $e_Y=\pm1$ therein still give the same fundamental nonchiral-flat U(4) irrep, see Eq.~(\ref{eq-SabeY})). 

\begin{table}[tbp]
\centering
\begin{tabular}{c|c|c|c|c|c}
\hline
TBG limit & $H_0$ & $w_0$ & $w_1$ & symmetry & PH/chiral  \\
\hline
nonchiral-nonflat & $\neq0$ & $>0$ & $>0$ & U(2)$\times$U(2) & ---  \\
\hline
nonchiral-flat & $=0$ & $>0$ & $>0$ & U(4) & $C_{2z}P$  \\
\hline
(1st) chiral-flat & $=0$ & $=0$ & $>0$ & U(4)$\times$U(4) & $C_{2z}P$, $C$  \\
\hline
(1st) chiral-nonflat & $\neq0$ & $=0$ & $>0$ & U(4) & $iCC_{2z}P$  \\
\hline
2nd chiral-flat & $=0$ & $>0$ & $=0$ & U(4)$\times$U(4) & $C_{2z}P$, $C'$  \\
\hline
2nd chiral-nonflat & $\neq0$ & $>0$ & $=0$ & U(4) & $iC'C_{2z}P$  \\
\hline
\end{tabular}
\caption{Symmetries in different limits. The last column are the contributing PH and chiral symmetries.}\label{tab-limits}
\end{table}

\section{U(4)$\times$U(4) symmetry in the second chiral-flat limit}\label{sec:2ndnonchiralflatlimit}

We find that there exists a \emph{second chiral limit} $w_1=0<w_0$ where the continuous symmetry of TBG is largely enhanced, similar to the situation in the first chiral limit discussed in Secs.~\ref{sec:U4symchiralflat} and~\ref{sec:U4symchiralnonflat}. Although this limit is far from the experimental reality of the TBG samples, its existence suggests the possibility of a possible hidden duality in the BM model and its interactions. 
For $w_1=0<w_0$, we can define a second chiral transformation $C'$ satisfying $C'^2=1$ and $C'\hat{H}_0C'^{-1}=-\hat{H}_0$, which acts as $C'c_{\mathbf{k},\mathbf{Q},\eta,\alpha, s}^{\dagger}C'^{-1}=(\sigma_z)_{\beta\alpha}\zeta_{\QQ}c_{\mathbf{k},\mathbf{Q},\eta,\beta, s}^{\dagger}$ with $\zeta_\QQ=\pm1$ for $\QQ\in\mathcal{Q}_\pm$. 
This new chiral symmetry has \emph{unusual} commutation relations with the 2-fold rotation $C_{2z}$, time-reversal $T$ and with the unitary particle hole symmetry $P$ (see App.~\ref{app:2nd-c-symmetry} and Ref.~\cite{ourpaper2} for details). It also satisfies (see App.~\ref{app:2ndchiral-flat}):
\begin{equation}\label{eq-2nd-Chiral}
\{C',H_0\}=0\ ,\qquad [C',H_I]=0\ ,
\end{equation}
similar to the first chiral symmetry $C$ (Eq. (\ref{eq-Chiral})). Note that the second chiral symmetry $C'$ preserves electron momentum $\kk$.
In the \emph{ second chiral-flat limit} with $w_1=0$ and $H_0=0$, similar to the first chiral-flat limit, we can define a symmetry 
\begin{equation}
S''^{z0}=\sum_{\kk,s} \sum_{nn^\pr \eta \eta^\pr} [B^{C'}(\kk)]_{n \eta, n^\pr \eta^\pr} c^\dg_{\kk,n,\eta,s} c_{\kk,n^\pr,\eta^\pr,s}, 
\end{equation}
where $B^{C'}(\kk)$ is the sewing matrix of $C'$. Together with $S^{ab}$ in Eq.~(\ref{eq-U(4)generator}), it generates a U(4)$\times$U(4) group with 32 generators $S'^{ab}_\pm$ (see App.~\ref{app:2ndchiral-flat}). Under the gauge fixings of Eqs.~(\ref{eq-fix1}) and~(\ref{eq-fix2}), and a further gauge fixing for $C'$ as $C'c^\dag_{\mathbf{k},n,\eta, s}C'^{-1}=i\text{sgn}(n)\eta c^\dag_{\mathbf{k},-n,\eta, s}$ (which is consistent with the continuous condition (\ref{eq-fix3}), see App.~\ref{app:2nd-full-symmetry}), 
we find $S'^{ab}_{\pm}=\sum_{\mathbf{k}}
(s_\pm'^{ab})_{m,\eta,s;n,\eta',s'}c_{\mathbf{k},m,\eta,s}^\dag c_{\mathbf{k},n,\eta',s'}$, where $s_\pm'^{ab}$ within each pair of bands $n=\pm n_\text{B}$ read
\begin{equation}\label{eq-Sabpmpr}
s_\pm'^{ab}=\left(\zeta^0\pm\zeta^y\right)\tau^a s^b/2\ .
\end{equation}
Again, we note that $s_\pm'^{ab}$ has no nonzero matrix elements between different pairs of PH symmetric bands $n_\text{B}\neq n_\text{B}'$. 

It is worthwhile to mention that, due to the unusual commutation relations of $C'$ with $C_{2z}$, $T$ and $P$ which flip $\kk$, one cannot further fix the sewing matrices of $C_{2z}$, $T$ and $P$ into a $\mathbf{k}$-independent form as in Eq. (\ref{eq:gauge-1}). Instead, the sewing matrices of these $\kk$ flipping symmetries have to be $\kk$-dependent, for instance, given by Eq. (\ref{eq-k-dependent-fixing}) in App. \ref{app:2nd-full-symmetry}. This is closely related to the topologically protected double degeneracies at $C_{2z}$-invariant points of the MBZ, as proved in Ref.~\cite{ourpaper2}. 

In this second chiral-flat limit, the 4 irrep band basis creation operators $d^{(n_\text{B})\dag}_{\mathbf{k},e_Y,\eta, s}$ ($\eta=\pm,s=\uparrow,\downarrow$) at a fixed $\mathbf{k}$ and $e_Y$ in Eq. (\ref{eq-irrepbasis}) occupy a fundamental irrep of the U(4) generated by $S'^{ab}_{e_Y}$, and a trivial identity irrep of the U(4) generated by $S'^{ab}_{-e_Y}$ ($e_Y=\pm1$). The corresponding representation matrices of $S'^{ab}_{\pm}$ are
\begin{equation}
s_\pm'^{ab}=\left(1\pm e_Y\right)\tau^a s^b/2\ ,
\end{equation}
which can be see by substituting matrix $\zeta^0$ ($\zeta^y$) by its eigenvalue $1$ ($e_Y$) in the irrep band basis $d^{(n_\text{B})\dag}_{\mathbf{k},e_Y,\eta, s}$. Therefore, the irrep basis $d^{(n_\text{B})\dag}_{\mathbf{k},+1,\eta, s}$ at a fixed $\mathbf{k}$ occupy an irrep $([1]_4,[0]_4)$ of the second chiral-flat U(4)$\times$U(4), while the irrep basis $d^{(n_\text{B})\dag}_{\mathbf{k},-1,\eta, s}$ at a fixed $\mathbf{k}$ occupy an irrep $([0]_4,[1]_4)$.

Furthermore, in App. \ref{app:2ndchiral-flat} we proved that (see Eq. (\ref{eq:M-para-2chiral})) the $C'$ symmetry restricts 
\begin{equation}\label{eq:2chiral-alpha13}
\alpha_1(\mathbf{k,q+G})=\alpha_3(\mathbf{k,q+G})=0
\end{equation}
in Eq.~(\ref{eq-Mmn}).

However, with $w_1=0<w_0$, there is barely an angle where a set of low energy bands become flat, and it is proved in Ref.~\cite{ourpaper2} that all the energy bands are topologically connected into a perfect metal (see Fig.~\ref{fig:band}(c),  App.~\ref{app:2ndchiral-flat} and  Refs.~\cite{ourpaper2,Mora2019-graphene}). This makes the second chiral-flat limit less related to experimental realities, although it can possibly be achieved by artificial patterning of the moir\'e lattice to enhance AA hopping. Besides, we note that for the lowest PH band pair of $n_B=1$, the ``Chern band basis" $d^{\dag}_{\mathbf{k},e_Y,\eta, s}$ in Eq. (\ref{eq-Chernbasis}) no longer have a well-defined Chern number, since the $n=\pm1$ bands are connected with all the higher bands.

We also note that although the representation matrices in the two chiral limits in Eqs. (\ref{eq-Sabpm}) and (\ref{eq-Sabpmpr}) are the same, their physical operations are different, since they are generated by the sewing matrices of the first chiral symmetry $C$ and the second chiral symmetry $C'$, respectively.

\section{U(4) symmetry in the second chiral-nonflat limit}

With the TBG bands in the second chiral limit poorly flat, the \emph{second chiral-nonflat limit} where $w_1=0<w_0$ and $H_0\neq0$ gives a more physical limit, which may be realized by artificial patterning of moir\'e lattices. 
In this limit, similar to the first chiral-nonflat limit, we can prove that (see App.~\ref{app:2nd-c-nf}) a U(4) symmetry remains, which is generated by the remaining $iC'C_{2z}P$ symmetry. The 16 U(4) generators are a subset of the generators $S'^{ab}_\pm$ in the second chiral-flat limit (Eq.~(\ref{eq-Sabpmpr})), which we denote by $\widetilde{S}'^{ab}=\sum_{\mathbf{k}} (\tilde{s}'^{ab})_{m,\eta,s;n,\eta',s'}c_{\mathbf{k},m,\eta,s}^\dag c_{\mathbf{k},n,\eta',s'}$, where $\tilde{s}'^{ab}$ within each pair of bands $n=\pm n_\text{B}$ are given by 
\begin{equation}
\tilde{s}'^{ab}=\zeta^0\tau^as^b\ .
\end{equation}
This simply gives the spin-valley rotations without affecting the space of energy band indices $n$. Accordingly, either the energy band basis $c_{\mathbf{k},n,\eta,s}^\dag$ or the irrep band basis $d_{\mathbf{k},e_Y,\eta,s}^{(n_\text{B})\dag}$ at a fixed $\kk$ and $n$ or $e_Y$ occupy a fundamental U(4) irrep, with the representation matrices of $\widetilde{S}'^{ab}$ given by 
\begin{equation}
\tilde{s}'^{ab}(n)=\tilde{s}'^{ab}(e_Y)=\tau^as^b\qquad (a,b=0,x,y,z). 
\end{equation}

\section{The Stabilizer Code Limit}

Generically, the projected interaction Hamiltonian $H_I$ in Eq. (\ref{eq-pHI}) cannot be analytically diagonalized, since generically $[O_{\qq,\GG},O_{\qq',\GG'}]\neq0$ for $\qq\neq\qq'$ or $\GG\neq\GG'$ (see Eq. (\ref{eq:OqG-commutator})), and thus the terms $O_{-\qq,-\GG}O_{\qq,\GG}$ in $H_I$ are non-commuting.

However, in the case we are only projecting into the lowest 8 bands with $n=\pm1$ (namely, $n_\text{max}=1$), there is limit which we call the \emph{stabilizer code limit}, where the Hamiltonian becomes similar to (but not strictly identical to, see App.~\ref{sec:stabilizer}) a stabilizer code Hamiltonian with all of its terms mutually commuting. The stabilizer code limit is defined in either the first chiral-flat limit (with first chiral symmetry $C$) or the second chiral-flat limit (with second chiral symmetry $C'$), where Eq. (\ref{eq:chiral-alpha13}) or (\ref{eq:2chiral-alpha13}) is satisfied, and the condition is that the form factors $M(\kk,\qq+\GG)$ in Eq. (\ref{eq-Mmn}) are $\kk$-independent for any $\qq,\GG$. In this limit, As we proved in App.~\ref{sec:stabilizer}, one would have $[O_{\qq,\GG},O_{\qq',\GG'}]=0$. Thus, all the terms $O_{-\qq,-\GG}O_{\qq,\GG}$ in the Hamiltonian $H=H_I$ in Eq. (\ref{eq-pHI}) will be commuting:
\begin{equation}
[O_{-\qq,-\GG}O_{\qq,\GG},O_{-\qq',-\GG'}O_{\qq',\GG'}]=0\ .
\end{equation}
This stabilizer code-like Hamiltonian have all of its many-body eigenstates exactly solvable, which will be solved in a separate paper \cite{ourpaper4}.

\section{Discussion}

\begin{figure}
\begin{centering}
\includegraphics[width=\linewidth]{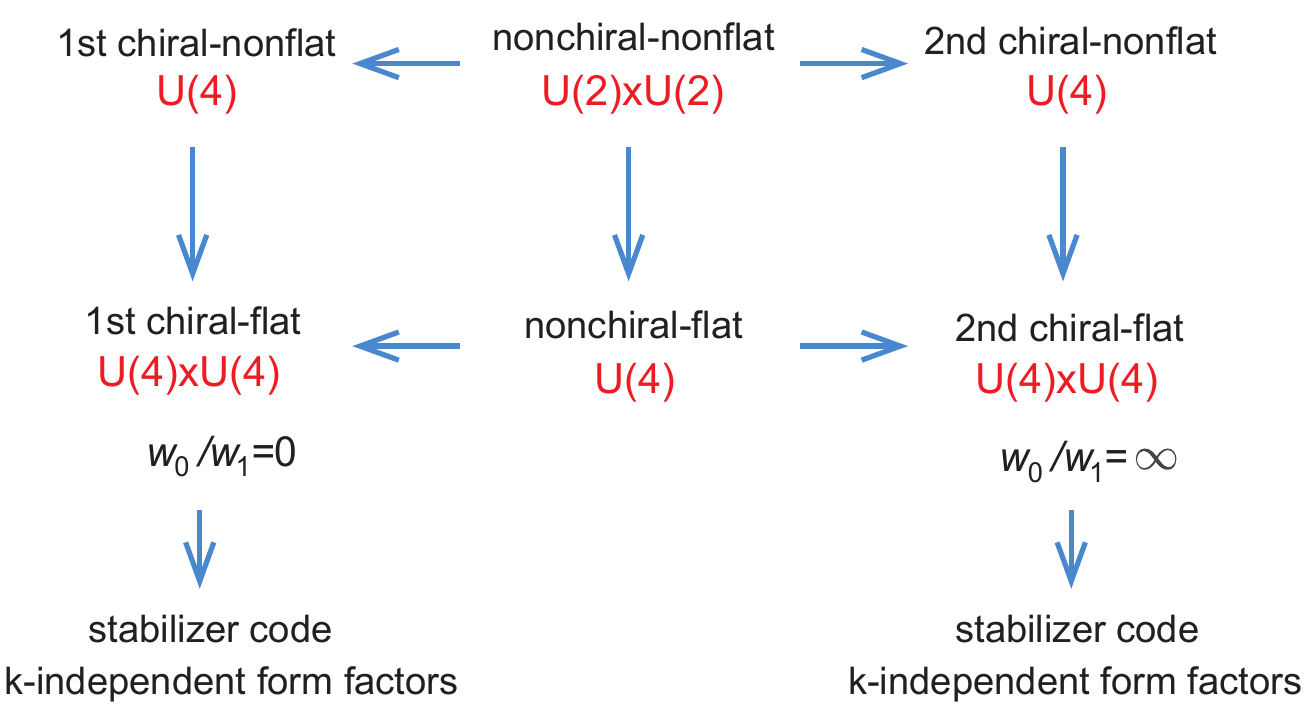}
\par\end{centering}

\protect\caption{\label{fig:sym} The relations between the symmetries of projected Hamiltonian within any set of PH symmetric bands of full spin-valley flavors in various limits. The arrows point along the directions along which the symmetry groups are enhanced into a larger one.}

\end{figure}

We have demonstrated that for the projected Hamiltonian with Coulomb interaction in the lowest $8n_\text{max}$ ($2n_\text{max}$ per spin-valley) bands of any $n_\text{max}\ge1$, there exists various different limits where a global U(4) or U(4)$\times$U(4) symmetry emerge. For $n_\text{max}$, there exists a stabilizer code limit for the Hamiltonian in either the first or the second chiral flat limit, where all the terms in the Hamiltonian are mutually commuting. Our conclusions are summarized in Tab. \ref{tab-limits} and Fig. \ref{fig:sym}. Near the first magic angle, the low energy physics is expected to be governed by the projected Hamiltonian with $n_\text{max}=1$. 
A projected Hamiltonian within higher number of bands could be a good approximation at higher magic angles, where more than 2 bands per spin-valley can become flat.

The U(4) symmetry in the nonchiral-flat limit in Eq.~(\ref{eq-sxyb}) and U(4)$\times$U(4) symmetry in the first chiral-flat limit in Eq.~(\ref{eq-Sabpm}) that  we prove here agree with those discussed in Ref.~\cite{bultinck_ground_2020} for the lowest 8 flat bands near the first magic angle. We note that, however, we show the symmetries are generic for the projection into any number of PH symmetric bands with full spin-valley degrees of freedom. Besides, we have identified a second chiral limit, which also enjoys a U(4)$\times$U(4) symmetry in a second chiral-flat limit. We have also derived the explicit irrep band basis of the symmetries in all the different limits. Furthermore, we showed that under a strong condition, the projected Hamiltonian in the lowest 8 bands in the first or second chiral-flat limit becomes similar to a stabilizer code Hamiltonian, thus allowing one to exactly solve all the many-body eigenstates, which we will study in Ref.~\cite{ourpaper4}.

A U(4) symmetry in the flat band limit is also discussed in Ref.~\cite{kang_strong_2019}, which is constructed based on a non-maximally-symmetric Wannier basis.
(These Wannier functions break the $C_{2z}T$ and $C_{2z}TP$ symmetries, which protect the fragile topology \cite{ahn_failure_2019,po_faithful_2019,song_all_2019} and stable topology \cite{ourpaper2} in TBG, respectively.)
The U(4) symmetry in Ref.~\cite{kang_strong_2019} is closest to our first chiral-nonflat U(4) symmetry that we introduce in Eq.~(\ref{eq-tilde-sab}) since they have the same generators $\tau^a s^b$ ($a,b=0,x,y,z$).
However, Ref.~\cite{kang_strong_2019} does not assume the $CC_{2z}P$ symmetry but requires the flatness of the two bands, which is in contradict with our first chiral-nonflat U(4), which assumes the $CC_{2z}P$ symmetry and does not require flat bands.
The reason Ref.~\cite{kang_strong_2019} needs flat bands is the absence of exact $CC_{2z}P$ symmetry.
We show in App.~\ref{app:kang-vafekU4} that, if the $CC_{2z}P$ symmetry is imposed to the Wannier functions, then the two U(4) symmetries become the same and do not require the flatness of bands.

The TBG interacting Hamiltonian, symmetries, and gauge fixings we derived here provide a solid ground for future theoretical studies. In the various limits we discussed, the many-body eigenstates of TBG should fall into irreps of U(4) or U(4)$\times$U(4) groups. Besides, the generic PSDH form of the projected interaction $H_I$ in Eq. (\ref{eq-pHI}) allows us to look for ground states of the Kang-Vafek type in the flat band limit. We will study the ground states and excitations of TBG in these limits  analytically and numerically in separate papers \cite{ourpaper4,ourpaper5,ourpaper6}. The existence of several limits with identical large continuous symmetry groups (but different generators) of the BM interacting Hamiltonian, as shown in Fig.~\ref{fig:sym} suggests the presence of a yet to be found duality of this model.

\begin{acknowledgments}
We thank Aditya Cowsik and Fang Xie for valuable discussions. B.A.B thanks Oskar Vafek for fruitful discussions, and for sharing their similar results on this problem before publication \cite{vafek2020hidden}. This work was supported by the DOE Grant No. DE-SC0016239, the Schmidt Fund for Innovative Research, Simons Investigator Grant No. 404513, the Packard Foundation, the Gordon and Betty Moore Foundation through Grant No. GBMF8685 towards the Princeton theory program, and a Guggenheim Fellowship from the John Simon Guggenheim Memorial Foundation. Further support was provided by the NSF-EAGER No. DMR 1643312, NSF-MRSEC No. DMR-1420541 and DMR-2011750, ONR No. N00014-20-1-2303, Gordon and Betty Moore Foundation through Grant GBMF8685 towards the Princeton theory program, BSF Israel US foundation No. 2018226, and the Princeton Global Network Funds.  B.L. acknowledge the support of Princeton Center for Theoretical Science at Princeton University during the early stage of this work.
\end{acknowledgments}

\bibliography{TBLGHexalogy,HexalogyInternalRefs}

\appendix

\onecolumngrid
\tableofcontents

\section{Review of the Single-particle Hamiltonian}\label{app:onebodyhamiltonian}

The quantitative and symmetry aspects of the  single-particle Hamiltonian of TBG are  discussed in details in Refs. \cite{ourpaper1, ourpaper2}. For completeness of a self-contained presentation, here we briefly review the notations and conclusions for the single-particle Hamiltonian.

\subsection{Bases}

We denote the fermion operator in the plane wave basis of graphene layer $l$ as $c_{\mathbf{p},\alpha,s,l}^{\dagger}$.
Here $\mathbf{p}$ is measured from the $\Gamma$ point of the monolayer graphene Brillouin zone BZ, $\alpha=A,B$ represents
the AB sublattice, $s=\uparrow,\downarrow$ is the spin index, and $l=\pm$
is the layer index. We define $\mathbf{K}_{+}$ as the K point in the top layer graphene BZ,
and $\mathbf{K}_{-}$ as the K point in the bottom
layer graphene BZ. $\mathbf{K}_+$ and $\mathbf{K}_-$ differ by a twist angle $\theta$ (Fig.~\ref{fig:MBZ}). For concreteness, we assume $\mathbf{K}_l$ is along the direction with an angle $-l \theta/2$ to the $p_x$ axis. Each graphene layer $l$ contains two valleys K and K' at momenta $\eta \mathbf{K}_l$, where $\eta=\pm$ denotes graphene valleys K and K', respectively.

For later use, we define the 2D momenta
\begin{equation}
\mathbf{q}_{1}=\left(\mathbf{K}_{-}-\mathbf{K}_{+}\right)=k_\theta (0,1)^T\ ,\qquad 
\mathbf{q}_{2}=C_{3z}\mathbf{q}_{1}=k_\theta (-\frac{\sqrt{3}}{2},-\frac{1}{2})^T\ ,\qquad
\mathbf{q}_{3}=C_{3z}^2\mathbf{q}_{1}=k_\theta (\frac{\sqrt{3}}{2},-\frac{1}{2})^T\ ,
\end{equation}
where $k_\theta=|\mathbf{K}_{-}-\mathbf{K}_{+}|=2|\mathbf{K}_{+}|\sin(\theta/2)$ for twist angle $\theta$. We can then define the moir\'e BZ (MBZ) for the TBG moir\'e lattice, which is generated by the moir\'e reciprocal vectors
\begin{equation}\label{eq-reciprocal}
\mathbf{b}_{M1}=\mathbf{q}_3-\mathbf{q}_1\ ,\qquad  \mathbf{b}_{M2}=\mathbf{q}_3-\mathbf{q}_2\ .
\end{equation}

\subsection{Single-particle Hamiltonian}

When the twist angle between the two graphene layers is small ($\theta \sim1^{\circ}$), an approximate
valley-U(1) symmetry, and an approximate moir\'e translation symmetry emerges. Accordingly, the single-particle Hamiltonian is decoupled between two valleys $\eta=\pm$.

To concentrate on the low energy physics of the two valleys, we define $\mathcal{Q}_{0}=\mathbb{Z}\mathbf{b}_{M1}+\mathbb{Z}\mathbf{b}_{M2}$ as the triangular moir\'e reciprocal lattice sites generated by the moir\'e reciprocal vectors $\mathbf{b}_{M1}$ and $\mathbf{b}_{M2}$ in Eq.~(\ref{eq-reciprocal}). We then define two shifted momentum lattices $\mathcal{Q}_{+}=\mathbf{q}_{1}+\mathcal{Q}_{0}$ and $\mathcal{Q}_{-}=-\mathbf{q}_{1}+\mathcal{Q}_{0}$. We then define the low energy fermion operators $c_{\mathbf{k},\mathbf{Q},\eta,\alpha,s}^{\dagger}$ at valley $\eta$ and $\mathbf{Q}\in\mathcal{Q}_\pm$ as
\begin{equation}
c_{\mathbf{k},\mathbf{Q},+,\alpha,s}^{\dagger}=\begin{cases}
c_{\mathbf{K}_{+}+\mathbf{k}-\mathbf{Q},\alpha,s,+}^{\dagger} & \qquad\mathbf{Q}\in\mathcal{Q}_{+}\\
c_{\mathbf{K}_{-}+\mathbf{k}-\mathbf{Q},\alpha,s,-}^{\dagger} & \qquad\mathbf{Q}\in\mathcal{Q}_{-}
\end{cases},\label{eq:ckQ+}
\end{equation}
\begin{equation}
c_{\mathbf{k},\mathbf{Q},-,\alpha,s}^{\dagger}=\begin{cases}
c_{-\mathbf{K}_{-}+\mathbf{k}-\mathbf{Q},\alpha,s,-}^{\dagger} & \qquad\mathbf{Q}\in\mathcal{Q}_{+}\\
c_{-\mathbf{K}_{+}+\mathbf{k}-\mathbf{Q},\alpha,s,+}^{\dagger} & \qquad\mathbf{Q}\in\mathcal{Q}_{-}
\end{cases},\label{eq:ckQ-}
\end{equation}
where $\mathbf{k}$ takes value in the MBZ, and $\mathbf{k=0}$ is chosen at the center ($\Gamma_M$ point) of the MBZ. In practice, we always take a finite cutoff for $\mathcal{Q}_{0,+,-}$; the largest $\mathbf{Q}$ in $\mathcal{Q}_{\pm}$ should have a norm much smaller than $|\mathbf{K}_{+}|$. We denote the number of points
in $\mathcal{Q}_{0,+,-}$ as $\left|\mathcal{Q}_{0,+,-}\right|$.

The single-particle Hamiltonian of TBG for small twist angle $\theta$ is given by \cite{bistritzer_moire_2011,ourpaper1,ourpaper2} 
\begin{equation}\label{eq:H0}
\hat{H}_{0}=\sum_{\mathbf{k}\in\text{MBZ}}\sum_{\eta\alpha\beta s}\sum_{\mathbf{Q}\mathbf{Q}^{\prime}}\left[ h_{\mathbf{Q},\mathbf{Q}^{\prime}}^{\left(\eta\right)}\left(\mathbf{k}\right)\right]_{\alpha\beta} c_{\mathbf{k},\mathbf{Q},\eta,\alpha, s}^{\dagger}c_{\mathbf{k},\mathbf{Q}^{\prime},\eta,\beta, s}\ ,
\end{equation}
where $h_{\mathbf{Q},\mathbf{Q}^{\prime}}^{\left(\eta\right)}\left(\mathbf{k}\right)$ is the first-quantized momentum space Hamiltonian at valley $\eta$ in the sublattice space, and $\QQ,\QQ'\in \mathcal{Q}_\pm$. At valley K ($\eta=+$), we have
\begin{equation}
h_{\mathbf{Q},\mathbf{Q}^{\prime}}^{\left(+\right)}\left(\mathbf{k}\right) =v_F(\mathbf{k-Q})\cdot\bm{\sigma}\delta_{\mathbf{Q,Q'}}+\sum_{j=1}^3 T_j\delta_{\mathbf{Q,Q'\pm q}_j}\ ,
\end{equation}
where $v_F$ is the graphene Fermi velocity, and the matrices
\begin{equation}\label{seq-Tj}
T_j=w_0\sigma_0+w_1\Big[\sigma_x\cos\frac{2\pi(j-1)}{3}+\sigma_y\sin\frac{2\pi(j-1)}{3}\Big]\ .
\end{equation}
Here $\sigma_0$ and $\bm{\sigma}=(\sigma_x,\sigma_y)$ are the $2\times2$ identity matrix and Pauli matrices in the space of sublattice indices, while $w_0\ge0$ and $w_1\ge0$ are the interlayer hoppings at the AA and AB stacking centers of TBG, respectively. Generically, in realistic systems $w_0<w_1$ due to the lattice relaxation. In the absence of lattice relaxation, one has $w_0=w_1$.

At valley K' ($\eta=-$), we have
\begin{equation}
h_{\mathbf{Q},\mathbf{Q}^{\prime}}^{\left(-\right)}\left(\mathbf{k}\right)=\sigma_x h_{-\mathbf{Q},-\mathbf{Q}^{\prime}}^{\left(+\right)}\left(-\mathbf{k}\right) \sigma_x 
=-v_F(\mathbf{k-Q})\cdot\bm{\sigma}^*\delta_{\mathbf{Q,Q'}}+\sum_{j=1}^3 (\sigma_xT_j\sigma_x)\delta_{\mathbf{Q,Q'\pm q}_j}\ ,
\end{equation}
where $\bm{\sigma}^*=(\sigma_x,-\sigma_y)$.

\subsection{Symmetries}\label{app:tbg-symmetry}

Here we summarize the symmetries of TBG, which can be found in Ref.~\cite{song_all_2019} and expanded on in Ref.~\cite{ourpaper2}.

1. \emph{Discrete symmetries}. Since graphene has zero spin-orbit coupling (SOC), we can define a set of spinless symmetries for TBG. In TBG, there are spinless unitary discrete rotational symmetries $C_{2z}$, $C_{3z}$ and $C_{2x}$, and the spinless anti-unitary time-reversal symmetry $T$, which satisfy
\begin{equation}
[C_{3z},\hat{H}_0]=[C_{2z},\hat{H}_0]=[C_{2x},\hat{H}_0]=[T,\hat{H}_0]=0\ .
\end{equation}
We denote the action of a spinless symmetry operator $g$ on the fermion basis $c_{\mathbf{k},\mathbf{Q},\eta,\alpha,s}^{\dagger}$ as
\begin{equation}
g c_{\mathbf{k},\mathbf{Q},\eta,\alpha,s}^{\dagger} g^{-1}=\sum_{\QQ'\eta'\beta} [D(g)]_{\QQ'\eta'\beta,\QQ\eta\alpha} c_{g\mathbf{k},\mathbf{Q}',\eta',\beta,s}^{\dagger}\ ,
\end{equation}
where $D(g)$ is the representation matrix of the symmetry operation $g$ in the space of indices $\{\QQ,\eta,\alpha\}$, and $g\kk$ is the momentum after acting $g$ on momentum $\kk$. In particular, $C_{2z}\kk=T\kk=-\kk$. The representation matrices for the discrete symmetries of TBG are given by
\beq
[D(C_{3z})]_{\QQ^\pr \eta^\pr \beta,\QQ \eta \alpha} = \delta_{\QQ^\pr, C_{3z} \QQ} \delta_{\eta^\pr, \eta} (e^{i\eta  \frac{2\pi}{3}\sigma_{z}})_{\beta\alpha} . \label{eq:C3}
\eeq
\beq
[D (C_{2x})]_{\QQ^\pr \eta^\pr \beta, \QQ \eta \alpha} = \delta_{\QQ^\pr, C_{2x}\QQ} \delta_{\eta^\pr, -\eta} (\sigma_x)_{\beta\alpha}, \label{eq:C2x}
\eeq
\beq
[D(C_{2z})]_{\QQ^\pr \eta^\pr \beta, \QQ\eta \alpha} = \delta_{\QQ^\pr,- \QQ} \delta_{\eta^\pr,-\eta} (\sigma_x)_{\beta\alpha}, \label{eq:C2}
\eeq
\beq
[D(T)]_{\QQ^\pr \eta^\pr \beta, \QQ \eta \alpha}= \delta_{\QQ^\pr,-\QQ} \delta_{\eta^\pr,-\eta} \delta_{\beta,\alpha, \label{eq:TRS}
}\eeq
Moreover, $T$ is anti-unitary, so $T iT^{-1}=-i$. 

In particular, the combined symmetry $C_{2z}T$ does not change $\mathbf{k}$, i.e., $C_{2z}T\kk=\kk$, and the representation matrix is
\beq
[D(C_{2z} T)]_{\QQ^\pr \eta^\pr \beta, \QQ \eta \alpha}= [D(C_{2z}) D(T)]_{\QQ^\pr \eta^\pr \beta, \QQ \eta \alpha} = \delta_{\QQ^\pr,\QQ} \delta_{\eta^\pr,\eta} (\sigma_x)_{\beta,\alpha}. \label{eq:C2T}
\eeq

2. \emph{U(2)$\times$U(2) spin-charge rotation symmetry}. The graphene has zero (negligible) spin-orbit  coupling (SOC). Since the single-particle Hamiltonian of TBG has two decoupled valleys $\eta=\pm$, and the SOC is zero, the electron SU(2) spins of each valley can be rotated freely. Each valley also has a charge U(1) rotation symmetry. This leads to a global U(2)$\times$U(2) symmetry. The 8 generators of the U(2)$\times$U(2) symmetry are given by
\begin{equation}\label{seq:U2U2-gene}
\hat{S}^{ a b}=\sum_{\mathbf{k}} (\tau^a)_{\eta\eta'}(s^b)_{ss'}c_{\mathbf{k},\mathbf{Q},\eta,\alpha,s}^{\dagger} c_{\mathbf{k},\mathbf{Q},\eta',\alpha,s'}\ , \qquad (a=0,z,\quad b=0,x,y,z)\ ,
\end{equation}
where we have defined $\tau^a$ and $s^a$ ($a=0,x,y,z$) as the $2\times2$ identity and Pauli matrices in the valley and spin spaces, respectively.

3. \emph{Particle-hole (PH) transformation $P$}. In addition to the above symmetries, TBG also has a unitary particle-hole (PH) ``symmetry" \cite{song_all_2019}, which satisfies the anti-commutation relation
\begin{equation}
\{P,\hat{H}_0\}=0\ .
\end{equation}
The action of $P$ is given by
\beq
P c^\dagger_{\kk,\QQ,\eta,\alpha,s} P^{-1} = \sum_{\QQ^\pr \eta^\pr \beta} 
[D(P)]_{\QQ^\pr \eta^\pr \beta, \QQ \eta \alpha} c^\dagger_{-\kk,\QQ^\pr,\eta^\pr,\beta,s}\ ,
\eeq
with the representation matrix
\beq
\quad[D(P)]_{\QQ^\pr \eta^\pr \beta, \QQ \eta \alpha} = \delta_{\QQ^\pr,-\QQ} \delta_{\eta^\pr, \eta} \delta_{\beta,\alpha} \zeta_{\QQ}\ , \label{eq:P}
\eeq
where $\zeta_{\QQ} = \pm 1$ for $\QQ\in\mcl{Q}_\pm$, respectively. 
Note that $P$ transforms creation operators to creation operators (rather than annihilation operators), and maps sites $\QQ\in \mathcal{Q}_\pm$ into $-\QQ \in\mathcal{Q}_\mp$. Since $P$ flips the single-particle Hamiltonian $\hat{H}_0$, it is not a commuting symmetry of TBG, but only reflects a relation between the positive and negative energy spectra. Furthermore, the PH transformation $P$ satisfies
\beq\label{seq:P-com}
P^2=-1,\qquad
[P,C_{3z}] = 0,\qquad
\{P,C_{2x}\} = 0,\qquad
\{P,C_{2z}\} = 0,\qquad
\{P,T\} = 0,\qquad
[P,C_{2z}T]=0.
\eeq

\subsection{Eigenstates}

The solutions to the single-particle Hamiltonian $\hat{H}_0$ in \cref{eq:H0} allows us to define the energy band basis
\begin{equation}
c_{\mathbf{k},n,\eta, s}^{\dagger}=\sum_{\mathbf{Q}\alpha}u_{\mathbf{Q}\alpha;n\eta}\left(\mathbf{k}\right)c_{\mathbf{k},\mathbf{Q},\eta,\alpha s}^{\dagger}\ ,\label{eq:solution}
\end{equation}
where $u_{\mathbf{Q}\alpha;n\eta}(\kk)$ is the eigenstate wave function of energy band $n$ of the first quantized single-particle Hamiltonian $h_{\mathbf{Q},\mathbf{Q}^{\prime}}^{\left(\eta\right)}\left(\mathbf{k}\right)$ in valley $\eta$. It satisfies 
\begin{equation}
\sum_{\QQ',\beta}[h_{\mathbf{Q},\mathbf{Q}^{\prime}}^{\left(\eta\right)}\left(\mathbf{k}\right)]_{\alpha\beta}u_{\mathbf{Q}'\beta;n\eta}(\kk)=\epsilon_{n, \eta}(\kk) u_{\mathbf{Q}\alpha;n\eta}(\kk)\ ,
\end{equation}
where $\epsilon_{n, \eta}(\kk)$ is the single-particle energy of eigenstate $u_{\mathbf{Q}\alpha;n\eta}(\kk)$. Note that the wave function $u_{\mathbf{Q}\alpha;n\eta}(\kk)$ and energy $\epsilon_{n, \eta}(\kk)$ are independent of spin $s$, because of the absence of SOC. In each valley and spin, we shall use integers $n>0$ to label the $n$-th conduction band, and use integer $n<0$ to label the $|n|$-th valence band (thus $n\neq0$). The lowest conduction and valence bands in each valley-spin flavor is thus labeled by $n=\pm1$. 

Since $c_{\mathbf{k}+\mathbf{b}_{Mi},\mathbf{Q},\eta\alpha s}^{\dagger}=c_{\mathbf{k},\mathbf{Q}-\mathbf{b}_{Mi},\eta\alpha s}^{\dagger}$ for reciprocal vector $\mathbf{b}_{Mi}$ ($i=1,2$), we generalize the eigenstate wave function to momenta $\kk$ outside the MBZ by the embedding relation for shifting momentum $\kk$ by a reciprocal vector $\mathbf{b}_{Mi}$:
\begin{equation}
u_{\mathbf{Q}\alpha; n\eta}\left(\mathbf{k}+\mathbf{b}_{Mi}\right)=u_{\mathbf{Q}-\mathbf{b}_{Mi},\alpha;n\eta}\left(\mathbf{k}\right)\ . \label{eq:U-embedding}
\end{equation}
This ensures our energy band basis is defined periodically in the MBZ, namely, $c_{\mathbf{k}+\mathbf{b}_{Mi},n\eta s}^{\dagger}=c_{\mathbf{k}n\eta s}^{\dagger}$. Besides, due to the $C_{2z}$ symmetry and PH symmetry $P$, the energy spectrum satisfies
\beq\label{seq:energy-relation}
\epsilon_{n,\eta}(\kk) = \epsilon_{n,-\eta}(-\kk)\ ,\qquad
\epsilon_{n,\eta}(\kk) =-\epsilon_{-n,\eta}(-\kk)\ .
\eeq

The single-particle Hamiltonian can then be rewritten in the energy band basis as
\beq
\hat{H}_0 = \sum_{\kk} \sum_{n \eta s} \epsilon_{n, \eta}(\kk) c_{\kk n \eta s}^\dagger c_{\kk n\eta s} .\label{eq:H0-energyband}
\eeq

\section{Gauge Fixing and the Chern Band Basis}\label{app:gaugefixing}
In this appendix, we fix the gauge for the energy band basis $c_{\mathbf{k}n\eta s}^{\dagger}$ in Eq.~(\ref{eq:solution}), so that we we are able to obtain an explicit form of the interaction Hamiltonian in App.~\ref{app:Mfixing}. We will also define a Chern band basis, whose gauge fixing was shown in Ref.~\cite{ourpaper2}, using the energy band basis. 

\subsection{Sewing matrices}\label{app:sewingmatrices}

The discrete symmetries in App.~\ref{app:tbg-symmetry} yield certain relations among the eigenstate wave functions related by these symmetries. For the purpose of gauge fixing, here we will discuss these relations among eigenstate wave functions for operators $C_{2z},T$ and $P$.

For notation simplicity, we denote the wave function $u_{\mathbf{Q}\alpha;n\eta}(\kk)$ as a column vector $u_{n\eta}(\kk)$ in the space of indices $\{\QQ,\alpha\}$. Furthermore, when a representation matrix $D(g)$ of an operation $g$ (defined in Eqs. \ref{eq:C3} to \ref{eq:P}) acts on a wave function $u_{n\eta'}(\kk)$, we denote the resulting wave function in valley $\eta$ for short as $[D(g)]_{\eta\eta'} u_{n\eta'}(\kk)$, the components of which are given by $\sum_{\mathbf{Q}'\beta\eta'}[D(g)]_{\mathbf{Q}\alpha\eta,\mathbf{Q}'\beta\eta'}u_{\mathbf{Q}'\beta;n\eta'}(\kk)$. Namely, we suppress the indices $\{\QQ,\alpha\}$ of the representation matrix $D(g)$ for short.

When $g$ is a symmetry operator satisfying $[\hat{H}_0,g]=0$ (or $\{\hat{H}_0,g\}=0$), if $u_{n\eta'}(\kk)$ is an eigenstate wave function at momentum $\kk$, the wave function $[D(g)]_{\eta\eta'} u_{n\eta'}(\kk)$ (an additional complex conjugation is needed if $g$ is anti-unitary) must also be an eigenstate wave function at momentum $g\kk$ at the same (or opposite) single-particle energy. For symmetries  $C_{2z},T$ and $P$, this allows us to define the sewing matrices $B^{g}(\kk)$ in the band and valley space connecting the symmetry related eigenstates by
\begin{equation}
[D(C_{2z})]_{\eta\eta'} u_{n\eta'}(\kk)= \sum_{m}[B^{C_{2z}}(\kk)]_{m\eta,n\eta'}u_{m\eta}(-\kk)\ ,
\end{equation}
\begin{equation}
[D(T)]_{\eta\eta'} u_{n\eta'}^*(\kk)= \sum_{m}[B^{T}(\kk)]_{m\eta,n\eta'}u_{m\eta}(-\kk)\ ,
\end{equation}
\begin{equation}
[D(P)]_{\eta\eta'} u_{n\eta'}(\kk)=\sum_{m}[B^{P}(\kk)]_{m\eta,n\eta'}u_{m\eta}(-\kk)\ .
\end{equation}
For non-degenerate wave function $u_{n\eta'}(\kk)$ in valley $\eta'$, since $C_{2z}$ and $T$ commute with the $\hat{H}_0$ and flips the valley $\eta$, while $P$ anti-commutes with $\hat{H}_0$ and preserves the valley $\eta$, we generically have
\begin{equation}\label{seq:sewing}
\begin{split}
&[B^{C_{2z}}(\kk)]_{m\eta,n\eta'}=\delta_{\eta,-\eta'}\delta_{m,n} e^{i\varphi^{C_{2z}}_{n,\eta'}(\kk)}\ , \quad [B^{T}(\kk)]_{m\eta,n\eta'}=\delta_{\eta,-\eta'}\delta_{m,n} e^{i\varphi^{T}_{n,\eta'}(\kk)}\ ,\\
& [B^{P}(\kk)]_{m\eta,n\eta'}=\delta_{\eta,\eta'}\delta_{-m,n} e^{i\varphi^{P}_{n,\eta'}(\kk)}\ .
\end{split}
\end{equation}
Accordingly, the action of a symmetry operator $g$ on the energy band fermion operators (defined in \cref{eq:solution}) is given by
\begin{equation}
g c_{\mathbf{k},n,\eta' ,s}^{\dagger} g^{-1}= \sum_{m\eta} [B^g(\kk)]_{m\eta,n\eta'} c_{g\mathbf{k},m,\eta, s}^{\dagger}\ .
\end{equation}
Since the three symmetries satisfy the relations
\begin{equation}
C_{2z}^2=1\ ,\ T^2=1\ ,\ P^2=-1\ ,\ \{P,C_{2z}\}=0\ ,\ \{P,T\}=0\ ,\ [C_{2z},T]=0\ ,
\end{equation}
With the above notations, the symmetries $C_{2z},T$ and $P$ allows us to define
\begin{equation}
\begin{split}
&B^{C_{2z}}(-\kk)B^{C_{2z}}(\kk)=B^{T}(-\kk)B^{T*}(\kk)=-B^{P}(-\kk)B^{P}(\kk)=I\ ,\qquad B^{P}(-\kk)B^{C_{2z}}(\kk)=-B^{C_{2z}}(-\kk)B^{P}(\kk)\ , \\
&\qquad\qquad B^{P}(-\kk)B^{T}(\kk)=-B^{T}(-\kk)B^{P*}(\kk)\ ,\qquad 
B^{T}(-\kk)B^{C_{2z}*}(\kk)=-B^{C_{2z}}(-\kk)B^{T}(\kk)\ ,
\end{split}
\end{equation}
where $B^{g*}(\kk)$ stands for the complex conjugation of matrix $B^g(\kk)$, and $I$ is the identity matrix in the $n,\eta$ space. More discussions on the sewing matrices can be found in Ref.~\cite{ourpaper2}.

The combination of the three symmetries yields two independent symmetry operations $C_{2z}T$ and $C_{2z}P$ which do not change $\kk$. Note that $C_{2z}T$ is anti-unitary, and $C_{2z}P$ is unitary. Their sewing matrices are defined by
\begin{equation}\label{seq:B-C2T}
[D(C_{2z})D(T)]_{\eta\eta'} u_{n\eta'}^*(\kk)=\sum_{m}[B^{C_{2z}T}(\kk)]_{m\eta,n\eta'}u_{m\eta}(\kk)\ ,
\end{equation}
\begin{equation}\label{seq:B-C2P}
[D(P)D(C_{2z})]_{\eta\eta'} u_{n\eta'}(\kk)= \sum_{m}[B^{C_{2z}P}(\kk)]_{m\eta,n\eta'}u_{m\eta}(\kk)\ .
\end{equation}
For non-degenerate eigenstates at momentum $\kk$ (non-degenerate within one valley), they are given by
\begin{equation}\label{seq:sewing-C2T-C2P}
[B^{C_{2z}T}(\kk)]_{m\eta,n\eta'}=\delta_{\eta,\eta'}\delta_{m,n} e^{i\varphi^{C_{2z}T}_{n,\eta'}(\kk)}\ , \qquad
[B^{C_{2z}P}(\kk)]_{m\eta,n\eta'}=\delta_{-\eta,\eta'}\delta_{-m,n} e^{i\varphi^{C_{2z}P}_{n,\eta'}(\kk)}\ ,
\end{equation}
where by definition we have $\varphi^{C_{2z}T}_{n,\eta'}(\kk)=\varphi^{T}_{n,\eta'}(\kk)+\varphi^{C_{2z}}_{n,-\eta'}(-\kk)$, and $\varphi^{C_{2z}P}_{n,\eta'}(\kk)=\varphi^{C_{2z}}_{n,\eta'}(\kk)+\varphi^{P}_{n,-\eta'}(-\kk)$. The sewing matrices of $C_{2z}T$ and $C_{2z}P$ are subject to the constraint that
\begin{equation}
(C_{2z}T)^2=(C_{2z}P)^2=1\ ,\qquad [C_{2z}T,C_{2z}P]=1\ ,
\end{equation}
thus they satisfy
\begin{equation}
B^{C_{2z}T}(\kk)B^{C_{2z}T*}(\kk)=[B^{C_{2z}P}(\kk)]^2=I\ , \qquad B^{C_{2z}P}(\kk)B^{C_{2z}T}(\kk)=B^{C_{2z}T}(\kk)B^{C_{2z}P*}(\kk)\ .\label{seq:sewing-C2T-C2P-relation}
\end{equation}

\subsection{Gauge fixing}\label{app:gauge-fixing}

We will now gauge fix the wave functions and sewing matrices of the $\kk$ preserving symmetry operations $C_{2z}T$ and $C_{2z}P$. By Eqs. (\ref{seq:sewing-C2T-C2P}) and (\ref{seq:sewing-C2T-C2P-relation}), we are able to choose the following $\kk$ independent choices for the sewing matrices:
\begin{equation}\label{eq:gauge-n}
[B^{C_{2z}T}(\kk)]_{m\eta,n\eta'}=\delta_{\eta,\eta'}\delta_{m,n} \ , \qquad
[B^{C_{2z}P}(\kk)]_{m\eta,n\eta'}=-\text{sgn}(n)\eta' \delta_{-\eta,\eta'}\delta_{-m,n}\ .
\end{equation}
Accordingly, the symmetry actions on the band basis fermion operators are given by
\begin{equation}
(C_{2z}T) c_{\mathbf{k},n,\eta ,s}^{\dagger} (C_{2z}T)^{-1}=c_{\mathbf{k},n,\eta ,s}^{\dagger}\ ,\qquad (C_{2z}P) c_{\mathbf{k},n,\eta ,s}^{\dagger} (C_{2z}P)^{-1}=-\text{sgn}(n)\eta c_{\mathbf{k},-n,-\eta ,s}^{\dagger}\ .
\end{equation}
This, however, does not yet fix the entire phases of the energy basis at momentum $\kk$, since the sewing matrices in Eq.~(\ref{eq:gauge-n}) are invariant under the unitary transformation of wave functions $u_{n\eta}(\kk)\rightarrow \text{sgn}(n)\eta u_{n\eta}(\kk)$ at each individual $\kk$. To further fix this gauge freedom for different $\kk\in\text{MBZ}$, we start by choosing a momentum $\kk=\kk_0$ where eigenstates within one valley are nondegenerate, and choose a fixing of the band basis at $\kk_0$ satisfying Eq.~(\ref{eq:gauge-n}). We then fix the band basis of bands $\pm n$ at other $\kk\neq\kk_0$ by requiring
\begin{equation}
f_{n,\eta}(\kk+\qq,\kk)= \left|u^\dag_{n,\eta}(\kk+\qq) u_{n,\eta}(\kk)- u^\dag_{-n,\eta}(\kk+\qq) u_{-n,\eta}(\kk)\right|
\end{equation} 
to be a continuous function of $\kk$ and $\qq$, and satisfies 
\begin{equation}\label{seq:c-continuous}
\lim_{\qq\rightarrow \mathbf{0}}f_{n,\eta}(\kk+\qq,\kk)=0
\end{equation} 
for all $\kk$. Meanwhile, we require the wave functions $u_{n,\eta}(\kk)$ at all $\kk$ to satisfy Eq.~(\ref{eq:gauge-n}). This fixes the relative sign between wave functions $u_{n,\eta}(\kk)$ and $u_{-n,\eta}(\kk)$ in a way that is continuous in $\kk$. Note that we do not require the wave function $u_{n,\eta}(\kk)$ itself to be globally continuous in $\kk$ of the entire MBZ, which is impossible when the band $n$ is topological. However, locally $u_{n,\eta}(\kk)$ can always be chosen to be continuous in $\kk$, provided $u_{n,\eta}(\kk)$ is non-degenerate at momentum $\kk$. We will see the importance of condition (\ref{seq:c-continuous}) in App.~\ref{app:chern} again.

We also note that, we could alternatively define the continuous condition between the same $n$ but opposite $\eta$ bands as $\lim_{\qq\rightarrow\mathbf{0}}\left|u^\dag_{n,\eta}(\kk+\qq) u_{n,\eta}(\kk)- u^\dag_{n,-\eta}(\kk+\qq) u_{n,-\eta}(\kk)\right|=0$. Together with Eq.~(\ref{eq:gauge-n}), this is equivalent to condition (\ref{seq:c-continuous}).
 
In particular, we see that all the sewing matrices in Eq. (\ref{eq:gauge-n}) are closed within each pair of bands $n=\pm n_\text{B}$ for any $n_\text{B}\ge1$. The same is true for all the sewing matrices we will consider in this paper, which are either commuting or anti-commuting with the single-particle Hamiltonian $\hat{H}_0$. Within the space of each pair of PH symmetric bands with band indices $n=\pm n_\text{B}$, if we use $\zeta^a$ and $\tau^a$ ($a=0,x,y,z$) to denote the identity and Pauli matrices in the energy band $n=\pm n_\text{B}$ space and the valley space, respectively, the sewing matrices in Eq.~(\ref{eq:gauge-n}) can be rewritten as
\begin{equation}\label{eq:gauge-0}
B^{C_{2z}T}(\kk)=\zeta^0\tau^0\ ,\qquad B^{C_{2z}P}(\kk)=\zeta^y\tau^y\ .
\end{equation}
We also mention that for $n_\text{B}=1$ (i.e., within the lowest conduction and valence bands $n=\pm1$ per spin per valley) when $\kk$ is at $K_M$ or $K_M'$ point of the MBZ, bands $n=+1$ and $n=-1$ are degenerate. In this case, we still choose the eigenstate basis at $K_M$ or $K_M'$ point such that Eqs. (\ref{eq:gauge-0}) and (\ref{seq:c-continuous}) are satisfied.

Lastly, we note that we can further fix the relative gauge between wave functions at momenta $\kk$ and $-\kk$ by fixing the sewing matrices of $C_{2z}$ and $P$. 
In particular, for $\kk$ not at the $P$-invariant momenta, which are $\Gamma_M$ and the three equivalent $M_M$ in TBG, one can choose the sewing matrices of $C_{2z}$, $T$ and $P$ between each pair of bands $n=\pm n_\text{B}$ as
\begin{equation}\label{eq:gauge-1}
B^{C_{2z}}(\kk)=\zeta^0\tau^x\ ,\qquad B^{T}(\kk)=\zeta^0\tau^x\ , \qquad B^{P}(\kk)=-i\zeta^y\tau^z\ .
\end{equation}
which are consistent with Eq.~(\ref{eq:gauge-0}). 
As proven in the next subsection, with the gauge condition \cref{seq:c-continuous}, the sewing matrix $B^P(\kk)$ must have additional minus signs, \ie $B^P(\kk)=i\zeta^y\tau^z$, at an odd (even) number of the four $P$-invariant momenta if the the two bands $n=\pm n_B$ have an odd (even) topological winding number protected by $C_{2z}T$; and at the other odd (even) $P$-invariant momenta $B^P(\kk)$ are $-i\zeta^y\tau^z$, same as those at generic momenta.
Accordingly, the sewing matrices $B^{C_{2z}}(\kk)$ and $B^{T}(\kk)$ also have the additional minus at momenta where $B^P(\kk)$ has the minus sign.
In this work, we choose $B^P(\kk_{\Gamma_M})=-i\zeta^y \tau^z$ and $B^P(\kk_{M_M})=i\zeta^y \tau^z$. It should be noticed that \cref{eq:gauge-1} is incompatible with the second chiral symmetry, which we explain in Sec. \ref{app:2ndchiral-flat}.

For the purpose of this paper, we will use the gauge conditions in Eqs.~(\ref{eq:gauge-0}) and~(\ref{eq:gauge-1}) for gauge fixing of the interaction Hamiltonian in App.~\ref{app:HI}.

\subsection{The Irrep band basis and Chern band basis}\label{app:chern}
After we have gauge fixed the wave functions as shown in Eqs.~(\ref{eq:gauge-0}) and~(\ref{seq:c-continuous}), we have defined a new basis $d^{(n_\text{B})\dag}_{\mathbf{k},e_Y,\eta, s}$ in Eq. (\ref{eq-irrepbasis}) within the band space of each pair of PH symmetric bands $n=\pm n_\text{B}$, which we call the irrep basis:
\begin{equation}\label{seq-irrepbasis}
d^{(n_\text{B})\dag}_{\mathbf{k},e_Y,\eta, s}=\frac{c^\dag_{\mathbf{k},n_\text{B},\eta, s}+ie_Y c^\dag_{\mathbf{k},-n_\text{B},\eta, s}}{\sqrt{2}}\ , \quad (e_Y=\pm1).
\end{equation}
In particular, for $n_\text{B}=1$, we call them the Chern band basis within the lowest two bands (in each valley-spin flavor), which we denote for simplicity as $d^{(1)\dag}_{\mathbf{k},e_Y,\eta, s}=d^\dagger_{\kk,e_Y,\eta,s}$, as given in Eq. (\ref{eq-Chernbasis}), 
where $e_Y=\pm1$. This basis will be useful when we discuss the symmetries in various limits in App.~\ref{app:various-limits}. 

In this appendix, we briefly show that the basis $d^{(n_\text{B})\dag}_{\mathbf{k},e_Y,\eta, s}$ defines a
band with well-defined Berry curvature, and for a fixed $e_Y,\eta,s$ gives a band with Chern number
\begin{equation}\label{seq:C-chern-band}
C^{n_\text{B}}_{e_Y,\eta,s} =e_Ye_{2,n_\text{B}}\ ,
\end{equation}
where $e_{2,n_\text{B}}\in\mathbb{Z}$ is the Wilson loop winding number of the two bands $n=\pm n_\text{B}$, provided the pair of bands $n=\pm n_\text{B}$ are disconnected with other bands. More details can be found in Ref.~\cite{ourpaper2}. 

The wave functions of the Chern band basis in Eq.~(\ref{seq-irrepbasis}) are given by (denoted by wave functions with a prime)
\begin{equation}
u_{e_Y,n_\text{B},\eta}'(\kk)=\frac{u_{+n_\text{B},\eta}(\kk)+ie_Yu_{-n_\text{B},\eta}(\kk)}{\sqrt{2}}\ . \label{szdeq:Chern-band-u}
\end{equation}
Due to the condition in Eq.~(\ref{seq:c-continuous}), we know that $\lim_{\qq\rightarrow \mathbf{0}} u^\dag_{+n_\text{B},\eta}(\kk+\qq) u_{+n_\text{B},\eta}(\kk)=\lim_{\qq\rightarrow \mathbf{0}} u^\dag_{-n_\text{B},\eta}(\kk+\qq) u_{-n_\text{B},\eta}(\kk)$. Therefore, we find the Chern band wave functions satisfy the continuous condition
\begin{equation}
\lim_{\qq\to 0} |u_{e_Y,n_\text{B},\eta}'^\dag(\kk+\qq)u_{e_Y',n_\text{B},\eta}'(\kk)| = \frac{1}{2}\lim_{\qq\to 0} |u^\dag_{+n_\text{B},\eta}(\kk+\qq) u_{+n_\text{B},\eta}(\kk)+e_Ye_Y'u^\dag_{-n_\text{B},\eta}(\kk+\qq) u_{-n_\text{B},\eta}(\kk)| = \delta_{e_Y,e_Y'}, \label{eq:Chern-band-gauge}
\end{equation} 
This continuous condition (which is due to condition (\ref{seq:c-continuous})) allows us to define a continuous Berry curvature for the Chern band wave function $u_{e_Y',n_\text{B},\eta}'(\kk)$.

We first focus in the valley $\eta=+$ sector. The sewing matrix for $C_{2z}T$ restricted in valley $\eta=+$ is given by $B^{C_{2z}T}(\kk)=\zeta^0$ (see Eq.~(\ref{eq:gauge-0})). Under this gauge, according to \cite{ahn_failure_2019}, the non-abelian Berry's connection $[\mbf{A}(\kk)]_{mn}=i u^\dag_{m,+}(\kk)\partial_\kk u_{n,+}(\kk)$ will take the form
\begin{equation}\label{seq:berry0}
\mbf{A}(\kk) = \begin{pmatrix}
0 & i\aa(\kk) \\
-i\aa(\kk) & 0
\end{pmatrix}
\end{equation}
in the energy band basis $u_{n,+}(\kk)$ of $n=\pm n_\text{B}$. The sign of wave functions $u_{n,+}(\kk)$ is fixed in such a way that $\aa(\kk)$ is globally continuous in the BZ excluding the Dirac nodes between the two bands $\pm n_\text{B}$ (recall that we assume the bands $\pm n_\text{B}$ are disconnected from other bands, thus there can be Dirac nodes between them only if $n_B=1$), which is always possible \cite{ahn_failure_2019}. In particular, this way of sign fixing is consistent with  \cref{seq:c-continuous}, since the vanishing of the diagonal Berry's connection requires $\lim_{\qq\to 0} |u^\dag_{m,\eta}(\kk+\qq) u_{n,\eta}(\kk)| =\delta_{m,n}$. 

It is known that the Wilson loop winding number of two bands isolated from other bands is given by the Euler class \cite{ahn_failure_2019}:
\begin{equation}\label{seq:e2}
e_{2,n_\text{B}} = \frac{1}{2\pi} \sum_i \oint_{\partial D_i} d\kk \cdot \aa(\kk)
= \frac1{2\pi} \int_{\mrm{MBZ}-\sum_i D_i } d^2\kk\ \Omega(\kk)\ ,
\end{equation}
where $D_i$ is a sufficiently small region containing the $i$th Dirac point in the BZ, and $\Omega(\kk) = \nabla_\kk \times \aa(\kk)$. 

With Eq.~(\ref{seq:berry0}), we can derive the Berry connection of the irrep band basis $d^\dg_{\kk,e_Y,+,s}$ at $\kk$ away from Dirac points as
\begin{align}
&\mbf{A}^{\pr}_{e_Y}(\kk) =i u'^\dag_{e_Y,n_\text{B},+}(\kk)\partial_\kk u_{e_Y,n_\text{B},+}'(\kk) \nonumber \\ 
&=\frac{i}{2} [u^\dag_{+n_\text{B},+}(\kk)\partial_\kk u_{+n_\text{B},+}(\kk) +i e_Y u^\dag_{+n_\text{B},+}(\kk)\partial_\kk u_{-n_\text{B},+}(\kk)-ie_Y u^\dag_{-n_\text{B},+}(\kk)\partial_\kk u_{+n_\text{B},+}(\kk)+u^\dag_{-n_\text{B},+}(\kk)\partial_\kk u_{-n_\text{B},+}(\kk)] \nonumber \\
&=e_Y \aa(\kk)\ .
\end{align}
Furthermore, the Berry curvature can be shown to be non-divergent at the Dirac points between the two bands $n=\pm n_\text{B}$ (see proof in \cite{ourpaper2}. If $n_\text{B}>1$, there are no Dirac points between bands $n=\pm n_\text{B}$). Therefore, by Eq.~(\ref{seq:e2}), we find the irrep basis $d^{(n_\text{B})\dg}_{\kk,e_Y,+,s}$ carries a Chern number given by Eq. (\ref{seq:C-chern-band}). 

Further, note that the $C_{2z}$ symmetry maps the irrep basis $d^{(n_\text{B})\dg}_{\kk,e_Y,+,s}$ into $d^{(n_\text{B})\dg}_{-\kk,e_Y,-,s}$ (see Eq.~(\ref{eq:gauge-1})). Since $C_{2z}$ does not change the Chern number, we conclude that the Chern number of the irrep basis $d^{(n_\text{B})\dg}_{\kk,e_Y,\eta,s}$ in the MBZ is simply given by Eq. (\ref{seq:C-chern-band}).

In particular, for the lowest two bands $n_\text{B}=1$, the bands are topological and carry a winding number $e_2=1$ \cite{song_all_2019,po_faithful_2019,ahn_failure_2019}. Therefore, for the Chern band basis (the irrep basis with $n_\text{B}=1$) $d^{\dg}_{-\kk,e_Y,-,s}$, we have Chern number
\begin{equation}
C_{e_Y,\eta,s} =e_Y\ ,
\end{equation}
thus the name ``Chern band basis" within the lowest two bands (see \cite{ourpaper2} for a more careful treatment at the Dirac points at CNP, which does not change the conclusion).

For $n_\text{B}>1$, if the two bands $n=\pm n_\text{B}$ are isolated from other bands, they will be trivial, and thus $e_{2,n_\text{B}}=0$ for $n_\text{B}>1$ \cite{song_all_2019,po_faithful_2019,ahn_failure_2019}. Therefore, they will have Chern number $C^{n_\text{B}}_{e_Y,\eta,s}=0$.

Now we show that if $e_{2,n_B}$ is odd, then the sign  of the sewing matrix $B^P(\kk)$ must be $\kk$-dependent: for $\eta=+$, $B^P(\kk)$ can be chosen as $-i\zeta^y$ at all the momenta except one or three of the $P$-invariant momenta, where $B^P(\kk)$ must be $i\zeta^y$.
To see this, we assume $B^P(\kk)=-i\chi(\kk) \zeta^y$, where $\chi(\kk)=\pm1$, and transform it into the Chern band basis \cref{szdeq:Chern-band-u}.
We obtain 
\begin{equation}
B^{P\prime}_{e_Y,e_Y'}(\kk) = u^{\prime\dagger}_{e_Y,n_B,+}(-\kk) D(P) u^\prime_{e_Y,n_B,+}(\kk) = -i \chi(\kk) e_Y \delta_{e_Y,e_Y'}.
\end{equation}
Therefore, $P$ leaves each branch of the Chern band basis, which has the Chern numbers $e_{2,n_B}e_Y$, invariant.
$iP$ can be equivalently thought as an inversion symmetry for each Chern band since it squares to 1 and changes $\kk$ to $-\kk$. 
The ``inversion'' eigenvalues of the Chern band $e_Y$ are given by $\chi(\kk) e_Y$ for $\kk$ being the $P$-invariant momentum. 
Due to the relation between Chern number and inversion eigenvalues, we have
\begin{equation}
    (-1)^{e_{2,n_B}} = \prod_{K} \chi(K), 
\end{equation}
where $K$ indexes the four $P$-invariant momenta. 
Therefore, the right hand side must be -1 (1) if $e_{2,n_B}$ is odd (even), implying $\chi(K)=-1$ at one or three (zero, two, or four) of the four $P$-invariant momenta. The sign of $B^{P}(\kk)$ in the other valley $\eta=-$ can be obtained from the constraint between $B^{C_{2z}P}(\kk)$ and $B^{P}(\kk)$.

In the case when a pair of bands $n=\pm n_\text{B}$ are not isolated, the Chern number $C^{n_\text{B}}_{e_Y,\eta,s}$ is not clearly well defined.
We leave this question for future studies.

\section{Interacting Hamiltonian with Coulomb Interaction}\label{app:HI}

In this appendix, we write down the interaction Hamiltonian of TBG for the Coulomb interaction with screening from the top and bottom gates.

\subsection{Low energy interaction}\label{app:interaction}

\begin{figure}
\begin{centering}
\includegraphics[width=0.5\linewidth]{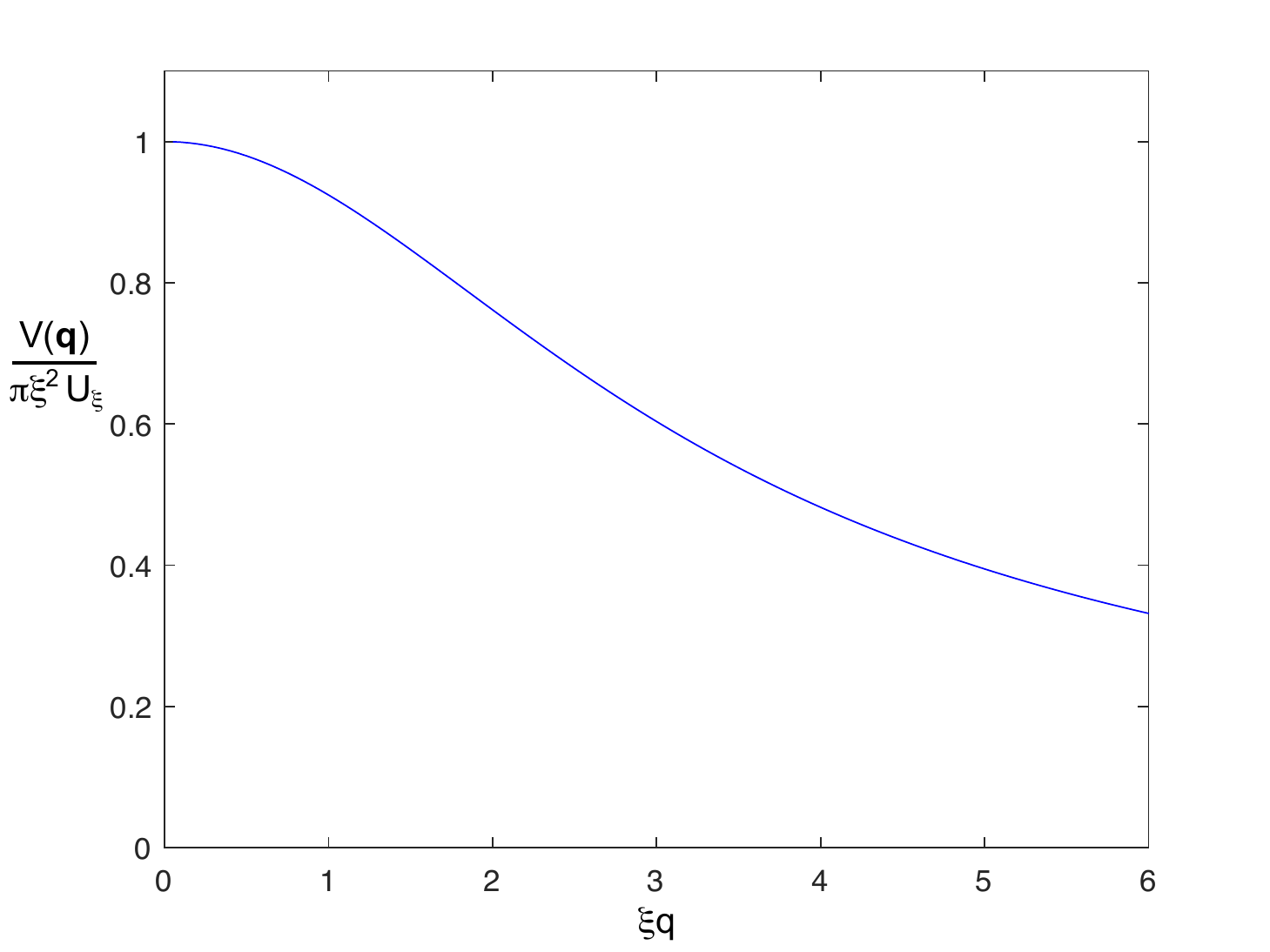}
\par\end{centering}

\protect\caption{\label{fig:Vq}The interaction $V(\qq)$ as a function of $\xi q$ given by Eq. \ref{seq:Vq}. }

\end{figure}

We denote the (screened) Coulomb interaction in TBG between two electrons of distance $\mathbf{r}$ as $V(\mathbf{r})$. Usually, TBG samples in experiments feel the Coulomb screenings from the top and bottom gates. Here we assume the TBG has a top gate plate and bottom gate plate which are distance $\xi$ away in the $z$ direction. The screened Coulomb interaction is then given by
\begin{equation}\label{seq:Vr}
\widetilde{V}\left(\mathbf{r}\right)=U_{\xi}\sum_{n=-\infty}^{\infty}\frac{\left(-1\right)^{n}}{\sqrt{\left(r/\xi\right)^{2}+n^{2}}}\ ,
\end{equation}
where $U_{\xi}=e^{2}/\left(\epsilon\xi\right)$, 
with $\epsilon$ being the dielectric constant, and $r=|\mathbf{r}|$. 
We call $\xi$ the screening length, which is usually around $10$nm and comparable to the moir\'e lattice constant. For $\epsilon\approx 6$ from typical hBN substrates, and $\xi\approx 10$nm, we have $U_{\xi}\approx 24$meV. Using the 2D Fourier transformation formula that
\begin{equation}
\begin{split}
&\int\frac{d^{2}\mathbf{q}}{\left(2\pi\right)^{2}}\cdot\frac{e^{-\xi q+i\mathbf{q}\cdot\mathbf{r}}}{q}=\int_0^{\infty} dq\int_0^{2\pi}d\theta e^{-\xi q+iqr\cos\theta} = \int_0^{2\pi}d\theta \frac{1}{\xi-ir\cos\theta} = \oint_{|z|=1} \frac{dz}{\xi z-ir(z^2+1)/2} \\
&=\frac{1}{2\pi}\frac{1}{\sqrt{\xi^{2}+r^{2}}}\ , \qquad\qquad\qquad\left(\xi\ge0\right), \label{eq:Fourier0}
\end{split}
\end{equation}
we find the Fourier transformation of the Coulomb interaction (\ref{seq:Vr}) is 
\begin{equation}\label{seq:Vq}
\begin{split}
V\left(\mathbf{q}\right)&=\int d^2\mathbf{r} e^{-i\mathbf{q\cdot r}} \widetilde{V}\left(\mathbf{r}\right)= \xi U_\xi \sum_{n=-\infty}^{\infty} \int d^2\mathbf{r}  \frac{\left(-1\right)^{n}e^{-i\mathbf{q\cdot r}}}{\sqrt{r^{2}+(n\xi)^{2}}} \\
&= 2\pi\xi U_\xi \sum_{n=-\infty}^{\infty} \int d^2\mathbf{r}\int \frac{d^2\qq'}{(2\pi)^2}  \left(-1\right)^{|n|}\frac{e^{-|n|\xi q'+i(\mathbf{q}'-\qq)\cdot\mathbf{r}}}{q'}  = 2\pi\xi U_\xi \sum_{n=-\infty}^{\infty} \int d^2\mathbf{q}'\delta^2(\qq-\qq')  \left(-1\right)^{n}\frac{e^{-|n|\xi q'}}{q'} \\
&= 2\pi\xi U_\xi \sum_{n=-\infty}^{\infty}  \left(-1\right)^{n}\frac{e^{-|n|\xi q}}{q} =\left(\pi\xi^{2}U_{\xi}\right)\frac{\tanh\left(\xi q/2\right)}{\xi q/2}=\frac{2\pi e^2}{\epsilon}\frac{\tanh\left(\xi q/2\right)}{q}\ ,
\end{split}
\end{equation}
where $q=|\mathbf{q}|$, and we have used the formula $\sum_{n=-\infty}^{\infty}e^{-|n|x}=1-\frac{2e^{-x}}{1+e^{-x}}=\tanh\left(\frac{x}{2}\right)$. Note that $V(-\qq)=V(\qq)$. The function $V(\qq)$ with respect to $\zeta q$ is plotted in Fig. \ref{fig:Vq}. 

The Coulomb interaction of the 2D TBG electrons can be written in the momentum space under the graphene plane wave basis as $c_{\mathbf{p},\alpha,s,l}^{\dagger}$ as
\begin{equation}\label{seq:HI-micro}
\hat{H}_I=\frac{1}{2\Omega_{\text{tot}}}\sum_{\mathbf{p,p',q}\in\text{GBZ}} \sum_{\alpha,\alpha',s,s',l,l'}V(\mathbf{q}) \left( c_{\mathbf{p+q},\alpha,s,l}^{\dagger} c_{\mathbf{p},\alpha,s,l} -\frac{1}{2}\delta_{\mathbf{q,0}} \right) \left(c_{\mathbf{p'-q},\alpha',s',l'}^{\dagger} c_{\mathbf{p'},\alpha',s',l'}-\frac{1}{2}\delta_{\mathbf{q,0}}\right)\ ,
\end{equation}
where $\mathbf{p,p',q}$ takes values in the microscopic graphene BZ, and $\Omega_{tot}$ is the total area of TBG. Note that we did not normal-order the interaction Hamiltonian $\hat{H}_I$ in Eq.~(\ref{seq:HI-micro}), and have subtracted a $\frac{1}{2}\delta_{\mathbf{q,0}}$ term in the two brackets of fermion operators. Normal-ordering or removing the term $\frac{1}{2}\delta_{\mathbf{q,0}}$ only shifts $\hat{H}_I$ by a chemical potential term of the form $\mu\sum_{\pp,\alpha,s,l} c_{\mathbf{p},\alpha,s,l}^{\dagger} c_{\mathbf{p},\alpha,s,l}$, which does not change the general physics. However, the advantage of the form in Eq.~(\ref{seq:HI-micro}), the Hamiltonian $\hat{H}_I$ is symmetric about the filling of the charge neutral point (CNP). In particular, this chemical potential shift allows us to easily obtain a many-body PH symmetric projected Hamiltonian, as we will derive below and discuss in more details in App.~\ref{app:manybody-cc}. The derived many-body PH symmetric projected Hamiltonian is the most appropriate one, as it effectively properly includes the Hartree-Fock contributions from the passive bands (App.~\ref{app:passivebandcontributions}).

The low energy physics of TBG is concentrated at microscopic electron momenta $\mathbf{p}$ around the two valleys $\pm \mathbf{K}_l$. Since $V(\mathbf{q})$ decays quickly when $q\gg 1/\xi$, and in TBG $|\mathbf{K}_l|\gg 1/\xi$, we can ignore the terms in Eq.~(\ref{seq:HI-micro}) with $|\qq|\sim |\mathbf{K}_l|$ connecting two valleys. After this approximation, at low energies we can assume $\pp$ and $\pp+\qq$ ($\pp'$ and $\pp'+\qq$) belong to the same graphene valley, namely, only intra-valley scattering is preserved. Rewriting the fermion operators using Eqs. (\ref{eq:ckQ+}) and (\ref{eq:ckQ-}), we can rewrite the low energy interaction Hamiltonian as
\begin{equation}\label{seq-HI}
\hat{H}_I=\frac{1}{2\Omega_{\text{tot}}}\sum_{\mathbf{G}\in\mathcal{Q}_0}\sum_{\mathbf{q}\in \text{MBZ}}V(\mathbf{q+G})\delta\rho_{\mathbf{-q-G}}\delta\rho_{\mathbf{q+G}}\ ,
\end{equation}
where 
\begin{equation}\label{eq:drho-0}
\delta\rho_{\mathbf{q+G}}=\sum_{\eta,\alpha,s}\sum_{\mathbf{k}\in\text{MBZ}}\sum_{\mathbf{Q}\in\mathcal{Q}_\pm} \left(c_{\mathbf{k+q},\mathbf{Q-G},\eta,\alpha, s}^\dag c_{\mathbf{k},\mathbf{Q},\eta,\alpha, s}-\frac{1}{2}\delta_{\mathbf{q,0}}\delta_{\mathbf{G,0}}\right)\ .
\end{equation} 
Physically, $\delta\rho_{\mathbf{q+G}}$ is the Fourier transform of the total electron density at momentum $\mathbf{q+G}$ relative to the filling of the graphene CNP (since the CNP of TBG when the two layers are decoupled is at half filling $\langle c_{\mathbf{k+q},\mathbf{Q-G},\eta,\alpha, s}^\dag c_{\mathbf{k},\mathbf{Q},\eta,\alpha, s}\rangle=\frac{1}{2}\delta_{\mathbf{q,0}}\delta_{\mathbf{G,0}}$ in both graphene layers).

\subsection{Projected Hamiltonian}\label{app:proj-H}

We now project the TBG Hamiltonian into the lowest $8n_\text{max}$ bands $|n|\le n_\text{max}$ in each spin and valley. 
When the twist angle $\theta$ is close to the magic angle $\theta_M\approx 1.1^\circ$, a reasonable projected Hamiltonian is with $n_\text{max}=1$. 
To distinguish with the unprojected Hamiltonians $\hat{H}_0$ and $\hat{H}_I$ in Eqs. (\ref{eq:H0}) and (\ref{seq-HI}) which have a hat, we denote the projected kinetic and interaction Hamiltonians as $H_0$ and $H_I$ (without a hat), and the total projected Hamiltonian as $H=H_0+H_I$.

From Eq.~(\ref{eq:H0-energyband}) we can easily write down the projected kinetic Hamiltonian into $|n|\le n_\text{max}$ bands as:
\beq
H_0 =\sum_{|n|\le n_\text{max}}\sum_{\eta s} \sum_{\kk\in\text{MBZ}}  \epsilon_{n, \eta}(\kk) c_{\kk n \eta s}^\dagger c_{\kk n\eta s}\ .\label{eq:H0-proj}
\eeq

To find the projected interaction Hamiltonian, we first note that due to \cref{eq:solution}, the density operator in \cref{eq:drho-0} can be written as
\beqs
\delta \rho_{\mathbf{G+q}}=&  \sum_{\eta \alpha s} \sum_{\kk} \sum_{\QQ\in\mcl{Q}_\pm} 
	\pare{ \pare{\sum_{mn} u^*_{\QQ-\GG,\alpha;m\eta}(\kk+\qq) u_{\QQ,\alpha;n\eta} (\kk)
	c^\dagger_{\kk+\qq,m,\eta,s} c_{\kk,n,\eta,s} }
	- \frac12 \delta_{\qq,0} \delta_{\GG,0}} \nono\\
=&\sum_{\eta \alpha s} \sum_{\kk} \sum_{\QQ\in\mcl{Q}_\pm} \sum_{m,n} 
	u^*_{\QQ-\GG,\alpha;m\eta}(\kk+\qq) u_{\QQ,\alpha;n\eta} (\kk)  
	\pare{ c^\dagger_{\kk+\qq,m,\eta,s} c_{\kk,n,\eta,s} 
	- \frac12 \delta_{\qq,0} \delta_{mn}}, \label{eq:drho-trans1}
\eeqs
where from the first line to the second line we have used the completeness relation 
\beq
\delta_{\GG,\mathbf{0}}=\sum_{n\eta }u^*_{\QQ-\GG,\alpha;n\eta}(\kk) u_{\QQ,\alpha;n\eta} (\kk).
\eeq
We then define the form factor (overlap) matrix as given in Eq. (\ref{eq:M-def0}), which we reprint here for convenience:
\begin{equation}
M_{m,n}^{\left(\eta\right)}\left(\mathbf{k},\mathbf{q}+\mathbf{G}\right)=\sum_{\alpha}\sum_{\substack{\mathbf{Q}\in\mathcal{Q}_{\pm}}
}u_{\mathbf{Q}-\mathbf{G},\alpha;m\eta}^{*}\left(\mathbf{k}+\mathbf{q}\right)u_{\mathbf{Q},\alpha;n\eta}\left(\mathbf{k}\right) . \label{eq:M-def}
\end{equation}
We note that if $\kk+\qq$ is outside the first BZ, it must be brought into the first BZ using the embedding relation in \cref{eq:U-embedding}.
This further simplifies Eq.~\ref{eq:drho-trans1} into
\beq
\delta \rho_{\GG+\qq} = \sum_{\kk \eta s} \sum_{m,n} M_{m,n}^\eta (\kk,\qq+\GG) 
	\pare{ c^\dagger_{\kk+\qq,m,\eta,s} c_{\kk,n,\eta,s} 
	- \frac12 \delta_{\qq,0} \delta_{mn}}. \label{eq:drho-trans2}
\eeq
We can then define a projected density operator $\overline{\delta\rho}_{\GG+\qq}$ by restricting $|m|,|n|\le n_{\text{max}}$ in Eq.~(\ref{eq:drho-trans2}): 
\beq
\overline{\delta \rho}_{\GG+\qq} = \sum_{\kk \eta s} \sum_{|m|,|n|\le n_{\text{max}}} M_{m,n}^\eta (\kk,\qq+\GG) 
	\pare{ c^\dagger_{\kk+\qq,m,\eta,s} c_{\kk,n,\eta,s} 
	- \frac12 \delta_{\qq,0} \delta_{mn}}\ , \label{eq:drho-trans3}
\eeq
and substitute $\overline{\delta\rho}_{\GG+\qq}$ into Eq.~(\ref{seq-HI}) 
to obtain the projected interaction Hamiltonian $H_I$ in the $n=\pm1$ bands. To simplify the form of the interaction Hamiltonian, we define a set of operators
\begin{equation}\label{seq-OqG}
O_{\mathbf{q,G}} =\sqrt{V(\mathbf{q+G})} \overline{\delta \rho}_{\GG+\qq}=\sum_{\mathbf{k}\eta s}\sum_{|m|,|n|\le n_{\text{max}}} \sqrt{V(\mathbf{q+G})} M_{m,n}^{\left(\eta\right)} \left(\mathbf{k},\mathbf{q}+\mathbf{G}\right) \left(\rho_{\mathbf{k,q},m,n,s}^\eta-\frac{1}{2}\delta_{\mathbf{q,0}}\delta_{m,n}\right)\ ,
\end{equation}
and the electron density operator within the flat bands
\begin{equation}
\rho_{\mathbf{k,q},m,n,s}^\eta=c^\dagger_{\kk+\qq,m,\eta,s} c_{\kk,n,\eta,s}\ .
\end{equation}
We can then write the projected interaction Hamiltonian as
\begin{equation}\label{seq-pHI}
\boxed{H_I=\frac{1}{2\Omega_{\text{tot}}}\sum_{\mathbf{q}\in\text{MBZ}}\sum_{\mathbf{G}\in\mathcal{Q}_0} O_{\mathbf{-q,-G}} O_{\mathbf{q,G}} }\ ,
\end{equation}
as given in the main text Eq. (\ref{eq-pHI}).
In particular, we have
\begin{equation}\label{eq:OqG-commutator}
\begin{split}
&[O_{\mathbf{q,G}},O_{\mathbf{q',G'}}]=\sum_{\mathbf{k},m,n,n',\eta,s} \sqrt{V(\mathbf{G}+\mathbf{q})V(\mathbf{G}'+\mathbf{q}')} \rho_{\mathbf{k,q+q'},m,n,s}^\eta \times  \\
&\qquad \qquad \left[M_{m,m'}^{\left(\eta\right)}\left(\mathbf{k+q'},\mathbf{q}+\mathbf{G}\right) M_{m',n}^{\left(\eta\right)}\left(\mathbf{k},\mathbf{q}'+\mathbf{G}'\right)- M_{m',n}^{\left(\eta\right)}\left(\mathbf{k},\mathbf{q}+\mathbf{G}\right) M_{m,m'}^{\left(\eta\right)}\left(\mathbf{k+q},\mathbf{q}'+\mathbf{G}'\right)\right]\ ,
\end{split}
\end{equation}
which in general does not vanish if $\qq\neq\qq'$ or $\GG\neq\GG'$. Therefore, different terms in the interaction Hamiltonian $H_I$ do not commute.

\subsection{Gauge fixing of the interaction}\label{app:Mfixing}
Eq.~(\ref{eq:M-def}) give the generic definition of the coefficient $M_{m,n}^{\left(\eta\right)}\left(\mathbf{k},\mathbf{q}+\mathbf{G}\right)$. Here we fix the form of this coefficient under the gauge fixing of Eq.~(\ref{eq:gauge-0}). Under this gauge, the following constraints must be satisfied:

(I) Hermiticity condition:
\beqs
M^{(\eta)*}_{mn}(\kk,\qq+\GG) = M_{nm}^{(\eta)} (\kk+\qq, -\qq-\GG)\ , \label{eq:Mcond-herm}
\eeqs
which is trivially satisfied by the definition in Eq.~(\ref{eq:M-def}).

(II) The $C_{2z}T$ symmetry yields the real condition
\beqs
M_{m,n}^{\left(\eta\right)}\left(\mathbf{k},\mathbf{q}+\mathbf{G}\right) 
=& \sum_{\alpha}\sum_{\mathbf{Q}\in\mathcal{Q}_{\pm}} 
[D(C_{2z}T) u_{m\eta}(\kk+\qq) ]_{\QQ-\GG,\alpha} 
[ D(C_{2z}T) u_{n\eta}^*\left(\mathbf{k}\right)]_{\QQ,\alpha}\nono\\
=& \sum_\alpha \sum_{\mathbf{Q}\in\mathcal{Q}_{\pm}}
u_{\QQ,\bar{\alpha}; n\eta}^*(\kk)  u_{\QQ-\GG \bar{\alpha}, m\eta}(\kk+\qq) 
= \sum_\alpha \sum_{\mathbf{Q}\in\mathcal{Q}_{\pm}}
u_{\QQ+\GG,\alpha; n\eta}^*(\kk) u_{\QQ \alpha, m\eta}(\kk+\qq) \nono\\
=&   M^{(\eta)*}_{mn}(\kk,\qq+\GG)\ . \label{eq:Mcond-C2T}
\eeqs

(III) Due to the combination operation $C_{2z} P$, which has the sewing matrix $D(C_{2z}P)=\zeta^y\tau^y$ in each pair of bands $n=\pm n_\text{B}$  (\cref{eq:gauge-0}), we have
\beqs
M_{mn} ^{(\eta)}(\kk,\qq+\GG) =& \sum_{\alpha}\sum_{\mathbf{Q}\in\mathcal{Q}_{\pm}} 
[D(C_{2z}P) u_{m\eta}^*(\kk+\qq) ]_{\QQ-\GG,\alpha} 
[ D(C_{2z}P) u_{n\eta}\left(\mathbf{k}\right)]_{\QQ,\alpha}\nono\\
=& \sum_\alpha \sum_{\mathbf{Q}\in\mathcal{Q}_{\pm}}
( \zeta_y)_{mm'} u_{\QQ-\GG \alpha, m',-\eta}^*(\kk+\qq)  u_{\QQ,\alpha; n',-\eta}(\kk) ( \zeta_y)_{n'n}  \nono\\
=&  [\zeta^y M^{(-\eta)} (\kk,\qq+\GG) \zeta^y]_{m, n},  \label{eq:Mcond-C2P}
\eeqs
where we write $M_{mn} ^{(\eta)}$ in short as a matrix $M^{(\eta)}$ in the band space, and $\zeta^\alpha$ means the Pauli matrix within each pair of bands $\pm n$.

(IV) For momenta $\kk$ and $\kk+\qq$ not at $M_M$ points, due to the $C_{2z}$ symmetry, which has the sewing matrix $B(C_{2z})(\kk)=\zeta^0\tau^x$ (\cref{eq:gauge-1}), we further have
\beq
M^{(\eta)} (\kk,\qq+\GG) = M^{(-\eta)}(-\kk,-\qq-\GG). \label{eq:Mcond-C2}
\eeq
For the case where $\kk$ is at $M_M$ and $\kk+\qq$ is not at $M_M$, the sewing matrices are given by $-B^{C_{2z}}(\kk)= B^{C_{2z}}(\kk+\qq+\GG)=i\zeta^0\tau^x$ due to the discussion in \cref{app:gauge-fixing}, hence the above condition changes to 
\beq
M^{(\eta)} (\kk,\qq+\GG) = - M^{(-\eta)}(-\kk,-\qq-\GG). \label{szdeq:Mcond-C2-2}
\eeq
For the case where $\kk$ is not at $M_M$ and $\kk+\qq$ is at $M_M$, the $M$ matrix also satisfies \cref{szdeq:Mcond-C2-2} for the same reason.
For the case where $\kk$ is at $M_M$ and $\qq=0$, the sewing matrices are given by $B^{C_{2z}}(\kk)= B^{C_{2z}}(\kk+\GG)=-i\zeta^0\tau^x$ and hence the $M$ matrix satisfies \cref{eq:Mcond-C2}.

We can generically parameterize $M^\eta_{m,n}(\kk,\qq)$ as
\beq
M^\eta_{m,n}(\kk,\qq+\GG) =\sum_{a=0,x,y,z} \sum_{b=0,z} (\zeta^a)_{mn} (\tau^b)_{\eta \eta} \alpha_{ab} (\kk,\qq+\GG)\ ,
\eeq
where only $b=0,z$ are allowed, since $M^\eta_{m,n}(\kk,\qq+\GG)$ is diagonal in valley $\eta$. We have assumed $\alpha_{ab} (\kk,\qq+\GG)$ are $n_\text{max}\times n_\text{max}$ matrices, and is tensor producted with $\zeta^a$ in each pair band basis $n=\pm n_\text{B}$ and the valley Pauli matrix $\tau^b$. Condition (III) requires $M$ to be commutative with $\zeta^y\tau^y$ in the band and valley indices, which restricts $M$ matrix to decompose into four terms
\beq \boxed{
M(\kk,\qq+\GG) = \zeta^0\tau^0 \alpha_0(\kk,\qq+\GG) + \zeta^x\tau^z \alpha_1(\kk,\qq+\GG) + i\zeta^y\tau^0 \alpha_2(\kk,\qq+\GG) + \zeta^z\tau^z \alpha_3 (\kk,\qq+\GG). } \label{eq:M-para}
\eeq 
We note that if $n_\text{max}=1$, $\alpha_{0,1,2,3}(\kk,\qq+\GG)$ are simply numbers, while if $n_\text{max}>1$, $\alpha_{0,1,2,3}(\kk,\qq+\GG)$ will be matrices.
Condition (II) requires $M^\eta_{m,n}(\kk,\qq+\GG)$ to be real, thus $\alpha_{0,1,2,3}(\kk,\qq+\GG)$ are all real (matrix) functions. We denote the matrix coefficient of $\alpha_j(\kk,\qq+\GG)$ in Eq.~(\ref{eq:M-para}) as $M_j$.
Besides, Condition (I) requires 
\beq
(1)\quad \boxed{ 
\alpha_a(\kk,\qq+\GG) = \alpha_a^T(\kk+\qq,-\qq-\GG)\quad \text{for }a=0,1,3,\qquad
\alpha_2(\kk,\qq+\GG) =-\alpha_2^T(\kk+\qq,-\qq-\GG).} \label{eq:alpha-cond1}
\eeq
Finally, for $\kk$ and $\kk+\qq$ not at $M_M$ points, Condition (IV) requires
\beq
(2)\quad\boxed{
\alpha_a(\kk,\qq+\GG) = \alpha_a(-\kk,-\qq-\GG)\quad \text{for }a=0,2,\qquad
\alpha_a(\kk,\qq+\GG) =-\alpha_a(-\kk,-\qq-\GG)\quad \text{for }a=1,3.}  \label{eq:alpha-cond2}
\eeq
In particular, the combination of Eqs. (\ref{eq:alpha-cond1}) and (\ref{eq:alpha-cond2}) implies that at $\qq=\mathbf{0}$, we have
\begin{equation} \label{szdeq:alpha-cond3}
\alpha_0(\kk,\GG)=\alpha_0^T(-\kk,\GG)\ ,\qquad \alpha_j(\kk,\GG)=-\alpha_j^T(-\kk,\GG),\quad (j=1,2,3).
\end{equation}

It is worth noting that, even though \cref{eq:alpha-cond2} is derived with assumption that $\kk$ and $\kk+\qq$ are not at the $M_M$ momentum, it is also true for $\kk$ at $M_M$ and $\qq=0$ because the Condition IV (\cref{eq:Mcond-C2}), from which \cref{eq:alpha-cond2} is derived, is true for $\kk$ at $M_M$ and $\qq=\mathbf{0}$. 
Therefore, \cref{szdeq:alpha-cond3}, which is the combination of \cref{eq:alpha-cond1,eq:alpha-cond2} at $\qq=\mathbf{0}$, is true for $\kk$ over the whole BZ.

\subsection{Many-body charge conjugation symmetry of the Projected Hamiltonian}\label{app:manybody-cc}

The full projected Hamiltonian $H=H_0+H_I$ has a many-body charge-conjugation symmetry, which ensures that all the physical phenomena is PH symmetric about the filling of the charge neutrality point (CNP) at $\nu=0$.

We define the \textit{many-body charge conjugation} $\CC_c$ as the single-particle transformation $C_{2z}TP$ followed by an interchange between electron annihilation operators $c$ and creation operators $c^\dag$, namely,
\beq
\CC_c c_{\kk,n,\eta,s}^\dg \CC_c^{-1}= c_{-\kk,m,\eta^\pr,s} [B^{C_{2z}TP}(\kk)]_{m\eta^\pr,n\eta}(\kk), \qquad
\CC_c c_{\kk,n,\eta,s} \CC_c^{-1}= c_{-\kk,m,\eta^\pr,s}^\dg [B^{C_{2z}TP*}(\kk)]_{m\eta^\pr,n\eta}\ .
\eeq
Under the gauge fixings of (\cref{eq:gauge-0}) and (\cref{eq:gauge-1}), one has $B^{C_{2z}TP}_{m\eta^\pr, n\eta} = B^{P}_{m\eta^\pr, n\eta} = (-i\zeta^y \tau^z)_{m \eta^\pr, n\eta}$ (\cref{eq:gauge-1}) within each pair of bands $n=\pm n_\text{B}$. We now show $\CC_c$ is a symmetry of the projected Hamiltonian.

Because of the relation $\ee_{n,\eta}(\kk) = - \ee_{-n,\eta}(-\kk)$, the kinetic Hamiltonian is invariant under $\CC_c$ up to a constant:
\beq
\CC_c H_0 \CC_c^{-1} = \sum_{\kk,n\eta,s} \ee_{n,\eta}(\kk) c_{-\kk,-n,\eta} c_{-\kk,-n,\eta}^\dg = \sum_{\kk,n\eta,s} \ee_{-n,\eta}(-\kk) c_{-\kk,-n,\eta}^\dg c_{-\kk,-n,\eta} + const. 
= H_0 + \text{const.} 
\eeq
Next, we note that the projected density operator $\overline{\delta \rho}_{\qq+\GG}$ in Eq.~(\ref{eq:drho-trans3}) satisfies
\beqs
\CC_c \overline{\delta\rho}_{\qq+\GG} \CC_c^{-1} =&
\sum_{\eta mn s} \sum_{\kk} (\zeta^y M^\eta (\kk,\qq+\GG) \zeta^y)_{mn}
	\pare{ c_{-\kk-\qq,m,\eta,s} c^\dg_{-\kk,n,\eta,s} 
	- \frac12 \delta_{\qq,0} \delta_{mn}} \nono\\
=& \sum_{\eta mn s} \sum_{\kk} (\zeta^y M^\eta (\kk,\qq+\GG) \zeta^y)_{mn}
	\pare{  -c^\dg_{-\kk,n,\eta,s}  c_{-\kk-\qq,m,\eta,s}
	+ \frac12 \delta_{\qq,0} \delta_{mn}} \nono\\
=& \sum_{\eta mn s} \sum_{\kk} (\zeta^y M^\eta (-\kk+\qq,\qq+\GG) \zeta^y)_{mn}
	\pare{ - c^\dg_{\kk+\qq,n,\eta,s}  c_{\kk,m,\eta,s}
	+ \frac12 \delta_{\qq,0} \delta_{mn}}
\eeqs
Due to \cref{eq:M-para,eq:alpha-cond2} we have
\beq
M_{m,n}^\eta(\kk,\qq+\GG) = \sum_{m^\pr n^\pr} \zeta^y_{m^\pr,m} M_{m^\pr,n^\pr}^\eta(-\kk,-\qq-\GG) \zeta^y_{n^\pr n}  \ ,
\eeq
and hence
\beq
\CC_c \overline{\delta\rho}_{\qq+\GG} \CC_c^{-1} =
\sum_{\eta mn s} \sum_{\kk} M^\eta_{mn} (\kk-\qq,-\qq-\GG)
	\pare{ - c^\dg_{\kk+\qq,n,\eta,s}  c_{\kk,m,\eta,s}
	+ \frac12 \delta_{\qq,0} \delta_{mn}}.
\eeq
Due to \cref{eq:Mcond-herm}, $M^\eta_{mn} (\kk-\qq,-\qq-\GG) = M^\eta_{nm} (\kk,\qq+\GG)$, thus
\beq
\CC_c \overline{\delta\rho}_{\qq+\GG}  \CC_c^{-1} =
\sum_{\eta mn s} \sum_{\kk} M^\eta_{mn} (\kk,\qq+\GG)
	\pare{ - c^\dg_{\kk+\qq,m,\eta,s}  c_{\kk,n,\eta,s}
	+ \frac12 \delta_{\qq,0} \delta_{mn}} = - \overline{\delta\rho}_{\qq+\GG} .
\eeq
Therefore, according to Eq.~(\ref{seq-OqG}), we have $\CC_c O_{\qq,\GG}\CC_c^{-1}=-O_{\qq,\GG}$, and thus the projected interaction in Eq.~(\ref{seq-pHI}) has the charge-conjugation symmetry $[\CC_c,H_I]=0$. In total, we have
\beq\label{seq-Pc-transformation}
\CC_cH\CC_c^{-1}=H-\sum_{|n|\le n_\text{max}}\sum_{\kk,\eta,s}\epsilon_{n,\eta}(\kk)=H
\eeq
for $H=H_0+H_I$, where we have used the fact that $\epsilon_{n,\eta}(\kk)=-\epsilon_{-n,\eta}(-\kk)$ in Eq. (\ref{seq:energy-relation}) due to the single-particle PH symmetry $P$. Note that $\CC_c$ maps a many-body state at filling $\nu$ to filling $-\nu$, where $\nu$ is the number of electrons per moir\'e unit cell relative to the CNP. Therefore, one expects the TBG ground states at $\nu$ and $-\nu$ to be PH symmetric.

\subsection{Contributions from the passive bands in the Projected Hamiltonian: Hartree-Fock Potential}\label{app:passivebandcontributions}

We note that the projected interaction Hamiltonian $H_I$ in Eq.~(\ref{seq-pHI}) is not normal ordered. We can rewrite $H_I$ into normal-ordered part and some quadratic fermionic terms as
\begin{equation}
\begin{split}
&H_{I}=H_I^{\text{norm}}+\Delta H^{(1)}+ \Delta H^{(2)}+\text{const.}\ .
\end{split}
\end{equation}
where $H_I^{\text{norm}}$ is the normal ordered Hamiltonian, and $H^{(1)}$ and $\Delta H^{(2)}$. are specified shortly below. By defining interaction parameters
\begin{equation}
U_{m^{\prime}n^{\prime};mn}^{\left(\eta' \eta\right)}\left(\mathbf{q};\mathbf{k}^{\prime}\mathbf{k}\right)
=\sum_{\mathbf{G}\in\mathcal{Q}_{0}}V\left(\mathbf{G}+\mathbf{q}\right)
M_{m^{\prime},n^{\prime}}^{\left(\eta^{\prime}\right)}\left(\mathbf{k}^{\prime},-\mathbf{q}-\mathbf{G}\right)
M_{m,n}^{\left(\eta\right)}\left(\mathbf{k},\mathbf{q}+\mathbf{G}\right)\ ,
\label{eq:U-def}
\end{equation}
we can rewrite each term as
\begin{equation}\label{seq:HInorm}
H_I^{\text{norm}}=\frac{1}{2\Omega_{\mathrm{tot}}}\sum_{\mathbf{q}\mathbf{k}\mathbf{k}^{\prime}\in\mathrm{MBZ}}\sum_{\eta \eta^{\prime}ss'}\sum_{m,n; m',n'}U_{m^{\prime}n^{\prime};mn}^{\left(\eta' \eta  \right)}\left(\mathbf{q};\mathbf{k}^{\prime}\mathbf{k}\right) c^\dag_{k+q,m,\eta,s} c^\dag_{k'-q,m',\eta',s'}c_{k',n',\eta',s'} c_{k,n,\eta,s}
\end{equation}
\begin{equation}\label{seq:HI1}
\Delta H^{(1)} =-\frac{1}{2\Omega_{\mathrm{tot}}}\sum_{\mathbf{kk'}}\sum_{\eta \eta^{\prime}ss'}\sum_{m,n; m'} U_{m^{\prime}m^{\prime};mn}^{\left(\eta' \eta  \right)}\left(\mathbf{0};\mathbf{k}^{\prime}\mathbf{k}\right) c_{k,m,\eta,s}^\dag c_{k,n,\eta,s}\ ,
\end{equation}
and
\begin{equation}\label{seq:HI2}
\Delta H^{(2)} =\frac{1}{2\Omega_{\mathrm{tot}}}\sum_{\mathbf{k,q}}\sum_{\eta,s }\sum_{m,n'; m'} U_{m^{\prime}n';mm'}^{\left(\eta \eta  \right)}\left(\mathbf{q};\mathbf{k},\mathbf{k-q}\right) c_{k,m,\eta,s}^\dag c_{k,n',\eta,s}\ ,
\end{equation}
where we have used the fact that $U_{m^{\prime}n^{\prime};mn}^{\left(\eta' \eta  \right)}\left(\mathbf{q};\mathbf{k}^{\prime}\mathbf{k}\right)=U_{mn;m^{\prime}n^{\prime}}^{\left(\eta \eta'  \right)}\left(-\mathbf{q};\mathbf{k}\mathbf{k}'\right)$ which trivially holds by exchanging the two $M$ matrices in the definition (\ref{eq:U-def}). By summing over only $|m|,|n|\le n_\text{max}$, $H_I$ gives the projected Hamiltonian. In the following, we prove that $H_I^{(1)}$ and $H_I^{(2)}$ can be heuristically understood as the Hartree and Fock potential of the higher passive bands $|n|>n_\text{max}$ which are projected out. We emphasize that the difference between the $H_I$ and its normal-ordered version is not just a simple chemical potential shift, contrary to the unprojected interaction Hamiltonian $\hat{H}_I$.

We first note that the full interaction Hamiltonian $\hat{H}_I$ before projection is simply given by Eqs~(\ref{seq:HInorm})-(\ref{seq:HI2}) with summation over all band indices $m,n,m',n'$. We now derive the Hartree-Fock Hamiltonian of the full Hamiltonian $\hat{H}_I$ at filling $\nu=-4n_\text{max}$ (number of electrons per moir\'e unit cell relative to the CNP). The occupied single-particle bands at $\nu=-4n_\text{max}$ produce a mean field
\begin{equation}
\langle c_{\kk,m,\eta,s}^\dag c_{\kk',n,\eta',s'}\rangle=\Theta(-n_\text{max}-m)\delta_{\kk,\kk'}\delta_{m,n}\delta_{\eta,\eta'}\delta_{s,s'}\ ,
\end{equation}
where we define $\Theta(x)=1$ if $x>0$, and $\Theta(x)=0$ if $x\le0$. We shall use the property of interaction parameter $U_{m^{\prime}n^{\prime};mn}^{\left(\eta' \eta  \right)}\left(\mathbf{q};\mathbf{k}^{\prime}\mathbf{k}\right)  = \text{sgn}(mn)U_{m^{\prime}n^{\prime}; -m,-n}^{\left(  \eta'- \eta  \right)}\left(\mathbf{q};\mathbf{k}^{\prime} \mathbf{k}\right)=  \text{sgn}(m'n')U_{-m^{\prime}-n^{\prime}; m,n}^{\left(  - \eta' \eta \right)}\left(\mathbf{q};\mathbf{k}^{\prime} \mathbf{k}\right)=U_{mn;m^{\prime}n^{\prime}}^{\left(\eta \eta'  \right)}\left(-\mathbf{q};\mathbf{k}\mathbf{k}'\right) $,
which can be verified by the properties of the $M$ matrices through Eq.~(\ref{eq:M-para})-(\ref{eq:alpha-cond2}). We then find the Hartree term
\begin{equation}
\begin{split}
&H_H^{\nu=-4n_\text{max}}=\frac{1}{2\Omega_{\mathrm{tot}}}\sum_{\mathbf{kk'}}\sum_{\eta \eta^{\prime}ss'}\sum_{m,n; m'} [2\Theta(-n_\text{max}-m')-1] U_{m^{\prime}m^{\prime};mn}^{\left(\eta' \eta  \right)}\left(\mathbf{0};\mathbf{k}^{\prime}\mathbf{k}\right) c_{k,m,\eta,s}^\dag c_{k,n,\eta,s} \\
&=-\frac{1}{2\Omega_{\mathrm{tot}}}\sum_{\mathbf{kk'}}\sum_{\eta \eta^{\prime}ss',m,n}\sum_{|m'|\le n_\text{max}} U_{m^{\prime}m^{\prime};mn}^{\left(\eta' \eta  \right)}\left(\mathbf{0};\mathbf{k}^{\prime}\mathbf{k}\right) c_{k,m,\eta,s}^\dag c_{k,n,\eta,s}\ ,
\end{split}
\end{equation}
and the Fock term
\begin{equation}
\begin{split}
&H_F^{\nu=-4n_\text{max}}=-\frac{1}{2\Omega_{\mathrm{tot}}}\sum_{\mathbf{k,q}}\sum_{\eta,s }\Big[ \sum_{m,n'; m'} \Theta(-n_\text{max}-m') U_{m^{\prime}n';mm'}^{\left(\eta \eta  \right)}\left(\mathbf{q};\mathbf{k+q},\mathbf{k}\right) c_{k+q,m,\eta,s}^\dag c_{k+q,n',\eta,s}\\
&\qquad +\sum_{m',n; m} \Theta(-n_\text{max}-m) U_{m'm;mn}^{\left(\eta \eta  \right)}\left(\mathbf{q};\mathbf{k+q},\mathbf{k}\right) c_{k,m',\eta,s}^\dag c_{k,n,\eta,s} \Big]\\
&=-\frac{1}{2\Omega_{\mathrm{tot}}}\sum_{\mathbf{k,q}}\sum_{\eta,s } \sum_{m,n'; m'} \Theta(-n_\text{max}-m') \Big[U_{m^{\prime}n;mm'}^{\left(\eta \eta  \right)}\left(\mathbf{q};\mathbf{k+q},\mathbf{k}\right) + U_{mm';m'n}^{\left(\eta \eta  \right)}\left(\mathbf{-q};\mathbf{k},\mathbf{k+q}\right) \Big] c_{k+q,m,\eta,s}^\dag c_{k+q,n,\eta,s} \\
&=-\frac{1}{2\Omega_{\mathrm{tot}}}\sum_{\mathbf{k,q}}\sum_{\eta,s } \sum_{m,n'; m'} 2\Theta(-n_\text{max}-m')  U_{m'n;mm'}^{\left(\eta \eta  \right)}\left(\mathbf{q};\mathbf{k},\mathbf{k-q}\right) c_{k,m,\eta,s}^\dag c_{k,n,\eta,s}\ .
\end{split}
\end{equation}
Similarly, one can show the Hartree term at $\nu=4n_\text{max}$ are given by
\begin{equation}
H_H^{\nu=4n_\text{max}}=-H_H^{\nu=-4n_\text{max}}\ ,
\end{equation}
and the Fock term at $\nu=4n_\text{max}$:
\begin{equation}
H_F^{\nu=4n_\text{max}}=-\frac{1}{2\Omega_{\mathrm{tot}}}\sum_{\mathbf{k,q}}\sum_{\eta,s } \sum_{m,n'; m'} 2\Theta(n_\text{max}+1-m')  U_{m'n;mm'}^{\left(\eta \eta  \right)}\left(\mathbf{q};\mathbf{k},\mathbf{k-q}\right) c_{k,m,\eta,s}^\dag c_{k,n,\eta,s}\ .
\end{equation}
Therefore, when projected into the lowest $8n_\text{max}$ bands ($2n_\text{max}$ per spin-valley), we find for the difference between our particle-hole symmetric Hamiltonian and its normal-ordered version:
\begin{equation}
\begin{split}
&\Delta H^{(1)}=H_H^{\nu=-4n_\text{max}}=-H_H^{\nu=4n_\text{max}}\ , \qquad\qquad \Delta H^{(2)}=\frac{1}{2}\Big(H_F^{\nu=-4n_\text{max}}-H_F^{\nu=4n_\text{max}}\Big)\ .
\end{split}
\end{equation}
Note that the interaction satisfies the orthonormal condition $\sum_{m'}U_{m'n;mm'}^{\left(\eta \eta  \right)}\left(\mathbf{q};\mathbf{k},\mathbf{k-q}\right)=\sum_{\mathbf{G}}V(\mathbf{q+G})\delta_{m,n}$, so under the single-particle PH transformation $P$ which takes $Pc^\dag_{k,m,\eta,s}P^{-1}=-\text{sgn}(m)\eta c^\dag_{-k,-m,\eta,s}$, we have
\begin{equation}
\begin{split}
&PH_F^{\nu=-4n_\text{max}}P^{-1}= -\frac{1}{2\Omega_{\mathrm{tot}}}\sum_{\mathbf{k,q}}\sum_{\eta,s } \sum_{m,n'; m'} 2\Theta(-n_\text{max}-m')  \text{sgn}(mn)U_{m'n;mm'}^{\left(\eta \eta  \right)}\left(\mathbf{q};\mathbf{k},\mathbf{k-q}\right) c_{-k,-m,\eta,s}^\dag c_{-k,-n,\eta,s} \\
&=-\frac{1}{2\Omega_{\mathrm{tot}}}\sum_{\mathbf{k,q}}\sum_{\eta,s } \sum_{m,n'; m'} 2\Theta(n_\text{max}+1+m')  U_{m'n;mm'}^{\left(\eta \eta  \right)}\left(\mathbf{q};\mathbf{k},\mathbf{k-q}\right) c_{k,m,\eta,s}^\dag c_{k,n,\eta,s}\\
&=-\frac{1}{2\Omega_{\mathrm{tot}}}\sum_{\mathbf{k,q}}\sum_{\eta,s } \sum_{m,n'; m'} 2[1-\Theta(n_\text{max}+1-m')]  U_{m'n;mm'}^{\left(\eta \eta  \right)}\left(\mathbf{q};\mathbf{k},\mathbf{k-q}\right) c_{k,m,\eta,s}^\dag c_{k,n,\eta,s}\\
&=-H_F^{\nu=4n_\text{max}}-\mu_{V}\sum_{\mathbf{k},m,\eta,s}c_{k,m,\eta,s}^\dag c_{k,m,\eta,s}\ ,
\end{split}
\end{equation}
where we have used the PH symmetry of interaction $U_{mn;m'n'}^{\left(\eta \eta'  \right)}\left(\mathbf{q};\mathbf{k},\mathbf{k'}\right)= \text{sgn}(mnm'n')U_{-m,-n;-m',-n'}^{\left(\eta \eta'  \right)}\left(\mathbf{-q};\mathbf{-k},\mathbf{-k'}\right)$. 
The constant $\mu_V$ is defined by $\mu_V=\frac{1}{\Omega_{\mathrm{tot}}}\sum_{\mathbf{q,G}}V(\mathbf{q+G})$, which is a coefficient of a chemical potential term.

\subsection{U(2)$\times$U(2) spin-charge rotational symmetry}

In Eq.~(\ref{seq:U2U2-gene}) we have given the generators of the U(2)$\times$U(2) symmetry of the single-particle Hamiltonian $H_0$ from the spin-charge rotational symmetry in each valley. Here we show that the projected interaction Hamiltonian also respects the U(2)$\times$U(2) symmetry. Hereafter, with the understanding that we assume the gauge fixing given by Eqs. (\ref{eq:gauge-0}) and (\ref{seq:c-continuous}) (We note that Eq. (\ref{seq:c-continuous}) is only used for defining the irrep band basis in Eq. (\ref{seq-irrepbasis}), which will be useful in the discussion of nonchiral-flat U(4) irreps in Sec. \ref{sec:nc-f-irrep} ), 
we shall use $\zeta^a$, $\tau^a$, $s^a$ to denote the identity matrix ($a=0$) and Pauli matrices ($a=x,y,z$) in the each pair of bands $n=\pm n_\text{B}$, valley $\eta=\pm$ and spin $s=\uparrow,\downarrow$ bases, respectively.

When projected into the $8n_\text{max}$ flat bands of $|n|\le n_\text{max}$, the 8 generators $S^{ab}$ ($a=0,z$, $b=0,x,y,z$) of the U(2)$\times$U(2) symmetry in Eq.~(\ref{seq:U2U2-gene}) take the form
\begin{equation}\label{seq-U2U2-Sab}
S^{ a b}=\sum_{\mathbf{k},m,\eta,s;n,\eta',s'} (s^{ab})_{m,\eta,s;n,\eta',s'}c_{\mathbf{k},m,\eta,s}^\dag c_{\mathbf{k},n,\eta',s'}\ ,\qquad (a=0,z,\quad b=0,x,y,z)\ ,
\end{equation}
where the matrices within each pair of bands $n=\pm n_\text{B}$ are given by
\begin{equation}\label{seq-s0zb}
s^{0b}=\zeta^0\tau^0 s^b,\qquad s^{zb}=\zeta^0\tau^z s^b, \qquad (b=0,x,y,z).
\end{equation}
In particular, $S^{0b}$ and $S^{zb}$ give the global spin-charge U(2) rotations and the valley spin-charge U(2) rotations, respectively.

It is easy to see that both $S^{0b}$ and $S^{zb}$ are diagonal in valley $\eta$, and only acts on spin $s$. Since the operator $O_{\qq,\GG}$ defined in Eq.~(\ref{seq-OqG}) is diagonal in valley $\eta$, and all the coefficients are independent of spin $s$, we conclude that
\begin{equation}
[O_{\qq,\GG},S^{0b}]=[O_{\qq,\GG},S^{zb}]=0\ .
\end{equation}
Accordingly, the interaction $H_I$ in Eq.~(\ref{seq-pHI}) respects the U(2)$\times$U(2) symmetry, and so does the full projected Hamiltonian $H=H_0+H_I$.

\section{Enhanced symmetries in various limits}\label{app:various-limits}

In this appendix, we will show that the U(2)$\times$U(2) symmetry (Eq.~(\ref{seq-U2U2-Sab})) of the full Hamiltonian $H=H_0+H_I$ is enhanced into higher symmetries in various limits of TBG. Since all these higher symmetries involve the U(4) group, we first briefly review the algebra of the U(4) group.

\subsection{Brief Review of the U(4) group}\label{app:rev-U4}

The U($N$) group is defined by all the $N\times N$ unitary matrices $U$ satisfying $U^\dag U=I_{N}$, where $I_N$ is the identity matrix. The matrices $U$ are generated by all the linearly independent $N\times N$ Hermitian matrices, thus the total number of generators is $N^2$. In particular, for the U(4) group, the 16 generators can be represented by the tensor product of two sets of $2\times2$ identity and Pauli matrices $\tau^a$ and $s^a$ ($a=0,x,y,z$) as
\begin{equation}\label{seq:U(4)-fundamental}
s_0^{ab}=\tau^a s^b\ ,\qquad (a,b=0,x,y,z)\ .
\end{equation}
We denote their commutation relations as
\begin{equation}\label{seq:U4-structure-const}
[s_0^{ab},s_0^{cd}]=f^{ab,cd}_{ef} s_0^{ef}\ .
\end{equation}
Then $f^{ab,cd}_{ef}$ are the group structure constants, which are the same for all representations of U(4) group.

The set of all the $4\times 4$ matrices $U$ defines the 4-dimensional fundamental irreducible representation (irrep) of the U($4$) group, the representation matrices of the generators are exactly given by Eq.~(\ref{seq:U(4)-fundamental}). There is also a 1-dimensional trivial identity irrep, in which the representation matrices of all generators $s_0^{ab}=0$. We shall use the following notation to denote the fundamental irrep and trivial identity irrep of the U(4) group:
\begin{equation}
\text{U(4) fundamental irrep:}\qquad [1]_4\ ,\qquad \qquad  \text{U(4) trivial identity irrep:}\qquad [0]_4\ .
\end{equation}
We will not explain the meaning of these notations, except that we mention they are consistent with the Young tableau notations for U(4) irreps we explain and adopt in Ref.~\cite{ourpaper4}.

\subsection{U(4) symmetry in the nonchiral-flat limit}\label{app:U4-nc-f}

\subsubsection{The symmetry}

We now assume the magic angle TBG is in the nonchiral-flat limit, where the projected kinetic Hamiltonian in Eq.~(\ref{eq:H0-proj}) becomes exactly $H_0=0$, while both $w_0>0$ and $w_1>0$ in Eq.~(\ref{seq-Tj}). In this case, the total projected Hamiltonian is $H=H_I$. We will show that there is an enhanced U(4) symmetry.

To see this, we first show that $C_{2z}P$ is a symmetry of $H=H_I$. With the sewing matrix of $C_{2z}P$ given by $B^{C_{2z}P}(\kk)=\zeta^y\tau^y$ in (Eq.~(\ref{eq:gauge-0})), we have
\begin{equation}
\begin{split}
&(C_{2z}P)O_{\qq,\GG}(C_{2z}P)^{-1}=\sum_{\mathbf{k}\eta s}\sum_{m,n=\pm1} \sqrt{V(\mathbf{q+G})} M_{m,n}^{\left(\eta\right)} \left(\mathbf{k},\mathbf{q}+\mathbf{G}\right) \left((C_{2z}P)c^\dagger_{\kk+\qq,m,\eta,s} c_{\kk,n,\eta,s}(C_{2z}P)^{-1}-\frac{1}{2}\delta_{\mathbf{q,0}}\delta_{m,n}\right)\\
&=\sum_{\mathbf{k}\eta s}\sum_{|m|,|n|\le n_\text{max}} \sqrt{V(\mathbf{q+G})} \left([\zeta^y M^{(-\eta)} \left(\mathbf{k},\mathbf{q}+\mathbf{G}\right)\zeta^y]_{mn} c^\dagger_{\kk+\qq,m,\eta,s} c_{\kk,n,\eta,s}-\frac{1}{2}M_{m,n}^{\left(\eta\right)} \left(\mathbf{k},\mathbf{q}+\mathbf{G}\right)\delta_{\mathbf{q,0}}\delta_{m,n}\right)\\
&=\sum_{\mathbf{k}\eta s}\sum_{|m|,|n|\le n_\text{max}} \sqrt{V(\mathbf{q+G})} M_{m,n}^{\left(\eta\right)} \left(\mathbf{k},\mathbf{q}+\mathbf{G}\right) \left(c^\dagger_{\kk+\qq,m,\eta,s} c_{\kk,n,\eta,s}-\frac{1}{2}\delta_{\mathbf{q,0}}\delta_{m,n}\right)=O_{\qq,\GG}\ ,
\end{split}
\end{equation}
where we have used Eq.~(\ref{eq:Mcond-C2P}). Therefore, we have $[C_{2z}P,O_{\qq,\GG}]=0$, and thus the interaction Hamiltonian $H_I$ in Eq.~(\ref{seq-pHI}) satisfies
\begin{equation}
[C_{2z}P,H_I]=0\ .
\end{equation}
Besides, since $[C_{2z},H_0]=0$ and $\{P,H_0\}=0$, we have $\{C_{2z}P,H_0\}=0$, which implies $\epsilon_{n,\eta}(\kk)=-\epsilon_{-n,-\eta}(\kk)$. If we want to have $[C_{2z}P,H_0]=0$, we would have to require $\epsilon_{n,\eta}(\kk)=\epsilon_{-n,-\eta}(\kk)$, which is only possible when $\epsilon_{n,\eta}(\kk)=0$, namely, only in the exact flat band limit with projected kinetic term $H_0=0$.

The $C_{2z}P$ symmetry allows us to define the following operator as a commuting symmetry of the projected Hamiltonian $H=H_I$:
\beq\label{seq:Sy0}
S^{y0}=\sum_{\kk,s} \sum_{nn^\pr \eta \eta^\pr} [B^{C_{2z}P}(\kk)]_{n \eta, n^\pr \eta^\pr} 
c^\dg_{\kk,n,\eta,s} c_{\kk,n^\pr,\eta^\pr,s}=\sum_{\kk,s} \sum_{nn^\pr \eta \eta^\pr} [\zeta^y\tau^y]_{n \eta, n^\pr \eta^\pr} 
c^\dg_{\kk,n,\eta,s} c_{\kk,n^\pr,\eta^\pr,s}\ ,
\eeq
where we have used the gauge fixing of Eq.~(\ref{eq:gauge-0}), and $\zeta^y$ only acts within each pair of bands $n=\pm n_\text{B}$. We note that when $S^{y0}$ acts on single-electron states $c^{\dag}_{\kk,n,\eta,s}|0\rangle$ where $|0\rangle$ is the vacuum, it is the same as the operation of $C_{2z}P$. To see this is a symmetry, we note that
\begin{equation}
[S^{y0},O_{\qq,\GG}]=\sum_{\kk,s} \sum_{nn^\pr \eta \eta^\pr} \left([\zeta^y\tau^y,M(\kk,\qq+\GG)]\right)_{n \eta, n^\pr \eta^\pr} 
c^\dg_{\kk,n,\eta,s} c_{\kk,n^\pr,\eta^\pr,s}=0\ ,
\end{equation}
where we have used the fact that the $M(\kk,\qq+\GG)$ matrix commutes with $\zeta^y\tau^y$ from condition (\ref{eq:M-para}), a result of the $C_{2z}P$ symmetry. Therefore, $S^{y0}$ is a commuting symmetry of the interaction Hamiltonian $H_I$ in Eq.~(\ref{seq-pHI}), namely,
\begin{equation}
[S^{y0},H_I]=0\ .
\end{equation}

Recall that $H_I$ has a U(2)$\times$U(2) symmetry with 8 generators $S^{0b}$ and $S^{zb}$ ($b=0,x,y,z$) in Eq.~(\ref{seq-U2U2-Sab}). The commutators of $S^{y0}$ in Eq.~(\ref{seq:Sy0}) with the 8 U(2)$\times$U(2) generators then yields 16 Hermitian operators in total:
\begin{equation}\label{eq:U4-generator0}
S^{ a b}=\sum_{\mathbf{k},m,\eta,s;n,\eta',s'} (s^{ab})_{m,\eta,s;n,\eta',s'}c_{\mathbf{k},m,\eta,s}^\dag c_{\mathbf{k},n,\eta',s'}\ ,\qquad (a, b=0,x,y,z)\ ,
\end{equation}
where within each pair of bands $n=\pm n_\text{B}$
\begin{equation}\label{eq:U4-generator}
\boxed{s^{ab}=\{\zeta^0\tau^0 s^b,\ \zeta^y\tau^x s^b,\ \zeta^y\tau^y s^b,\ \zeta^0\tau^z s^b\}, \qquad (a, b=0,x,y,z)\ .}
\end{equation}
More specifically, the new generators are given by
\begin{equation}
S^{xb}=-\frac{i}{2}[S^{y0},S^{zb}]\ ,\qquad  S^{yb}=\frac{i}{2}[S^{x0},S^{zb}]\ .
\end{equation}
It is then easy to see that the 16 operators $S^{ab}$ satisfy the commutation relations of U(4) generators:
\begin{equation}
[S^{ab},S^{cd}]=f^{ab,cd}_{ef} S^{ef}\ ,
\end{equation}
where $f^{ab,cd}_{ef}$ are the U(4) structure constants defined in Eq.~(\ref{seq:U4-structure-const}). Therefore, we find in the nonchiral-flat limit, the projected interaction Hamiltonian $H_I$ has an enhanced U(4) symmetry.

The Cartan subalgebra of the U(4) generators in Eq.~(\ref{eq:U4-generator}) can be chosen as
\beq  \text{Cartan:}\qquad 
\zeta^0 \tau^0 s^0, \quad \zeta^0 \tau^0 s^z,\quad \zeta^0 \tau^z s^0, \quad \zeta^0 \tau^z s^z. \label{eq:U4-cartan}
\eeq

We note that although we proved the symmetry of $S^{y0}$ under the fixed gauge (\ref{eq:gauge-0}), the definition of $S^{y0}=\sum_{\kk,s} \sum_{nn^\pr \eta \eta^\pr} [B^{C_{2z}P}(\kk)]_{n \eta, n^\pr \eta^\pr} 
c^\dg_{\kk,n,\eta,s} c_{\kk,n^\pr,\eta^\pr,s}$ in Eq.~(\ref{seq:Sy0}) is gauge invariant. This can be seen by noting that under a gauge transformation $c^\dg_{\kk,n,\eta,s}\rightarrow e^{i\phi_{n,\eta}}c^\dg_{\kk,n,\eta,s}$, the sewing matrix elements change according to $[B^{C_{2z}P}(\kk)]_{n \eta, n^\pr \eta^\pr}\rightarrow e^{-i\phi_{n,\eta}+i\phi_{n',\eta'}} [B^{C_{2z}P}(\kk)]_{n \eta, n^\pr \eta^\pr}$.

\subsubsection{The single-electron irreps}\label{sec:nc-f-irrep}

The $\kk$-independent representation matrices of Eq.~(\ref{eq:U4-generator}) at each momentum $\kk$ can be decomposed into fundamental U(4) irreps. This can be done by transforming into a new basis where $\zeta^y$ is diagonalized. This turns out to be exactly the irrep band basis $d^{(n_\text{B})\dag}_{\kk,e_Y,\eta,s}=\frac{1}{\sqrt{2}}(c^\dag_{\kk,+n_\text{B},\eta,s}+ie_Yc^\dag_{\kk,-n_\text{B},\eta,s})$ we defined earlier in Eq.~(\ref{eq-irrepbasis}) (see also Ref.~\cite{ourpaper2}). For $n_\text{B}=1$, $e_Y=\pm1$ gives the Chern number of the band basis. The single-electron state in irrep band $e_Y$
\begin{equation}\label{seq-1electron-1}
d^{(n_\text{B})\dag}_{\kk,e_Y,\eta,s}|0\rangle 
\end{equation}
has eigenvalue $\zeta^y=e_Y$. It is then easy to see that the representations of the U(4) generators $S^{ab}$ for the single-electron state (\ref{seq-1electron-1}) are given by 
\begin{equation}\label{seq-SabeY}
\boxed{ s^{ab}(e_Y)=\{ \tau^0s^b,\ e_Y\tau^x s^b,\ e_Y\tau^y s^b,\ \tau^zs^b \}. }
\end{equation}
Therefore, the single-electron state (\ref{seq-1electron-1}) for a fixed $e_Y$, or equivalently the irrep band fermion operator $d^\dag_{\kk,e_Y,\eta,s}$ for a fixed $e_Y$, occupies a fundamental irrep $[1]_4$ of the U(4) group. However, we note that the $e_Y=+1$ and $e_Y=-1$ irreps differ by a $\pi$ valley rotation $e^{i\pi\tau^z/2}$ about the $z$ axis.

For many-body Fock states created by multiple $d^{(n_\text{B})\dag}_{\kk,e_Y,\eta,s}$, the U(4) representation is given by the tensor product of the U(4) fundamental irreps $[1]_4$ of each $d^{(n_\text{B})\dag}_{\kk,e_Y,\eta,s}$. Such tensor product representations can be further decomposed into U(4) irreps, which we will not discuss here, but in our upcoming paper of the many-body states of the PSDHs\cite{ourpaper4}.

\subsection{U(4)$\times$U(4) symmetry in the (first) chiral-flat limit}\label{app:chiral-flat}

In this appendix, we demonstrate that by setting $w_0=0<w_1$ (the chiral condition), and set the projected kinetic Hamiltonian $H_0$ to zero (flat condition), the system has a unitary U(4)$\times$U(4) symmetry. We call this limit the first chiral-flat limit.

\subsubsection{The chiral symmetry at $w_0=0$}

In the first chiral-flat limit, since $w_0=0$, the single-particle Hamiltonian of TBG acquires an additional unitary chiral symmetry $C$, which satisfies the anti-commutation relation with the full single-particle Hamiltonian $\hat{H}_0$ in Eq.~(\ref{eq:H0}):
\begin{equation}
\{C,\hat{H}_0\}=0\ .
\end{equation}
The action of $C$ is given by
\beq
C c^\dagger_{\kk,\QQ,\eta,\alpha,s} C^{-1} = \sum_{\QQ^\pr \eta^\pr \beta} 
[D(C)]_{\QQ^\pr \eta^\pr \beta, \QQ \eta \alpha} c^\dagger_{\kk,\QQ^\pr,\eta^\pr,\beta,s}\ ,
\eeq
with the representation matrix
\beq
\quad[D(C)]_{\QQ^\pr \eta^\pr \beta, \QQ \eta \alpha} = \delta_{\QQ^\pr,\QQ} \delta_{\eta^\pr, \eta} (\sigma_z)_{\beta,\alpha}\ . \label{eq:C}
\eeq
Note that $C$ preserves the electron momentum $\kk$. 
Since $C$ flips the single-particle Hamiltonian $\hat{H}_0$, it is not a commuting symmetry of TBG, but only reflects a relation between the positive and negative energy spectra. The transformation $C$ satisfies
\beq
C^2=1,\qquad
\{C,C_{2z}\} = 0,\qquad
[C,T] = 0,\qquad
[C,P]=0, \qquad
\{C,C_{2z}T\}=0,\qquad
\{C,C_{2z}P\}=0.
\eeq

\subsubsection{The full symmetry}

When transformed into the energy band basis, the chiral symmetry $C$ implies
\begin{equation}\label{seq:C-H0energy}
\ee_{n,\eta}(\kk)=-\ee_{-n,\eta}(\kk)\ ,
\end{equation}
\begin{equation}\label{seq:sewing-C-wf}
[D(C)]_{\eta\eta'} u_{n\eta'}(\kk)= \sum_{m}[B^{C}(\kk)]_{m\eta,n\eta'}u_{m\eta}(\kk)\ ,
\end{equation}
where
\begin{equation}\label{seq:sewing-C0}
[B^{C}(\kk)]_{m\eta,n\eta'}=\delta_{\eta,\eta'}\delta_{-m,n} e^{i\varphi^{C}_{n,\eta'}(\kk)}
\end{equation}
This implies the transformation
\begin{equation}\label{seq:CcC}
C c_{\mathbf{k},n,\eta' ,s}^{\dagger} C^{-1}= \sum_{m\eta} [B^C(\kk)]_{m\eta,n\eta'} c_{\mathbf{k},m,\eta, s}^{\dagger}\ .
\end{equation}
By the relations $\{C,C_{2z}T\}=\{C,C_{2z}P\}=0$, the sewing matrix of $C$ satisfies
\beq\label{seq:sewing-C1}
B^C(\kk) B^{C_{2z}T}(\kk) = - B^{C_{2z}T}(\kk) B^{C*}(\kk)\ ,\qquad B^C(\kk) B^{C_{2z}P}(\kk) = - B^{C_{2z}P}(\kk) B^{C}(\kk)\ .
\eeq
Under the gauge fixing of Eq.~(\ref{eq:gauge-0}), we have $B^{C_{2z}T}(\kk)=\zeta^0\tau^0$, and $B^{C_{2z}P}(\kk)=\zeta^y\tau^y$. The only $\kk$-independent gauge for sewing matrix of $C$ in consistency with Eqs. (\ref{seq:sewing-C0}) and (\ref{seq:sewing-C1}) within each pair of bands $n=\pm n_\text{B}$ is then (up to a global minus sign)
\begin{equation}\label{seq:sewing-C}
B^C(\kk)=\zeta^y\tau^z\ .
\end{equation}
In particular, this $\kk$-independent gauge fixing (\ref{seq:sewing-C}) of $C$ automatically ensures the continuous gauge fixing condition (\ref{seq:c-continuous}), which is crucial for defining the irrep band basis in Eq.~(\ref{eq-irrepbasis}). To see this, note that Eq.~(\ref{seq:sewing-C}) tells us that $u_{-n,\eta}(\kk)=i\text{sgn}(n)\eta u_{n,\eta}(\kk)$ for band $n=\pm n_\text{B}$, and thus we have
\begin{equation}
f_{n,\eta}(\kk+\qq,\kk)= \left|u^\dag_{n,\eta}(\kk+\qq) u_{n,\eta}(\kk)- u^\dag_{-n,\eta}(\kk+\qq) u_{-n,\eta}(\kk)\right|=\left|u^\dag_{n,\eta}(\kk+\qq) u_{n,\eta}(\kk)[1-\text{sgn}(n)^2\eta^2]\right|=0
\end{equation}
for any $\kk$ and $\qq$, satisfying Eq.~(\ref{seq:c-continuous}).

We also note that this gauge fixing of $C$ is consistent with the gauge fixings of both $C_{2z}$ and $P$ separately in Eq.~(\ref{eq:gauge-1}). Basically, the relations $\{C,C_{2z}\}=0$ and $[C,P]=0$ requires
\begin{equation}
B^C(-\kk) B^{C_{2z}}(\kk) = - B^{C_{2z}}(\kk) B^{C}(\kk)\ , \qquad B^C(-\kk) B^{P}(\kk) = B^{P}(\kk) B^{C}(\kk)\ ,
\end{equation}
which is satisfied by Eq.~(\ref{seq:sewing-C}). 

For the projected Hamiltonian $H=H_0+H_I$, we now show that $C$ is a symmetry of the interaction Hamiltonian $H_I$, and further constraints the matrix $M(\kk,\qq+\GG)$ in Eq.~(\ref{eq:M-para}). To see this, we note that with the relation (\ref{seq:sewing-C-wf}) due to $C$ symmetry, the definition of $M(\kk,\qq+\GG)$ in Eq.~(\ref{eq:M-def}) satisfies (written as a matrix in the $n$, $\eta$ space)
\begin{equation}\label{seq:M-C-constraint0}
\begin{split}
&M_{m,n}^{\left(\eta\right)}\left(\mathbf{k},\mathbf{q}+\mathbf{G}\right)=\sum_{\alpha}\sum_{\substack{\mathbf{Q}\in\mathcal{Q}_{\pm}}
}u_{\mathbf{Q}-\mathbf{G},\alpha;m\eta}^{*}\left(\mathbf{k}+\mathbf{q}\right)u_{\mathbf{Q},\alpha;n\eta}\left(\mathbf{k}\right) \\
&=\sum_{\alpha}\sum_{\substack{\mathbf{Q}\in\mathcal{Q}_{\pm}}
} [u_{m\eta}^{\dag}\left(\mathbf{k}+\mathbf{q}\right)D^\dag(C)]_{\mathbf{Q}-\mathbf{G},\alpha} [D(C)u_{n\eta}\left(\mathbf{k}\right)]_{\mathbf{Q},\alpha} \\
&=[B^{C}(\kk+\qq)^\dag]_{m\eta,m'\eta'} \sum_{\alpha}\sum_{\substack{\mathbf{Q}\in\mathcal{Q}_{\pm}}
}u_{\mathbf{Q}-\mathbf{G},\alpha;m'\eta'}^{*}\left(\mathbf{k}+\mathbf{q}\right)u_{\mathbf{Q},\alpha;n'\eta'}\left(\mathbf{k}\right) [B^{C}(\kk)]_{n'\eta',n\eta}\ ,
\end{split}
\end{equation}
 or in matrix form,
 \begin{equation}\label{seq:M-C-constraint}
M(\kk,\qq+\GG)=B^{C}(\kk+\qq)^\dag M(\kk,\qq+\GG) B^{C}(\kk)\ .
 \end{equation}
 We note that Eq.~(\ref{seq:M-C-constraint}) is independent of gauge fixings. If we take the gauge fixed form of $M(\kk,\qq+\GG)$ in Eq.~(\ref{eq:M-para}), and the gauge fixing of $C$ in Eq.~(\ref{seq:sewing-C}), we find $M(\kk,\qq+\GG)$ has to commute with $\zeta^y\tau^z$. Thus when there is the chiral symmetry $C$, the gauge fixed $M(\kk,\qq+\GG)$ has to take the form
\begin{equation} \boxed{
M(\kk,\qq+\GG) = \zeta^0\tau^0 \alpha_0(\kk,\qq+\GG) + i\zeta^y\tau^0 \alpha_2(\kk,\qq+\GG).  } \label{eq:M-para-chiral}
\end{equation}
In particular, the functions $\alpha_1(\kk,\qq+\GG)=\alpha_3(\kk,\qq+\GG)=0$. 

By Eqs. (\ref{seq:CcC}) and (\ref{seq:M-C-constraint}), it is easy to see that
\begin{equation}
\begin{split}
&CO_{\qq,\GG}C^{-1}=\sum_{\mathbf{k}\eta s}\sum_{m,n=\pm1} \sqrt{V(\mathbf{q+G})} M_{m,n}^{\left(\eta\right)} \left(\mathbf{k},\mathbf{q}+\mathbf{G}\right) \left(Cc^\dagger_{\kk+\qq,m,\eta,s} c_{\kk,n,\eta,s}C^{-1}-\frac{1}{2}\delta_{\mathbf{q,0}}\delta_{m,n}\right)\\
&=\sum_{\mathbf{k}\eta s}\sum_{m,n=\pm1} \sqrt{V(\mathbf{q+G})} \left([B^{C}(\kk+\qq)^\dag M(\kk,\qq+\GG) B^{C}(\kk)]_{m\eta;n\eta} c^\dagger_{\kk+\qq,m,\eta,s} c_{\kk,n,\eta,s}-\frac{1}{2}M_{m,n}^{\left(\eta\right)} \left(\mathbf{k},\mathbf{q}+\mathbf{G}\right)\delta_{\mathbf{q,0}}\delta_{m,n}\right)\\
&=\sum_{\mathbf{k}\eta s}\sum_{m,n=\pm1} \sqrt{V(\mathbf{q+G})} M_{m,n}^{\left(\eta\right)} \left(\mathbf{k},\mathbf{q}+\mathbf{G}\right) \left(c^\dagger_{\kk+\qq,m,\eta,s} c_{\kk,n,\eta,s}-\frac{1}{2}\delta_{\mathbf{q,0}}\delta_{m,n}\right)=O_{\qq,\GG}\ .
\end{split}
\end{equation}
Therefore, $[C,O_{\qq,\GG}]=0$, and accordingly the projected interaction $H_I$ satisfies
\begin{equation}
[C,H_I]=0\ ,
\end{equation}
implying $C$ is a symmetry of $H_I$.

The $C$ symmetry allows us to define the following operator as a commuting symmetry of $H_I$:
\beq\label{seq:S'z0}
S'^{z0}=\sum_{\kk,s} \sum_{nn^\pr \eta \eta^\pr} [B^{C}(\kk)]_{n \eta, n^\pr \eta^\pr} 
c^\dg_{\kk,n,\eta,s} c_{\kk,n^\pr,\eta^\pr,s}=\sum_{\kk,s} \sum_{nn^\pr \eta \eta^\pr} [\zeta^y\tau^z]_{n \eta, n^\pr \eta^\pr} 
c^\dg_{\kk,n,\eta,s} c_{\kk,n^\pr,\eta^\pr,s}\ ,
\eeq
where we have gauge fixed its representation by Eq.~(\ref{seq:sewing-C}). We note that when $S^{y0}$ acts on single-electron states $c^{\dag}_{\kk,n,\eta,s}|0\rangle$ where $|0\rangle$ is the vacuum, it is the same as the operation of $C$. To see this is a symmetry, we note that
\begin{equation}
[S'^{z0},O_{\qq,\GG}]=\sum_{\kk,s} \sum_{nn^\pr \eta \eta^\pr} \left([\zeta^y\tau^z,M(\kk,\qq+\GG)]\right)_{n \eta, n^\pr \eta^\pr} 
c^\dg_{\kk,n,\eta,s} c_{\kk,n^\pr,\eta^\pr,s}=0\ .
\end{equation}
Therefore, $S'^{z0}$ is a commuting symmetry of the interaction Hamiltonian $H_I$ in Eq.~(\ref{seq-pHI}), namely,
\begin{equation}
[S'^{z0},H_I]=0\ .
\end{equation}
Note that $S'^{z0}$ does not commute with the single-particle Hamiltonian $H_0$ unless $\ee_{n,\eta}(\kk)=\ee_{-n,\eta}(\kk)$. Due to Eq.~(\ref{seq:C-H0energy}), this is only possible when $\ee_{n,\eta}(\kk)=0$, namely, in the exact flat band limit $H_0=0$.

Recall that $H_I$ already has a U(4) symmetry generated by $S^{ab}$ in Eq.~(\ref{eq:U4-generator0}). The commutation of $S^{ab}$ with $S'^{z0}$ then produces another 16 Hermitian operators:
\begin{equation}\label{eq:U4U4-generator0}
S'^{ a b}=\sum_{\mathbf{k},m,\eta,s;n,\eta',s'} (s'^{ab})_{m,\eta,s;n,\eta',s'}c_{\mathbf{k},m,\eta,s}^\dag c_{\mathbf{k},n,\eta',s'}\ ,\qquad (a, b=0,x,y,z)\ ,
\end{equation}
where for each pair of bands $n=\pm n_\text{B}$
\begin{equation}\label{eq:U4U4-generator1}
s'^{ab}=\{\zeta^y\tau^0 s^b,\ \zeta^0\tau^x s^b,\ \zeta^0\tau^y s^b,\ \zeta^y\tau^z s^b\}, \qquad (a, b=0,x,y,z)\ .
\end{equation}
In summary, the single-particle representation matrices of all the generators $S^{ab}$ and $S'^{ab}$ can be reorganized into
\begin{equation}\label{eq:U4U4-generator2}
\boxed{
\{ \zeta^0 \tau^a s^b\ ,\quad \zeta^y \tau^a s^b\ \}\ ,\qquad (a, b=0,x,y,z)\ .}
\end{equation}

It is more convenient to linear combine the U(4)$\times$U(4) generators as
\begin{equation}
S^{ab}_{\pm}=\sum_{\mathbf{k},m,\eta,s;n,\eta',s'} (s_\pm^{ab})_{m,\eta,s;n,\eta',s'}c_{\mathbf{k},m,\eta,s}^\dag c_{\mathbf{k},n,\eta',s'}\ ,
\end{equation}
where we define
\begin{equation} \boxed{
s^{ab}_\pm=\frac{1}{2}(\zeta^0\pm\zeta^y) \tau^a s^b\ , \qquad
(a,b=0,x,y,z).} \label{eq:U4U4-generator}
\end{equation}
In this form, it is easier to see that the 16 generators $S_+^{ab}$ generates one U(4), the 16 generators $S_-^{ab}$ generates another U(4), and $[S_+^{ab},S_-^{cd}]=0$. Therefore, in total they give a U(4)$\times$U(4) symmetry in the first chiral-flat limit.

We note that the U(4) group in the nonchiral-flat limit in Eq.~(\ref{eq:U4-generator}) is a subgroup of the U(4)$\times$U(4) group in the first chiral-flat limit in Eq.~(\ref{eq:U4U4-generator}), but it is not one of the two tensor-producted U(4) groups.

The Cartan subalgebra of the first chiral-flat $U(4)\times U(4)$ generators in Eq.~(\ref{eq:U4U4-generator2}) can be chosen as:
\begin{eqnarray}
& \text{Cartan of first Chiral } U(4) \times U(4): 
\nonumber \\ &\zeta^0 \tau^0s^0, \quad  \zeta^0 \tau^0 s^z, \quad \zeta^0 \tau^z s^0,\quad  \zeta^0 \tau^z s^z,\quad \zeta^y \tau^0 s^0,\quad  \zeta^y \tau^0 s^z,\quad  \zeta^y \tau^z s^0,\quad \zeta^y \tau^z s^z
\end{eqnarray}

\subsubsection{The single-electron irreps}

The irreps of the U(4)$\times$U(4) group can be obtained by the tensor product of the irreps of the first U(4) and the second U(4), respectively. We shall use 
\begin{equation}
([\lambda_1]_4,[\lambda_2]_4) 
\end{equation}
to represent a U(4)$\times$U(4) irrep which is the tensor product of an irrep $[\lambda_1]_4$ of the first U(4) and an irrep $[\lambda_2]_4$ of the second U(4).

At each momentum $\kk$, the $\kk$-independent representation matrices in Eq.~(\ref{eq:U4U4-generator}) can be decomposed into U(4)$\times$U(4) irreps. This can be done again by transforming into a new basis where $\zeta^y$ is diagonalized, which is exactly the irrep band basis $d^{(n_\text{B})\dag}_{\kk,e_Y,\eta,s}=\frac{1}{\sqrt{2}}(c^\dag_{\kk,+n_\text{B},\eta,s}+ie_Yc^\dag_{\kk,-n_\text{B},\eta,s})$ we defined earlier in Eq.~(\ref{eq-irrepbasis}), where $e_Y=\pm1$ gives the irrep number of the band basis. The single-electron state in irrep band $e_Y$
\begin{equation}\label{seq-1electron-2}
d^{(n_\text{B})\dag}_{\kk,e_Y,\eta,s}|0\rangle 
\end{equation}
has eigenvalue $\zeta^y=e_Y$. It is then easy to see that the representation matrices of the U(4)$\times$U(4) generators $S_\pm^{ab}$ for the single-electron state (\ref{seq-1electron-2}) are given by the $4\times4$ matrices
\begin{equation}\label{seq-Sabpm}
\boxed{ s_\pm^{ab}=\frac{1}{2}\left(1\pm e_Y\right)\tau^a s^b\ . }
\end{equation}
Therefore, the single-electron state (\ref{seq-1electron-2}) for a fixed $e_Y$, or equivalently the irrep band fermion operator $d^{(n_\text{B})\dag}_{\kk,e_Y,\eta,s}$ for a fixed $e_Y$, occupies an irrep of the U(4)$\times$U(4) group. The U(4)$\times$U(4) irrep of $d^{(n_\text{B})\dag}_{\kk,+1,\eta,s}$ is given by $([1]_4,[0]_4)$, while the U(4)$\times$U(4) irrep of $d^\dag_{\kk,-1,\eta,s}$ is $([0]_4,[1]_4)$, where we recall that $[1]_4$ and $[0]_4$ are the 4-dimensional fundamental irrep and the 1-dimensional trivial identity irrep of U(4) group, respectively.

We also note that the operator $O_{\qq,\GG}$ in the first chiral limit can be rewritten under irrep band basis as
\begin{equation}\label{eq:chiral-OqG}
O_{\mathbf{q,G}}=\sum_{\mathbf{k},e_Y,\eta,s} \sqrt{V(\mathbf{G}+\mathbf{q})} [M_{e_Y}(\kk,\qq+\GG)]_{n_\text{B},n_\text{B}'} \sum_{\eta, s}\left(d^{(n_\text{B})\dagger}_{\kk+\qq,e_Y,\eta,s} d_{\kk,e_Y,\eta,s}^{(n_\text{B}')}-\frac{1}{2}\delta_{\mathbf{q,0}}\delta_{n_\text{B},n_\text{B}'}\right)\ ,
\end{equation}
where
\begin{equation}
M_{e_Y}(\kk,\qq+\GG)=\alpha_0 \left(\mathbf{k},\mathbf{q}+\mathbf{G}\right)+ie_Y\alpha_2 \left(\mathbf{k},\mathbf{q}+\mathbf{G}\right)\ .
\end{equation}
Therefore, the interaction $H_I$ in Eq. (\ref{eq-pHI}) is diagonal in the index $e_Y$.

For many-body Fock states created by multiple $d^{(n_\text{B})\dag}_{\kk,e_Y,\eta,s}$, the U(4)$\times$U(4) representation is given by the tensor product of the U(4)$\times$U(4) irreps $([1]_4,[0]_4)$ or $([0]_4,[1]_4)$ of each $d^{(n_\text{B})\dag}_{\kk,e_Y,\eta,s}$. Such tensor product representations can be further decomposed into U(4)$\times$U(4) irreps, which will be discussed in a separate paper \cite{ourpaper4}.

\subsection{U(4) symmetry in the (first) chiral-nonflat limit}\label{app:chiral-nonflat}
We have seen that the first chiral-flat limit has a U(4)$\times$U(4) symmetry in the projected Hamiltonian $H=H_I$. Here we show that if $w_0=0<w_1$ but $H_0\neq0$, which we define as the nonchiral-flat limit, there is still a remaining U(4) symmetry.

\subsubsection{The symmetry}\label{app:symmetry}

When $H_0\neq0$, namely, when $\epsilon_{n\eta}(\kk)$ is not constantly zero, we have $H=H_0+H_I$, and neither $C_{2z}P$ nor $C$ is a commuting symmetry of $H$. However, their combination $CC_{2z}P$ is still a commuting symmetry, namely, 
\begin{equation}
[CC_{2z}P,H]=[CC_{2z}P,H_0]+[CC_{2z}P,H_I]=0\ .
\end{equation}
Therefore, the symmetry is still enhanced compared to the nonchiral-nonflat case. This can be most easily seen as follows: among the 32 generators in Eq.~(\ref{eq:U4U4-generator2}), only those with a single-particle representation matrix proportional $\zeta^0$ is still a symmetry when $H_0\neq0$. This is because the kinetic Hamiltonian in the first chiral limit (denoted by $H_0^+$) can be written as
\begin{equation}
H_0=H_0^+=\sum_{\mathbf{k}} \epsilon_{|n|,\eta}(\mathbf{k}) (\zeta^z\tau^0 s^0)_{m,\eta,s;n,\eta',s'}c_{\mathbf{k},m,\eta,s}^\dag c_{\mathbf{k},n,\eta',s'}\ ,
\end{equation}
where we have used the constraint $\epsilon_{n,\eta}(\mathbf{k})=-\epsilon_{-n,\eta}(\mathbf{k})$ due to the chiral symmetry $C$. It is then clear that the generators in Eq.~(\ref{eq:U4U4-generator2}) proportional to $\zeta^y$ will flip the pair of single-particle bands $n=\pm n_\text{B}$, and do not commute with $H_0$. Therefore, we are left with $16$ generators commuting with $H=H_0+H_I$. We redefine their notations as follows:
\begin{equation}
\widetilde{S}^{ab}=\sum_{\mathbf{k},m,\eta,s;n,\eta',s'} (\tilde{s}^{ab})_{m,\eta,s;n,\eta',s'}c_{\mathbf{k},m,\eta,s}^\dag c_{\mathbf{k},n,\eta',s'}\ ,
\end{equation}
where for each pair of bands $n=\pm n_\text{B}$
\begin{equation}\label{eq:U4-generator-chiral-nonflat}
\tilde{s}^{ab}=\boxed{ \zeta^0 \tau^a s^b,\qquad (a,b=0,x,y,z) \ .}
\end{equation}
They form the generators of a U(4) symmetry group. In particular, $\zeta^0 \tau^x s^0$ is the sewing matrix of $iCC_{2z}P$.

We note that this U(4) symmetry group in the first chiral-nonflat limit is different from the U(4) symmetry group in the nonchiral-flat limit (Eq.~(\ref{eq:U4-generator})). Here the generators $\widetilde{S}^{ab}$ are simply the full unitary rotations in the valley-spin space, while the band space is not transformed.

\subsubsection{The single-electron irreps}

Since the generators in Eq.~(\ref{eq:U4-generator-chiral-nonflat}) is proportional to $\zeta^0$, any fixed band basis of all valleys and spins form a fundamental U(4) irrep. For example, we still consider the single-electron state in the irrep band basis
\begin{equation}\label{seq-1electron-3}
d^{(n_\text{B})\dag}_{\kk,e_Y,\eta,s}|0\rangle \ .
\end{equation}
For a fixed $\kk$ and $e_Y$, the states in Eq.~(\ref{seq-1electron-3}) occupies a fundamental irrep $[1]_4$ of the first chiral-nonflat U(4), and the representation matrices of the generators are given by 
\begin{equation}
\tau^as^b \ ,\qquad (a,b=0,x,y,z)
\end{equation}
for either $e_Y=\pm1$. Similarly, the many-body Fock states created by $d^{(n_\text{B})\dag}_{\kk,e_Y,\eta,s}$ are given by the tensor product of the fundamental irreps $[1]_4$ of each $d^{(n_\text{B})\dag}_{\kk,e_Y,\eta,s}$ \cite{ourpaper4}.

\subsection{U(4)$\times$U(4) symmetry in the second chiral-flat limit}\label{app:2ndchiral-flat}

We now consider an opposite limit where $w_1=0<w_0$, which we define as the second chiral limit. Although this limit is far from experimental reality, and the band structure contains no flat bands over the full MBZ (but they are flat in some directions of the MBZ) and is a perfect metal (see Fig. \ref{fig:band}, proof is given in Ref.~\cite{ourpaper2}), the interaction Hamiltonian enjoys a enhanced U(4)$\times$U(4) symmetry of different physical origin from the first chiral limit. One cannot help but hope there is some hidden duality in the TBG problem.

\subsubsection{The second chiral symmetry}\label{app:2nd-c-symmetry}

When $w_1=0$, we can define a second chiral transformation $C'$, which anti-commutes with the full single-particle Hamiltonian $\hat{H}_0$ in Eq.~(\ref{eq:H0}):
\begin{equation}
\{C',\hat{H}_0\}=0\ .
\end{equation}
The operation of $C'$ is given by
\beq
C' c^\dagger_{\kk,\QQ,\eta,\alpha,s} C'^{-1} = \sum_{\QQ^\pr \eta^\pr \beta} 
[D(C')]_{\QQ^\pr \eta^\pr \beta, \QQ \eta \alpha} c^\dagger_{\kk,\QQ^\pr,\eta^\pr,\beta,s}\ ,
\eeq
with the representation matrix
\beq
\quad[D(C')]_{\QQ^\pr \eta^\pr \beta, \QQ \eta \alpha} = \zeta_\QQ\delta_{\QQ^\pr,\QQ} \delta_{\eta^\pr, \eta} (\sigma_z)_{\beta,\alpha}\ , \label{eq:P1}
\eeq
where $\zeta_{\QQ}=\pm1$ for $\QQ\in\mathcal{Q}_{\pm}$. Note that $C'$ preserves the electron momentum $\kk$. 
Since $C'$ flips the single-particle Hamiltonian $\hat{H}_0$, it is not a commuting symmetry of TBG, but only reflects a relation between the positive and negative energy spectra. The transformation $C'$ satisfies
\beq
C'^2=1,\qquad
[C',C_{2z}] = 0,\qquad
\{C',T\} = 0,\qquad
\{C',P\}=0, \qquad
\{C',C_{2z}T\}=0,\qquad
\{C',C_{2z}P\}=0.
\eeq

\subsubsection{The full symmetry}\label{app:2nd-full-symmetry}
When transformed into the energy band basis, the second chiral symmetry $C'$ implies
\begin{equation}\label{seq:P1-H0energy}
\ee_{n,\eta}(\kk)=-\ee_{-n,\eta}(\kk)\ ,
\end{equation}
\begin{equation}\label{seq:sewing-P1-wf}
[D(C')]_{\eta\eta'} u_{n\eta'}(\kk)= \sum_{m}[B^{C'}(\kk)]_{m\eta,n\eta'}u_{m\eta}(\kk)\ ,
\end{equation}
where
\begin{equation}\label{seq:sewing-P10}
[B^{C'}(\kk)]_{m\eta,n\eta'}=\delta_{\eta,\eta'}\delta_{-m,n} e^{i\varphi^{C'}_{n,\eta'}(\kk)}
\end{equation}
This implies the transformation
\begin{equation}\label{seq:P1cP1}
C' c_{\mathbf{k},n,\eta' ,s}^{\dagger} C'^{-1}= \sum_{m\eta} [B^{C'}(\kk)]_{m\eta,n\eta'} c_{\mathbf{k},m,\eta, s}^{\dagger}\ .
\end{equation}
By the relations $\{C',C_{2z}T\}=\{C',C_{2z}P\}=0$, the sewing matrix of $C'$ satisfies
\beq\label{seq:sewing-P11}
B^{C'}(\kk) B^{C_{2z}T}(\kk) = - B^{C_{2z}T}(\kk) B^{C'*}(\kk)\ ,\qquad B^{C'}(\kk) B^{C_{2z}P}(\kk) = - B^{C_{2z}P}(\kk) B^{C'}(\kk)\ .
\eeq
Note that this constraint for the sewing matrix of $C'$ is exactly the same as that for $C$ in Eq.~(\ref{seq:sewing-C1}). Therefore, within each pair of bands $n=\pm n_\text{B}$, if we impose the gauge fixing of Eq.~(\ref{eq:gauge-0}), we similarly find the only consistent $\kk$-independent gauge for sewing matrix of $C'$ is (up to a global minus sign)
\begin{equation}\label{seq:sewing-P1}
B^{C'}(\kk)=\zeta^y\tau^z\ .
\end{equation}
This $\kk$-independent gauge fixing (\ref{seq:sewing-P1}) of $C'$ also automatically ensures the continuous gauge fixing condition (\ref{seq:c-continuous}), which is crucial for defining the irrep band basis in Eq.~(\ref{eq-irrepbasis}). This is because Eq.~(\ref{seq:sewing-P1}) tells us that $u_{-n,\eta}(\kk)=i\text{sgn}(n)\eta u_{n,\eta}(\kk)$ for band $n=\pm n_\text{B}$, which implies
\begin{equation}
f_{n,\eta}(\kk+\qq,\kk)= \left|u^\dag_{n,\eta}(\kk+\qq) u_{n,\eta}(\kk)- u^\dag_{-n,\eta}(\kk+\qq) u_{-n,\eta}(\kk)\right|=\left|u^\dag_{n,\eta}(\kk+\qq) u_{n,\eta}(\kk)[1-\text{sgn}(n)^2\eta^2]\right|=0
\end{equation}
for any $\kk$ and $\qq$, satisfying Eq.~(\ref{seq:c-continuous}).

However, the gauge fixing of $C'$ in Eq.~(\ref{seq:sewing-P1}) is inconsistent with the $\kk$-independent gauge fixings of both $C_{2z}$ and $P$ separately in Eq.~(\ref{eq:gauge-1}). This is because $[C',C_{2z}]=0$ and $\{C',P\}=0$ requires
\begin{equation}
B^{C'}(-\kk) B^{C_{2z}}(\kk) =  B^{C_{2z}}(\kk) B^{C'}(\kk)\ , \qquad B^{C'}(-\kk) B^{P}(\kk) = B^{P}(\kk) B^{C'}(\kk)\ ,
\end{equation}
which is, however, not satisfied by the simultaneous gauge fixings of Eqs. (\ref{seq:sewing-P1}) and (\ref{eq:gauge-1}). If we fix the sewing matrix of $C'$ to be $\kk$-independent as given in Eq.~(\ref{seq:sewing-P1}), the sewing matrices of $C_{2z}$ and $P$ have to be $\kk$ dependent. In this appendix we shall choose the gauge fixing of Eq.~(\ref{seq:sewing-P1}) and give up the separate gauge fixing of $C_{2z}$ and $P$ in Eq.~(\ref{eq:gauge-1}), since only their combination $C_{2z}P$ is used for the U(4) symmetries discussed here. 

However, we note that if a momentum $\kk$ is $C_{2z}$ invariant (the $\Gamma_M$ point and the 3 $M_M$ points in MBZ), the above gauge fixing problem appears to imply the absence of well-defined sewing matrices of $C_{2z}$ and $P$. In fact, this is because at $w_1=0$, the TBG band structure is protected to be doubly degenerate at $C_{2z}$ invariant momenta, which leads to a perfect metal (see Fig. \ref{fig:band}, and see \cite{ourpaper2} for proof). Therefore, the pair of bands $n=\pm n_\text{B}$ are connected with the other bands at $\Gamma_M$ and $M_M$ points, where the projection within the two bands $n=\pm n_\text{B}$ is ill-defined. The sewing matrices of $C_{2z}$ and $P$ at such $C_{2z}$ invariant momenta can only be written down when the additional degenerate states at these momenta  from other bands are included. We shall not discuss this matter here, since we will not use the sewing matrices of $C_{2z}$ and $P$ in this appendix.

Nevertheless, we note that one could fix the gauge of $C_{2z}$ and $P$ in a simple $\kk$ dependent way, provided $\kk$ is not a $C_{2z}$ invariant point. First, we divide all the $C_{2z}$-non-invariant $\kk$ points into two sets $\mathcal{K}_1,\mathcal{K}_2$ related by $C_{2z}$, namely, $C_{2z}\mathcal{K}_1=\mathcal{K}_2$. For instance, $\mathcal{K}_1$ and $\mathcal{K}_2$ can be two half-MBZs related by $C_{2z}$. Then we can fix the sewing matrices of $C_{2z}$ and $P$ (and $T$ given that $C_{2z}$ and $C_{2z}T$ are fixed) at $C_{2z}$-non-invariant $\kk$ within each pair of bands $n=\pm n_\text{B}$ as
\begin{equation}\label{eq-k-dependent-fixing}
B^{C_{2z}}(\kk)=(-1)^j\zeta^0\tau^x\ ,\qquad B^{T}(\kk)=-(-1)^j\zeta^0\tau^x\ ,\qquad B^{P}(\kk)=i(-1)^j\zeta^y\tau^z\ ,\qquad (\text{for  }\kk\in\mathcal{K}_j).
\end{equation}

Since the gauge fixed sewing matrix of $C'$ in Eq.~(\ref{seq:sewing-P1}) is exactly the same as that of $C$ in Eq.~(\ref{seq:sewing-C}), we can follow a similar derivation as that from Eqs. (\ref{seq:M-C-constraint0}) to (\ref{eq:U4U4-generator}), which gives us the following.

First, $C'$ is a symmetry of $H_I$ satisfying $[C',H_I]=0$, and the $M(\kk,\qq+\GG)$ matrix is restricted to have the form
\begin{equation} \boxed{
M(\kk,\qq+\GG) = \zeta^0\tau^0 \alpha_0(\kk,\qq+\GG) + i\zeta^y\tau^0 \alpha_2(\kk,\qq+\GG).  } \label{eq:M-para-2chiral}
\end{equation}
The Hermitian condition of the $M(\kk,\qq+\GG)$ (\cref{eq:Mcond-herm}) requires that 
\begin{equation}
\alpha_0(\kk,\qq+\GG) = \alpha_0^T(\kk+\qq,-\qq-\GG),\qquad
\alpha_2(\kk,\qq+\GG) =-\alpha_2^T(\kk+\qq,-\qq-\GG).
\end{equation}
For $\qq=0$, the $B^{C_{2z}}(\kk)$ sewing matrix implies $M^{\eta}(\kk,\GG)=M^{-\eta} (-\kk,-\GG)$ and hence 
\begin{equation}
\alpha_0(\kk,\GG) = \alpha_0(-\kk,-\GG),\qquad \alpha_2(\kk,\GG) = \alpha_2(-\kk,-\GG).
\end{equation}
Combining the above two constraints, we obtain
\begin{equation}
\alpha_0(\kk,\GG) = \alpha_0^T(-\kk,\GG),\qquad 
\alpha_2(\kk,\GG) =-\alpha_2^T(-\kk,\GG).
\end{equation}

Second, the $C'$ symmetry yields a U(4)$\times$U(4) symmetry with generators
\begin{equation}
S'^{ab}_{\pm}=\sum_{\mathbf{k},m,\eta,s;n,\eta',s'} (s_\pm'^{ab})_{m,\eta,s;n,\eta',s'}c_{\mathbf{k},m,\eta,s}^\dag c_{\mathbf{k},n,\eta',s'}\ ,
\end{equation}
where we define
\begin{equation} \boxed{
s'^{ab}_\pm=\frac{1}{2}(\zeta^0\pm\zeta^y) \tau^a s^b\ , \qquad
(a,b=0,x,y,z).} \label{eq:U4U4-2-generator}
\end{equation}
We note, however, although these generators take the same gauge-fixed form 
as those in the first chiral-flat limit (Eq.~(\ref{eq:U4U4-2-generator})), their physical origins are different: here the U(4)$\times$U(4) generators are generated by the sewing matrix of the second chiral symmetry $C'$, while in the first chiral-flat limit, the U(4)$\times$U(4) generators are generated by the sewing matrix of the first chiral symmetry $C$.

\subsubsection{The single-electron irreps}

We have shown that under the gauge fixings (\ref{eq:gauge-0}) and (\ref{seq:sewing-P1}), the U(4)$\times$U(4) generators of the second chiral-flat limit is exactly the same as that of the first chiral-flat limit. Therefore, exactly parallel to the first chiral-flat limit, the single-electron irreps in the second chiral-flat limit are given by the irrep band basis $d^{(n_\text{B})\dag}_{\kk,e_Y,\eta,s}=\frac{1}{\sqrt{2}}(c^\dag_{\kk,+n_\text{B},\eta,s}+ie_Yc^\dag_{\kk,-n_\text{B},\eta,s})$ we defined earlier in Eq.~(\ref{eq-irrepbasis}), where $e_Y=\pm1$ gives the irrep number of the band basis. The single-electron state
\begin{equation}\label{seq-1electron-4}
d^{(n_\text{B})\dag}_{\kk,e_Y,\eta,s}|0\rangle 
\end{equation}
with a fixed $e_Y$ and $\kk$ occupies a U(4)$\times$U(4) irrep of $([1]_4,[0]_4)$ if $e_Y=+1$, and $([0]_4,[1]_4)$ if $e_Y=-1$. The representation matrices of the U(4)$\times$U(4) generators $S_\pm^{ab}$ for the single-electron state (\ref{seq-1electron-2}) are given by the $4\times4$ matrices
\begin{equation}\label{seq-Sabpmpr}
\boxed{ s_\pm'^{ab}=\frac{1}{2}\left(1\pm e_Y\right)\tau^a s^b\ . }
\end{equation}

However, in this second chiral limit, we note that the basis $d^\dag_{\kk,e_Y,\eta,s}=d^{(1)\dag}_{\kk,e_Y,\eta,s}$ when $n_\text{B}=1$ no longer give a well-defined Chern band in the MBZ with a definite Chern number as illustrated in Sec. \ref{app:chern}, since the lowest two bands $n=\pm1$ are gapless with the higher bands when $w_1=0$ (see Fig. \ref{fig:band}). Neither are the bands flat, possibly giving rise to interesting, gapless phases.

\subsection{U(4) symmetry in the second chiral-nonflat limit}\label{app:2nd-c-nf}

If $w_1=0<w_0$, taking into account the kinetic term $H_0\neq0$, we are still left with a U(4) symmetry. We call this limit the second chiral-nonflat limit. Since the U(4)$\times$U(4) generators in the second chiral-flat limit are exactly the same as those in the first chiral-flat limit, the case here is mathematically exactly the same as the first chiral-nonflat limit in App.~\ref{app:chiral-nonflat}. Therefore, we conclude that the second chiral-nonflat limit has a remaining U(4) symmetry with generators given by
\begin{equation}
\widetilde{S}'^{ab}=\sum_{\mathbf{k},m,\eta,s;n,\eta',s'} (\tilde{s}'^{ab})_{m,\eta,s;n,\eta',s'}c_{\mathbf{k},m,\eta,s}^\dag c_{\mathbf{k},n,\eta',s'}\ ,
\end{equation}
where within each pair of bands $n=\pm n_\text{B}$
\begin{equation}\label{eq:U4-generator-2chiral-nonflat}
\tilde{s}'^{ab}=\boxed{ \zeta^0 \tau^a s^b,\qquad (a,b=0,x,y,z) \ ,}
\end{equation}
under the gauge fixings of Eqs. (\ref{eq:gauge-0}) and (\ref{seq:sewing-P1}). The only difference is that here the U(4) symmetry is generated by the sewing matrix of $iC'C_{2z}P$, which reads $\zeta^0 \tau^x s^0$.

This second chiral-nonflat limit is more physical, since when $w_1=0<w_0$, the bands are never too flat (Fig. \ref{fig:band}).

\section{The stabilizer Code Limit}\label{sec:stabilizer}

The projected interacting Hamiltonian in Eq.~(\ref{eq-pHI}) is generically a quantum Hamiltonian, where the terms $O_{-\qq,-\GG}O_{\qq,\GG}$ do not commute, since the commutator $[O_{\mathbf{q,G}},O_{\mathbf{q',G'}}]$ given in Eq. (\ref{eq:OqG-commutator}) 
does not vanish for generic form factors (overlaps) $M_{m,n}^{\left(\eta\right)}\left(\mathbf{k},\mathbf{q}+\mathbf{G}\right)$. Thus, although it gives a PSDH, which allows us to find exact ground states at certain fillings in the flat band limit (for which $H=H_I$) as we will demonstrate in a separate paper \cite{ourpaper4}, it is impossible to analytically solve all the many-body eigenstates of $H=H_I$. 

However, in the case where we are projecting only into the 8 lowest $n=\pm1$ bands (i.e., $n_\text{max}=1$), in the first (or second) chiral-flat limit $w_0=0$ (or $w_1=0$) and $H_0=0$, if we further has $M_{m,n}^{\left(\eta\right)}\left(\mathbf{k},\mathbf{q+G}\right)$ being independent of $\mathbf{k}$, we would have $[O_{\mathbf{q,G}},O_{\mathbf{q',G'}}]=0$. We call this limit the \emph{stabilizer code limit}: 
\begin{equation}
\text{stabilizer code limit}\quad= \text{1st/2nd chiral-flat limit}\quad +\quad  \text{$\kk$-independent form factors } M(\kk,\qq+\GG).
\end{equation}
Indeed, Eq. (\ref{eq:M-para-chiral}) or (\ref{eq:M-para-2chiral}), and our $\kk$-independent assumption lead to a $\kk$-independent form factor matrix
\begin{equation}
M(\kk,\qq+\GG) = M(\mathbf{0},\qq+\GG) = \zeta^0\tau^0 \alpha_0(\mathbf{0},\qq+\GG) + i\zeta^y\tau^0 \alpha_2(\mathbf{0},\qq+\GG) .
\end{equation}
In particular, if $n_\text{max}=1$, both $\alpha_0(\mathbf{0},\qq+\GG)$ and $\alpha_2(\mathbf{0},\qq+\GG)$ are not matrices but just numbers, thus they commute among each other. 
Therefore by Eq. (\ref{eq:OqG-commutator}), we have
\begin{equation}
[O_{\mathbf{q,G}},O_{\mathbf{q',G'}}]\propto M(\kk+\qq',\qq+\GG)M(\kk,\qq'+\GG')- M(\kk+\qq,\qq'+\GG') M(\kk,\qq+\GG)=0\ .
\end{equation}
This yields a Hamiltonian similar to a stabilizer code Hamiltonian
\begin{equation}
H=H_I=\frac{1}{2\Omega_{\text{tot}}}\sum_{\mathbf{q}\in\text{MBZ}}\sum_{\mathbf{G}\in\mathcal{Q}_0} O_{\mathbf{-q,-G}} O_{\mathbf{q,G}}\ ,\label{eq:stabilizerham}
\end{equation}
where all the terms commute:
\begin{equation}
[O_{\mathbf{-q,-G}} O_{\mathbf{q,G}},O_{\mathbf{-q',-G'}} O_{\mathbf{q',G'}}]=0\ .
\end{equation}
Therefore, all the terms $O_{\mathbf{-q,-G}} O_{\mathbf{q,G}}$ can be simultaneously diagonalized, which makes all the many-body eigenstates of the Hamiltonian exactly solvable. Note that Eq.~\ref{eq:stabilizerham} is not strictly a stabilizer code Hamiltonian since the terms $O_{\mathbf{-q,-G}} O_{\mathbf{q,G}}$ do not have a spectrum equal to $0$ or $1$ (moreover their spectrum depends on $\mathbf{q}$ and $\mathbf{G}$). Nevertheless, Eq.~\ref{eq:stabilizerham} has the crucial feature that makes the spectrum of a stabilizer code solvable (namely a sum of commuting operators), thus its name.

As we will prove in Ref.~\cite{ourpaper4}, the Hamiltonian $H=H_I$ in the stabilizer code limit is an extended Hubbard model with extended interactions and zero hoppings. Therefore, although far from physical, the stabilizer code limit provides a Hubbard-model understanding of the TBG physics, as suggested by the recent experimental observations \cite{wong_cascade_2020,zondiner_cascade_2020}.

We will solve the stabilizer code limit Hamiltonian in Ref.~\cite{ourpaper4}.

\section{Comparison with the U(4) symmetry of Ref.  \cite{kang_strong_2019}}\label{app:kang-vafekU4}

In this appendix, we discuss the interaction Hamiltonian of Ref.~\cite{kang_strong_2019}, and compare with ours. In Ref.~\cite{kang_strong_2019}, Kang and Vafek were the first to show the appearance of a U(4) approximate symmetry in their Hamiltonian, which is a type of PSDH obtained by projecting into a Wannier basis.

\subsection{The Wannier gauge}

The $s=\up\down$ sectors are related by an SU(2) rotation.
Thus we only need to construct Wannier functions in the $s=\up$ sector, the Wannier functions in the $s=\down$ sector can then be symmetrically generated.
Before we introduce the Wannier functions, let us first write the Bloch states of TBG as linear combintations of plane waves
\beq
\ket{\psi_{\kk,n,\eta}} = \frac{1}{\sqrt{N}} \sum_{\QQ\in\mcl{Q}_\pm} \sum_{\RR \alpha} u_{\QQ,\alpha;n\eta}(\kk) e^{i(\kk+\QQ)\cdot(\RR+\tt_\alpha)} \ket{\RR \alpha},
\eeq
where summations over $\RR,\alpha$ are limited to sites in the top layer (bottom layer) graphene for $\QQ\in\mcl{Q}_+$ ($\QQ\in\mcl{Q}_-$), $\ket{\RR \alpha}$ is the atomic orbital at $\RR +\tt_\alpha$, and $N$ is the number of unit cells in each of the two graphene layers.
Generally, the Wannier functions are linear combinations of the Bloch states
\begin{equation}
\ket{w_{\RR_M,\mu,\eta}} = \frac{1}{\sqrt{N_M}} \sum_{\kk}  e^{-i\kk\cdot\RR_M} \ket{\td{\psi}_{\kk,\mu,\eta}},\qquad 
    \ket{\td{\psi}_{\kk,\mu,\eta}} = \sum_{n=\pm1}\ket{\psi_{\kk,n,\eta}}  W_{n,\mu}^{\eta}(\kk),
\end{equation}
where $N_M$ is the number of moir\'e unit cells and $W^\eta(\kk)$ at each $\kk$ is a two-by-two matrix.
We denote the center of the Wannier function $\ket{w_{\RR_M,\mu,\eta}} $ as $\RR_M+\tt_{M,\mu}$, with
$\RR_M= l_1 \mbf{a}_{M1} + l_2 \mbf{a}_{M2}$ being a moir\'e lattice and $\tt_{M,\mu}$ ($\mu=1,2$) the sublattice vectors. 
Here we take the unit cell basis as $\aa_{M1}=(0,-1)$ and $\aa_{M2}=(\frac{\sqrt3}2,\frac12)$.
Notice that the Bloch states are periodic in momentum space, so the transformation coefficient is $e^{-i\kk\cdot\RR}$ rather than $e^{-i\kk\cdot(\RR+\tt_\mu)}$.
The sewing matrices of $\ket{\td{\psi}_{\kk,\mu,\eta}}$ are defined as 
\beq
B^g_{\mu\eta, \nu \eta^\pr}(\kk) = \bra{\td{\psi}_{g\kk,\mu,\eta}} g \ket{\td{\psi}_{\kk,\nu,\eta^\pr}}.
\eeq
The two bands in each valley have a fragile topology protected by $C_{2z}T$ symmetry and a stable topology protected by the $PC_{2z}T$ symmetry \cite{ourpaper2}. 
Thus, in order to obtain the Wannier functions, we have to abandon smooth $C_{2z}T$ gauge and smooth $P$ gauge of $\ket{\td{\psi}_{\kk,\mu,\eta}}$.
It is possible to choose smooth gauge for the remaining symmetries $g=C_{3z}, C_{2y}, T$ since they do not protect a topology.
According to Kang and Vafek \cite{kang_symmetry_2018}, the two Wannier states ($\mu=1,2$) in each valley locate at the honeycomb lattice, \ie $\tt_{M,1}=\frac13 \mbf{a}_{M,1}+\frac23 \mbf{a}_{M,2}=(\frac1{\sqrt3},0)$, and $\tt_{M,2}=-\tt_{M,1}$, and one can choose the Wannier functions to satisfy
\begin{equation}
C_{3z}^\pr \ket{w_{\RR_M,\mu,\eta}} = e^{i\eta \frac{2\pi}3} \ket{w_{\RR^\pr_M,\mu,\eta}},\qquad 
(\RR^\pr_M+\tt_{M,\mu} = C_{3z}^\pr (\RR_M+\tt_{M,\mu})),
\end{equation}
\begin{equation}
C_{2y} \ket{w_{\RR_M,\mu,\eta}} = \sum_{\beta} \gamma^x_{\nu \alpha} \tau^x_{\eta^\pr,\eta}\ket{w_{\RR^\pr_M,\nu,\eta^\pr}},\qquad 
(\RR^\pr_M+\tt_{M,\nu} = C_{2y} (\RR_M+\tt_{M,\mu})),
\end{equation}
\begin{equation}
T \ket{w_{\RR_M,\mu,\eta}} = \ket{w_{\RR_M,\mu,-\eta}},
\end{equation}
where $\gamma^x$ is the first Pauli matrix in the moir\'e sublattice space, and $\tau^x$ is the first Pauli matrix in the valley space.
Kang and Vafek's $\ket{w_{1,2,3,4}}$ are our $\ket{w_{\RR_M,1,+}}$, $\ket{w_{\RR_M,1,-}}$, $\ket{w_{\RR_M,2,-}}$, $\ket{w_{\RR_M,2,+K}}$, respectively.
Here we have used $C_{3z}'$ to represent the $2\pi/3$ rotation microscopically centered at honeycomb vertex of graphene.
In this work, We use $C_{3z}$ to denote the $2\pi/3$ rotation microscopically centered at the honeycomb center of graphene.
One should notice that the $C_{3z}^\pr$ eigenvalues, which are $e^{i\eta\frac{2\pi}3}$ at $\Gamma_M$, are different from the $C_{3z}$ in the BM model, which are $1$ at $\Gamma_M$ \cite{song_all_2019}.
We will discuss the relation between $C_{3z}$ and $C_{3z}^\pr$ in the end of this subsection.

The sewing matrices of $C_{3z}, C_{2y}, T$ on $\ket{\td{\psi}_{\kk,\mu,\eta}}$ can be obtained from the actions of $C_{3z}, C_{2y}, T$ on the Wannier functions.
We have
\begin{align}
C_{3z}^\pr \ket{\td{\psi}_{\kk,\mu,\eta}} =& \frac1{\sqrt{N_M}} \sum_{\RR_M} 
	e^{i\kk\cdot\RR_M}  C_{3z}^\pr\ket{W_{\RR_M,\mu,\eta}}
= e^{i\eta\frac{2\pi}3} \frac1{\sqrt{N_M}} \sum_{\RR_M} 
	e^{i\kk\cdot\RR_M}  \ket{W_{\RR^\pr_M,\mu,\eta}} \nono\\
=& e^{i\eta\frac{2\pi}3} \frac1{\sqrt{N_M}} \sum_{\RR^\pr_M} 
e^{i\kk\cdot(C_{3z}^{\pr-1}\RR^\pr_M + C_{3z}^{\pr-1}\tt_{M,\mu}- \tt_{M,\mu} )}  \ket{W_{\RR^\pr_M,\mu,\eta}}\nono\\
=& e^{i\eta\frac{2\pi}3} e^{i(C_{3z}^\pr\kk-\kk)\cdot\tt_{M,\mu}} \ket{\td{\psi}_{C_{3z}^\pr\kk,\mu,\eta}},
\end{align}
where $\RR_M'=C_{3z}'( \RR_M + \tt_{M,\mu} ) - \tt_{M,\mu}$.
Thus the $C_{3z}'$ sewing matrix is 
\begin{equation}
B^{C_{3z}^\pr}_{\mu\eta,\nu\eta^\pr}(\kk) = \delta_{\mu\nu} \delta_{\eta\eta^\pr} 
	e^{i\eta\frac{2\pi}3} e^{i(C_{3z}\kk-\kk)\cdot\tt_\mu}.\label{eq:Wannier-C3p}
\end{equation}
We also have
\begin{align}
C_{2y} \ket{\td{\psi}_{\kk,\mu,\eta}} =& \frac1{\sqrt{N_M}} \sum_{\RR_M} 
	e^{i\kk\cdot\RR_M}  C_{2y} \ket{w_{\RR_M,\mu,\eta}}
= \sum_{\beta\eta^\pr} \gamma^{x}_{\nu,\mu} \tau^x_{\eta^\pr \eta} 
	\frac1{\sqrt{N_M}} \sum_{\RR_M} 
	e^{i\kk\cdot\RR_M}  \ket{w_{\RR^\pr_M,\nu,\eta^\pr}} \nono\\
=& \sum_{\beta\eta^\pr} \gamma^{x}_{\beta,\alpha} \tau^x_{\eta^\pr \eta} 
	\frac1{\sqrt{N_M}} \sum_{\RR^\pr_M} 
	e^{i\kk\cdot(C_{2y}^{-1}\RR^\pr_M + C_{2y}^{-1}\tt_{M,\nu}- \tt_{M,\mu} )}
	\ket{w_{\RR^\pr_M,\nu,\eta^\pr}}\nono\\
=& \sum_{\nu\eta^\pr} \gamma^{x}_{\nu,\mu} \tau^x_{\eta^\pr \eta} 
 	\ket{\td{\psi}_{C_{2y}\kk,\mu,\eta^\pr}},
 \end{align}
where $C_{2y}\tt_{M,\nu} = \tt_{M,\mu}$ and $\RR^\pr_M = C_{2y} (\RR_M+\tt_{M,\mu})-\tt_{M,\nu}$.
Thus the $C_{2y}$ sewing matrix is
\begin{equation}\boxed{
B^{C_{2y}}_{\mu\eta,\nu\eta^\pr}(\kk) = \gamma^x_{\mu\nu}\tau^x_{\eta\eta^\pr}. }\label{eq:Wannier-C2y}
\end{equation}
For the time-reversal, we have
\begin{equation}
T \ket{\td{\psi}_{\kk,\mu,\eta}} = \frac1{\sqrt{N_M}} \sum_{\RR_M} 
	e^{-i\kk\cdot\RR_M}  \ket{w_{\RR,\mu,-\eta}} = \ket{\td{\psi}_{-\kk,\mu,-\eta}}.
\end{equation}
Thus the time-reversal sewing matrix is
\begin{equation}\boxed{	
B^T_{\mu\eta,\nu\eta^\pr}(\kk) = \delta_{\mu\nu} \tau^x_{\eta,\eta^\pr}.} \label{eq:Wannier-T}
\end{equation}

In this gauge, the $C_{2z}T$ sewing matrix $B^{C_{2z}T}(\kk)$ and the $P$ sewing matrix $B^{(P)}(\kk)$ must have be discontinuous at some momenta due to the topology protected by $C_{2z}T$ and/or $PC_{2z}T$ of the two lowest bands.
Correspondingly, in the Wannier basis the $C_{2z}T$ and $P$ representations must be non-local. 
Usually, an Wannier function at $\mbf{r}$ ($\mbf{r}\neq \mbf{0}$) would be transformed to another Wannier function at $-\mbf{r}$ under $C_{2z}T$ or $P$.
In the non-local case, the Wannier function at $\mbf{r}$ ($\mbf{r}\neq \mbf{0}$) will be transformed to a linear combination of all the Wannier functions in the whole 2D space under $C_{2z}T$ or $P$.
Any tight-binding model in this Wannier representation that have finite-range hopping will break the $C_{2z}T$ and the $P$ symmetries.

\paragraph{Re-choice of $C_{3z}$ center}
We find that the $C_{3z}^\pr$ operation, which is a $2\pi/3$ rotation at the honeycomb vertex of graphene, is the $2\pi/3$ centered at honeycomb center of graphene followed by a microscopic translation, \ie $C_{3z}^\pr = \{1|-\mbf{a}_1\} C_{3z}$.
Here $\mbf{a}_1$ is the lattice basis of single layer graphene.
The microscopic model of TBG cannot have both $C_{3z}^\pr$ and $C_{3z}$.
For example, we choose the twisting center at the honeycomb center, then $C_{3z}$ is an exact symmetry but $C_{3z}^\pr$ is only an approximate symmetry; however, the microscopic error of $C_{3z}^\pr$ should be negligible, diminishing at small angle.
The translation $-\mbf{a}_1$ will leads to factors $e^{i\frac{2\pi}3}$ and $e^{-i\frac{2\pi}3}$ for the two valleys $K$ and $K^\pr$, respectively.
Thus the representation matrix of $C_{3z}^\pr$ in the BM model is given by
\begin{equation}
D(C_{3z}^\pr) = e^{i\frac{2\pi}3\tau^z} D(C_{3z}),
\end{equation}
where $D(C_{3z})$ is given by \cref{eq:C3}.
Thus $C_{3z}$ acts on the Wannier functions as
\begin{equation}
C_{3z} \ket{w_{\RR_M,\mu,\eta}} =  \ket{w_{\RR^\pr_M,\mu,\eta}},\qquad 
(\RR^\pr_M+\tt_{M,\mu} = C_{3z} (\RR_M+\tt_{M,\mu})).
\end{equation}
It follows that the $C_{3z}$ sewing matrix is
\begin{equation}\boxed{
B^{C_{3z}}_{\mu\eta,\nu\eta^\pr}(\kk) = \delta_{\mu\nu} \delta_{\eta\eta^\pr} 
	 e^{i(C_{3z}\kk-\kk)\cdot\tt_\mu}.} \label{eq:Wannier-C3}
\end{equation} 
Notice that the $C_{2y}$ axis of Kang and Vafek's model is same as ours, so we do not need to change the $C_{2y}$ sewing matrix.

\subsection{Interaction}
Now that we have implement the Kang and Vafek Wannier symmetries, we transform their interaction \cite{kang_strong_2019} into momentum space.
Let us denote the fermion annihilation operator of the Wannier states as $c_{\RR_M,\mu,\eta,s}$.
Then the Kang-Vafek interaction has the form
\begin{equation}
H_I = \frac{V_0}2 \sum_{\RR_M} O_{\RR_M} O_{\RR_M},
\end{equation}
\begin{equation}
O_{\RR_M} = \frac13 Q_{\RR_M} + \kappa T_{\RR_M}
\end{equation}
where $\RR_M$ sums over all the lattice vectors (honeycomb centers), and $Q_{\RR_M}$ and $T_{\RR_M}$ are given by
\begin{equation}
Q_{\RR_M} = \sum_{\eta,s} \sum_{j\in \varhexagon} c_{\RR_M+\dd_{M,j} - \tt_{M,[j]}, [j],\eta,s}^\dagger c_{\RR_M+\dd_{M,j}-\tt_{M,[j]}, [j],\eta,s},
\end{equation}
\begin{equation}
T_{\RR_M} = \sum_{\eta,s} \sum_{j\in \varhexagon}
\left( (-1)^{j-1} e^{i (-1)^{j-1} \eta  \vartheta} 
 	c_{\RR_M+\dd_{M,j+1}-\tt_{M,[j+1]}, [j+1],\eta,s}^\dagger c_{\RR_M+\dd_{M,j}-\tt_{M,[j]}, [j],\eta,s} + h.c.\right).
\end{equation}
Here $j$ sums over the six hexagon vertex around the triangle site $\RR_M$, $[j]=j\;\rm mod\; 2$ is the sublattice index, $\vartheta$ is a phase factor.
The vectors are given by $\dd_{M,1}= \tt_{M,1}$, $\dd_{M,4}=\tt_{M,2}=-\tt_{M,1}$, $\dd_{M,j+2} = C_{3z} \dd_{M,j}$.
To match our convention of Wannier functions, we have decomposed the position ($\RR_M+\dd_{M,j}$) of the operator $c_{\RR_M+\dd_{M,j} - \tt_{M,[j]}, [j],\eta,s}^\dagger$ into a lattice vector $\RR_M+\dd_{M,j} - \tt_{M,[j]}$ and a sublattice vector $\tt_{M,[j]}$.  
$Q_{\RR_M}$ is the total charge on the six vertices of the honeycomb centered at $\RR_M$.
$T_{\RR_M}$ is a hopping-like term where each term annihilates an electron at the vertex $j$ and create an electron at the vertex $j+1$ or $j-1$.
The phase factor associated with the hopping is $e^{i\eta\vartheta}$ if $j=1,3,5$ and is $-e^{-j\eta\vartheta}$ if $j=2,4,6$.
$\kappa$ is a factor determining the strength of the hopping-like term and is estimated as 0.16 in Ref. \cite{kang_strong_2019}.
($\kappa$ is originally denoted as $\alpha_1$ in Ref. \cite{kang_strong_2019}. We changed the notation to avoid confusion with the $\alpha_1(\kk,\qq+\GG)$ function (\cref{eq:M-para}) in this manuscript.)
We can write $O_{\RR_M}$ as
\begin{align}
O_{\RR_M} =&  \sum_{\eta,s} \sum_{j\in \varhexagon} \frac13 c_{\RR_M+\dd_{M,j} - \tt_{M,[j]}, [j],\eta,s}^\dagger c_{\RR_M+\dd_{M,j}-\tt_{M,[j]}, [j],\eta,s} \nono\\
&\quad + 
    \kappa\left( (-1)^{j-1} e^{i (-1)^{j-1} \eta  \vartheta} 
 	c_{\RR_M+\dd_{M,j+1}-\tt_{M,[j+1]}, [j+1],\eta,s}^\dagger c_{\RR_M+\dd_{M,j}-\tt_{M,[j]}, [j],\eta,s} + h.c.\right).
\end{align}
Now we apply the transformation
\begin{equation}
O_{\RR_M} = \frac1{N_M} \sum_{\qq} e^{-i\qq\cdot\RR_M} O_{\qq}
\end{equation}
such that the interaction can be written as our interaction form (already present in Vafek and Kang)
\begin{equation}
H_I = \frac{V_0}{2N_M} \sum_{\qq} O_{-\qq} O_\qq.
\end{equation}
We transform the three terms, \ie the $Q_{\RR_M}$ term, the first term in $T_{\RR_M}$ term, and the second term in $T_{\RR_M}$, in $O_{\RR_M}$ one by one.
First we have
\begin{align}
O_{\qq}^1 =& \frac1{N_M} \sum_{\RR_M} e^{i\RR_M\cdot\qq} O_{\RR_M}^1 
= \frac1{3N_M} \sum_{\RR_M} e^{i\RR_M\cdot\qq} \sum_{\eta,s} 
	\sum_{j\in \varhexagon} c_{\RR_M+\dd_{M,j} - \tt_{M,[j]}, [j],\eta,s}^\dagger 
	c_{\RR_M+\dd_{M,j}-\tt_{M,[j]}, [j],\eta,s} \nono\\
=& \frac1{3N_M} \sum_{\RR_M} e^{i\RR_M\cdot\qq} \sum_{\eta,s} \sum_{j\in \varhexagon} 
	\sum_{\pp\kk} e^{-i(\RR_M+\dd_{M,j} - \tt_{M,[j]})\cdot \pp} e^{i(\RR_M+\dd_{M,j} -\tt_{M,[j]})\cdot\kk}
	c^\dg_{\pp,[j],\eta,s} c_{\kk,[j],\eta,s} \nono\\
=& \frac13 \sum_{\eta s} \sum_{j\in \varhexagon } \sum_{\kk} c^\dg_{\kk+\qq,[j],\eta,s} c_{\kk,[j],\eta,s} 
= \sum_{\mu \eta s} \sum_{\kk} c^\dg_{\kk+\qq,\mu,\eta,s} c_{\kk,\mu,\eta,s} .
\end{align}
Second we have
\begin{align}
O_\qq^2 =& \frac{\kappa}{N_M} \sum_{\RR_M} e^{i\RR_M\cdot\qq} \sum_{\eta,s} 
	\sum_{j\in \varhexagon} (-1)^{j-1} e^{i (-1)^{j-1} \eta  \vartheta} 
	c_{\RR_M+\dd_{M,j+1} - \tt_{M,[j+1]}, [j+1],\eta,s}^\dagger 
	c_{\RR_M+\dd_{M,j}-\tt_{M,[j]}, [j],\eta,s} \nono\\
=& \frac{\kappa}{N_M} \sum_{\RR_M} e^{i\RR_M\cdot\qq} \sum_{\eta,s} 
	\sum_{j\in \varhexagon}  (-1)^{j-1} e^{i (-1)^{j-1} \eta  \vartheta}  \sum_{\pp\kk}
	e^{-i(\RR_M+\dd_{M,j+1} - \tt_{M,[j+1]})\cdot \pp} e^{i(\RR_M+\dd_{M,j}-\tt_{M,[j]}) \cdot \kk}
	c_{\pp, [j+1],\eta,s}^\dagger 
	c_{\kk, [j],\eta,s} \nono\\
=& \kappa\sum_{\eta,s} \sum_{j\in \varhexagon}  (-1)^{j-1} e^{i (-1)^{j-1} \eta  \vartheta}  \sum_{\kk}
	e^{-i(\dd_{M,j+1} - \tt_{M,[j+1]})\cdot (\kk+\qq)} e^{i(\dd_{M,j}-\tt_{M,[j]}) \cdot \kk}
	c_{\kk+\qq, [j+1],\eta,s}^\dagger 
	c_{\kk, [j],\eta,s}.
\end{align}
We split the summation $\sum_{j\in\varhexagon}$ into $\sum_{j=1,3,5}$ and $\sum_{j=2,4,6}$, then
\begin{align}
O_\qq^2 =& \kappa\sum_{\eta,s} \sum_{j=1,3,5}  e^{i \eta \vartheta}  \sum_{\kk}
 	e^{i(-\dd_{M,j+1}+\tt_{M,2})\cdot\qq} e^{i(\dd_{M,j}-\tt_{M,1} - \dd_{M,j+1} + \tt_{M,2}) \cdot \kk}
	c_{\kk+\qq,2,\eta,s}^\dagger 
	c_{\kk, 1,\eta,s} \nono\\
-& \kappa \sum_{\eta,s} \sum_{j=2,4,6}  e^{-i \eta \vartheta}  \sum_{\kk}
	e^{i(-\dd_{M,j+1}+\tt_{M,1})\cdot\qq} e^{i(\dd_{M,j}-\tt_{M,2} - \dd_{M,j+1} + \tt_{M,1}) \cdot \kk}
	c_{\kk+\qq,1,\eta,s}^\dagger 
	c_{\kk, 2,\eta,s}.
\end{align}
Since $\tt_{M,1} = -\tt_{M,2}$ and $\dd_{M,j+3}=-\dd_{M,j}$, the phase factors of the second term are the complex conjugations of those of the first term, thus we can rewrite $O^2_\qq$ as
\begin{equation}
O^2_\qq = \kappa \sum_{\eta s} \sum_{\kk} e^{i\eta\vartheta} \omega(\kk,\qq) c_{\kk+\qq,2,\eta,s}^\dagger c_{\kk, 1,\eta,s}
	-e^{-i\eta\vartheta} \omega^*(\kk,\qq) c_{\kk+\qq,1,\eta,s}^\dagger c_{\kk, 2,\eta,s}
\end{equation}
with 
\begin{equation}
\omega(\kk,\qq) = \sum_{j=1,3,5} 
	e^{i(-\dd_{M,j+1}+\tt_{M,2})\cdot\qq} e^{i(\dd_{M,j}-\tt_{M,1} - \dd_{M,j+1} + \tt_{M,2}) \cdot \kk}.
\end{equation}
Now we list all the involved vectors in the phase factors ($\tt_{M,1}=\dd_{M,1}$, $\tt_{M,2}=\dd_{M,4}$):
\begin{equation}
j=1,\qquad -\dd_{M,j+1}+\dd_{M,4}=-{\mbf{a}}_{M2},\qquad \dd_{M,j} - \dd_{M,1} -\dd_{M,j+1}+\dd_{M,4}=-{\mbf{a}}_{M2}.
\end{equation}
\begin{equation}
j=3, \qquad -\dd_{M,j+1}+\dd_{M,4}=0,\qquad \dd_{M,j} - \dd_{M,1} -\dd_{M,j+1}+\dd_{M,4}=-\mbf{a}_{M1}-\mbf{a}_{M2}.
\end{equation}
\begin{equation}
j=5, \qquad -\dd_{M,j+1}+\dd_{M,4}=-\td{\aa}_{1}-\mbf{a}_{M2},\qquad \dd_j - \dd_1 -\dd_{M,j+1}+\dd_{M,4}=-\mbf{a}_{M1}-2\mbf{a}_{M2}.
\end{equation}
Thus we have
\begin{equation}\boxed{
\omega(\kk,\qq) = e^{-i\mbf{a}_{M2}\cdot\qq} e^{-i\mbf{a}_{M2}\cdot\kk}
	+ e^{-i(\mbf{a}_{M1}+\mbf{a}_{M2})\cdot\kk}
	+ e^{-i(\mbf{a}_{M1}+\mbf{a}_{M2})\cdot\qq} e^{i(\mbf{a}_{M1}-2\mbf{a}_{M2})\cdot\kk}. }
\end{equation}
Since the third term in $O_{\RR_M}$ is the Hermitian conjugation of the second term, we have
\begin{align}
O^3_\qq =& Q^{2\dagger}_{-\qq} = \kappa\sum_{\eta s}\sum_{\kk} \pare{ 
	- e^{i\eta\vartheta} \omega(\kk,-\qq) c_{\kk,2,\eta,s}^\dg c_{\kk-\qq,1,\eta,s}  
	+ e^{-i\eta\vartheta} \omega^*(\kk,-\qq) c_{\kk,1,\eta,s}^\dg c_{\kk-\qq,2,\eta,s} } \nono\\
=& \kappa\sum_{\eta s}\sum_{\kk} \pare{ 
	- e^{i\eta\vartheta} \omega(\kk+\qq,-\qq) c_{\kk+\qq,2,\eta,s}^\dg c_{\kk,1,\eta,s}  
	+ e^{-i\eta\vartheta} \omega^*(\kk+\qq,-\qq) c_{\kk+\qq,1,\eta,s}^\dg c_{\kk,2,\eta,s} }
\end{align}
We define
\begin{equation}\boxed{
\beta(\kk,\qq) = \omega(\kk,\qq) - \omega(\kk+\qq,-\qq) .}
\end{equation}
Thus $O^2_\qq +O^3_\qq $ can be written as
\begin{equation}
O^2_\qq + O^3_\qq= \sum_{\eta s} \sum_{\kk} e^{i\eta\vartheta} \beta(\kk,\qq) c_{\kk+\qq,2,\eta,s}^\dagger c_{\kk, 1,\eta,s}
	-e^{-i\eta\vartheta} \beta^*(\kk,\qq) c_{\kk+\qq,1,\eta,s}^\dagger c_{\kk, 2,\eta,s}.
\end{equation}

Now we write the total $O_\qq$ operator as
\begin{equation}
O_{\qq} = \sum_{\eta s} \sum_{\kk} M_{\mu,\nu}^\eta(\kk,\qq) c_{\kk+\qq,\mu,\eta,s}^\dagger
	c_{\kk,\nu,\eta,s},
\end{equation}
where $M$ is
\begin{equation}
M^\eta(\kk,\qq) = \gamma^0 - i \kappa \gamma^y \Re[ e^{i\eta\vartheta} \beta(\kk,\qq)] + i\kappa \gamma^x \Im[ e^{i\eta\vartheta} \beta(\kk,\qq) ]. \label{eq:M-wannier}
\end{equation}
We can also express $M$ as a 4 by 4 matrix (in the sublattice and valley spaces) as
{\small
\begin{align}
M(\kk,\qq) =& \gamma^0 \tau^0 
	- i \kappa \gamma^y \Re[ (\cos\vartheta \tau^0 + i \sin\vartheta \tau^z) \beta(\kk,\qq)] 
	+ i \kappa \gamma^x \Im[ (\cos\vartheta \tau^0 + i \sin\vartheta \tau^z)  \beta(\kk,\qq)] \nono\\
=& \gamma^0 \tau^0 
 -  i \kappa \gamma^y \tau^0 \Re[ \cos\vartheta \beta(\kk,\qq)]
 +  i \kappa \gamma^x \tau^0 \Im[ \cos\vartheta \beta(\kk,\qq)]
 +  i \kappa \gamma^y \tau^z \Im[ \sin\vartheta \beta(\kk,\qq)]
 +  i \kappa \gamma^x \tau^z \Re[ \sin\vartheta \beta(\kk,\qq)]. \label{eq:Wannier-M}
\end{align}}
With this, we have brought the Kang-Vafek interaction to the same form as our momentum-space interactions. 

\subsection{The Kang-Vafek U(4) symmetry}
It is obvious that the Kang-Vafek interaction have spin-valley U(2)$\times$U(2) symmetry, whose generators are
\begin{equation}
	\gamma^0 \tau^0 s^a, \qquad
	\gamma^0 \tau^z s^a, \qquad a=0,x,y,z.
\end{equation}
Now we show that it indeed has a U(4) symmetry. Our proof is the momentum-space version of the original proof \cite{kang_strong_2019}.
We introduce two matrices
\begin{equation}
\Sigma^x = \gamma^x\tau^0 \cos\vartheta + \gamma^y \tau^z \sin\vartheta,\qquad
\Sigma^y = -\gamma^x\tau^z \sin\vartheta + \gamma^y \tau^0 \cos\vartheta,
\end{equation}
and rewrite the $M$ matrix as
\begin{equation}
M(\kk,\qq) = \gamma^0\tau^0 +i \kappa \Sigma^x \Im[\beta(\kk,\qq)] + i \kappa \Sigma^y \Re[\beta(\kk,\qq)].
\end{equation}
One can verify that $\{\Sigma^x,\Sigma^y\}=0$.
We then apply a \textit{$\kk$-independent} gauge transformation $e^{i\frac{\vartheta}2\gamma^z\tau^z}$ such that $ e^{i\frac{\vartheta}2\gamma^z\tau^z} \Sigma^x e^{-i\frac{\vartheta}2\gamma^z\tau^z}=\gamma^x\tau^0$ and $e^{i\frac{\vartheta}2\gamma^z\tau^z} \Sigma^y e^{-i\frac{\vartheta}2\gamma^z\tau^z} = \gamma^y\tau^0$.
After the transformation, $M$ becomes 
\begin{equation}
M(\kk,\qq) = \gamma^0\tau^0 +i \kappa \gamma^x \tau^0 \Im[\beta(\kk,\qq)] - i \kappa  \gamma^y \tau^0 \Re[\beta(\kk,\qq)].
\end{equation}
Therefore, the $M$ matrix is invariant under the U(4) generators
\begin{equation}
 \gamma^0 \tau^a s^b,\qquad a,b=0,x,y,z. \label{eq:U4-Wannier1}
\end{equation}
This gauge transformation seems equivalent to setting $\vartheta=0$ \cite{kang_strong_2019}.
However, after the gauge transformation, the sewing matrices might change.

\subsection{ Relation between Kang-Vafek U(4) and the $C_{2z}P$-implied U(4) symmetry}
Let us first fix the $C_{2z}P$ gauge of the Wannier functions.
According to \cref{seq:P-com}, we have
\begin{equation} \label{commutationrelationsforkangvafek1}
(C_{2z}P)^2=1,\qquad
[C_{3z},C_{2z}P] = 0,\qquad
[C_{2y},C_{2z}P] = 0,\qquad
\{T,C_{2z}P\}=0,
\end{equation}
and hence
\begin{equation}
(B^{C_{2z}P}(\kk) )^2=1,\qquad
B^{C_{3z}}(\kk) B^{C_{2z}P}(\kk) = B^{C_{2z}P}(C_{3z} \kk) B^{C_{3z}}(\kk)
\end{equation}
\begin{equation}
B^{C_{2y}}(\kk) B^{C_{2z}P}(\kk) = B^{C_{2z}P}(C_{2y} \kk) B^{C_{2y}}(\kk),\qquad
B^{T}(\kk) B^{C_{2z}P*}(\kk) = -B^{C_{2z}P}(-\kk) B^{T}(\kk)
\end{equation}
Since both $T$ and $C_{2z}TP$ (the charge-conjugation) are local in real space, as shown in App.~\ref{app:manybody-cc}, $C_{2z}P$ must also be a local operator in real space. 
Thus we want $C_{2z}P$ to be local in the Wannier representation.
However, this is incompatible with the crystalline and time-reversal symmetries.
In order to be local in the Wannier representation, $C_{2z}P$ must leave the center of each Wannier function invariant and hence will be $\kk$-independent.
Since $C_{2z}P$ does not change the sublattice,  the sewing matrix $B^{C_{2z}P}(\kk)$ should be diagonal in the sublattice index and thus does not contain $\gamma^{x,y}$ terms.
Since $C_{2z}P$ changes valley, it must not contain $\tau^0$ and $\tau^z$.
Thus  $B^{C_{2z}P}$ can only have four possible terms, \ie $\gamma^{0,z} \tau^{x,y} $.
All the four terms commute with $B^{C_{3z}}(\kk)$, which only contains the terms $\gamma^{0,z}\tau^0$.
In order to commute with $B^{C_{2y}}(\kk)$ ($\gamma^x\tau^x$), only $\gamma^0\tau^x$ and $\gamma^z \tau^y$ are possible.
However, both commute with $T=\tau^xK$, whereas \cref{commutationrelationsforkangvafek1} shows that  ${C_{2z}P}$  anti-commutes with $T$.
Thus a local representation of $C_{2z}P$ is not compatible with the $C_{3z}, C_{2y}, T$ symmetries.
In other words, $C_{2z}P$ in the Wannier representation that respects  $C_{3z}, C_{2y}, T$ symmetries must be non-local.

The above analysis leads to two conclusions: (i) our charge-conjugation symmetry must be non-local in the Kang-Vafek Wannier representation, (ii) Kang and Vafek U(4) symmetry, which is local in the Wannier representation, is not equivalent to the ${C_{2z}P}$-implied U(4) symmetry, which is non-local in their Wannier representation.

\subsection{ Kang-Vafek U(4) as our U(4) chiral-nonflat limit symmetry }

We now ask:  is the Kang and Vafek U(4) consistent with our U(4) implied by the $C_{2z}PC$? We assume that the Kang and Vafek  model (at least approximately) preserves the $C_{2z}PC$ symmetry.
Here $C$ is the chiral symmetry
\begin{equation}
D(C) h(\kk) D^{-1}(C) = - h(\kk),\;\;\; 
D(C)_{\QQ, \alpha; \QQ', \alpha' }=\delta_{\QQ,\QQ'} \sigma^z_{\alpha \alpha'}, \;\;\; C^2=1.
\end{equation}
Commutations between $C$ and $T,C_{3z},C_{2y}, C_{2z}P$ are
\begin{equation}
[T,C] =0,\qquad
[C_{3z},C]=0,\qquad 
[C_{2y},C]=0,\qquad
\{C,C_{2z}P\}=0.
\end{equation}
Thus we have
\begin{equation}
(C_{2z}PC)^2=-1,\qquad
[C_{3z},C_{2z}PC]=0,\qquad
[C_{2y}, C_{2z}PC]=0,\qquad
\{T,C_{2z}PC\}=0,
\end{equation}
and hence
\begin{equation}
(B^{C_{2z}PC}(\kk) )^2=-1,\qquad
B^{C_{3z}}(\kk) B^{C_{2z}PC}(\kk) = B^{C_{2z}PC}(C_{3z} \kk) B^{C_{3z}}(\kk)
\end{equation}
\begin{equation}
B^{C_{2y}}(\kk) B^{C_{2z}PC}(\kk) = B^{C_{2z}PC}(C_{2y} \kk) B^{C_{2y}}(\kk),\qquad
B^{T}(\kk) B^{C_{2z}PC*}(\kk) = -B^{C_{2z}PC}(-\kk) B^{T}(\kk)
\end{equation}
We try to find a $\kk$-independent solution, which means $C_{2z}PC$ is local in the Wannier representation.
Since $C_{2z}PC$ preserves the sublattice (local) and changes valley, $B^{C_{2z}PC}$ can only have four terms $i\gamma^{0,z}\tau^{x,y}$, each of which squares to -1.
All the four terms commute with $C_{3z}$ ($\gamma^{0,z}$).
Two terms commute with $C_{2y}$ ($\gamma^x\tau^x$): $i\gamma^0\tau^x$, $i\gamma^z\tau^y$.
And the two terms also anti-commute with $T$.
Therefore, there are two solutions of $B^{C_{2z}PC}$:
\begin{equation}
	B^{C_{2z}PC(1)} = i\gamma^0\tau^x,\qquad
	B^{C_{2z}PC(2)} = i\gamma^z\tau^y.
\end{equation}
If we can understand $M$ (\cref{eq:M-wannier}) as the inner product of periodic part of Bloch wave functions, \ie
\begin{equation}
M^\eta_{\mu\nu} (\kk,\qq) \sim \sqrt{V(\qq)} \inn{u_{\kk+\qq,\mu,\eta}|u_{\kk,\nu,\eta}},
\end{equation}
then $M$ must commute with $B^{C_{2z}PC}$. 
Applying $B^{C_{2z}PC(1)}$ and $B^{C_{2z}PC(2)}$ to \cref{eq:Wannier-M}, we obtain $\vartheta=0,\pi$ and $\vartheta=\pm\frac{\pi}2$, respectively.
For $\vartheta=0,\pi$, the U(4) generators are
\begin{equation}
\gamma^0\tau^a s^b,\qquad a,b=0,x,y,z; \label{eq:U4-Wannier11}
\end{equation}
for $\vartheta=\pm\frac{\pi}2$, the U(4) generators are
\begin{equation}
\gamma^z\tau^{x,y} s^a,\qquad \gamma^0\tau^{0,z} s^a,\qquad a=0,x,y,z. \label{eq:U4-Wannier2}
\end{equation}

Now we show that the two representations \cref{eq:U4-Wannier11,eq:U4-Wannier2} are equivalent.
Under the gauge transformation 
$\begin{pmatrix} \tau^0 & 0\\ 0 & i\tau^z \end{pmatrix}$,
\cref{eq:U4-Wannier2} becomes \cref{eq:U4-Wannier11} and $B^{C_{3z}}$, $B^{C_{2y}}$, $B^T$ remain unchanged.

To summarize: (i) The $C_{2z}PC$ can be chosen as local in the Wannier representation.
(ii) If Kang and Vafek's model does not have an exact $C_{2z}PC$ symmetry (which remains to be checked), then, if we continuously recover the $C_{2z}PC$ symmetry, their U(4) continuously changes to our U(4) implied by the $C_{2z}PC$ symmetry; this U(4) is implied by the chiral, non-flat limit. Hence we conjecture that the Kang and Vafek U(4) is also invariant to the addition of \emph{some} kinetic terms. 
(iii) The U(4) symmetry implied by $C_{2z}PC$ is also local in the Wannier representation because the U(2)$\times$U(2) part is already local, and the additional generator is just the $C_{2z}PC$ operation.

If we impose the $C C_{2z} P$ symmetry to the Kang and Vafek's tight-binding model, their U(4) symmetry would become the chiral-nonflat U(4) symmetry, since the two U(4) symmetries share the same generators $\tau^a s^b$ ($a,b=0,x,y,z$).

\end{document}